\documentclass[twocolumn]{mn2e}
\usepackage{mnras_cite}
\usepackage{color}   
\usepackage[dvips]{graphicx}
\usepackage{amsmath}

\newcommand{\beq}{\begin{equation}}
\newcommand{\eeq}{\end{equation}}

\newcommand{\beqa}{\begin{eqnarray}}
\newcommand{\eeqa}{\end{eqnarray}}
\newcommand{\xips}{\xi(\pi,\sigma)}

\newcommand{\nc}{\newcommand}   
\nc{\citealt}{\cite}

\def\vk{{\hbox{\bf k}}}

\def\vx{{\hbox{\bf x}}}

\def\vn{{\hbox{$\hat{{\bf n}}$}}}

\def\Mpc{\, h^{-1} \, {\rm Mpc}}

\def\invMpccube{\, h^{3} \, {\rm Mpc}^{-3}}
\def\Gpccube{\, h^{-3} \, {\rm Gpc}^3}

\def\Msun{\,h^{-1}\,{\rm M_{\odot}}}
\def\ltsima{$\; \buildrel < \over \sim \;$}   
\def\gtsima{$\; \buildrel > \over \sim \;$}   
\def\simlt{\lower.5ex\hbox{\ltsima}}   
\def\simgt{\lower.5ex\hbox{\gtsima}}   
\def\etal{{et al. }}

\input epsf

\begin{document}

\title[Modeling the angular correlation function and its full covariance in Photometric Galaxy Surveys]
{Modeling the angular correlation function and its full covariance in Photometric Galaxy Surveys}

\author[Crocce \etal]{Mart\'in Crocce$^{1}$, Anna Cabr\'e$^{2}$ \&
  Enrique Gazta\~{n}aga$^{1}$ \\ \\
$^{1}$ Institut de Ci\`encies de l'Espai, IEEC-CSIC, Campus UAB, Facultat de
Ci\`encies, Torre C5 par-2, Barcelona 08193, Spain \\
$^{2}$ Center for Particle Cosmology, University of Pennsylvania, 209, South $33^{rd}$ Street, Philadelphia, PA, 19104, USA
}
\maketitle

\begin{abstract}

Near future cosmology will see the advent of wide area photometric
galaxy surveys, like the Dark Energy Survey (DES), that extent to high redshifts ($z\sim 1 - 2$) but with
poor radial distance resolution. In such cases splitting the
data into redshift bins and using the angular
correlation function $w(\theta)$, or the $C_{\ell}$ power spectrum, will become the standard
approach to extract cosmological information or to study the nature of
dark energy through the Baryon Acoustic Oscillations (BAO) probe.
In this work we present a detailed model for $w(\theta)$ at large scales
as a function of redshift and bin width, including all relevant
effects, namely nonlinear gravitational
clustering, bias, redshift space distortions and photo-z uncertainties. 
We also present a model for the full covariance matrix characterizing the angular
correlation measurements, that takes into account the same effects as
for $w(\theta)$ and also the possibility of a shot-noise component and
partial sky coverage. 
Provided with a large volume N-body simulation from the MICE
collaboration we built several ensembles of mock redshift bins with a sky
coverage and depth typical of forthcoming photometric surveys.
The model for the angular correlation and the one for the covariance
matrix agree remarkably well with the mock measurements in all
configurations. The prospects for a full shape analysis of
$w(\theta)$ at BAO scales in forthcoming photometric surveys
such as DES
are thus very encouraging. 
\end{abstract}

\maketitle

\section{Introduction}

The statistical analysis of the distribution of structure at large
astronomical scales has played a key role in advancing the field of
Cosmology over the last 20 years. From shaping our understanding of
complex processes driving galaxy formation and evolution to
constraining the energy density content of the Universe.

The completion of large extra-galactic surveys such at the Sloan
Digital Sky Survey (SDSS, \pcite{york00}) and the 2dF Galaxy Redshift
Survey (2dFGRS, \pcite{colless03}) have bolstered our general knowledge
in the field. Particularly more so when combined with the precise measurements
of the Cosmic Microwave Background or the increasingly reach data from
Supernova data \cite{sanchez09,percival10,reid10,komatsu10}
.
One of the most promising, and eventually rewarding, challenges for
the field of large scale structure today is the prospect for determining
what drives the late time acceleration of the Universe \cite{riess98,perlmutter99}. This
is probed by the presence, in the clustering pattern of galaxies, of 
remanent features from the coupling of baryon and photons prior to 
recombination known as the Baryon Acoustic Oscillations (BAO). The BAO
have already been detected in the spectroscopic samples of Luminous
Red Galaxies (LRGs) in both SDSS and 2dFGRS \cite{cole05,eisenstein05}, and studied in the early
imagining data of SDSS \cite{padmanabhan07}.

But the observational quest has only started.
Several of the next-generation surveys will gain in area and depth, in
exchange for a poorer determination of radial positions. In turn this
imposes the need for angular clustering analysis in redshift bins of
width few times larger that of the photometric error uncertainty at
the given redshifts. 
The difficulty lies in that the projection in redshift bins lowers the
clustering amplitude, erasing any particular feature and increasing
the noise-to-signal ratio. The achievable precision of our photometrically
estimated redshift will play a crucial role. We thus need to
understand what affects the angular clustering pattern more severely.

The aim of our work is to tackle this problem, providing a well
calibrated model for the clustering signal at large-scales as a
function of angle, radial distance and bin width, deepening the
available literature in the subject
(e.g. \pcite{padmanabhan07}, \pcite{blake07} and references therein). 
We put particular effort in stressing the most relevant effects,
redshift distortions and photo-z uncertainties, and how they interplay.

An equally important problem is to have the capability of estimating
the full errors of the measurements. We thus provide a well tested
description  of the complex error matrix characterizing the
measurements of the correlation function in real situations,
i.e. including effects of partial sky coverage, photo-z, redshift
distortions, bias and shot-noise. 

Both, the model for the correlation and the one for the error matrix, 
will be extensively tested against a very rich set of mocks redshift bins.
This work should therefore be relevant for ongoing projects that use
photometric redshift estimates like the Dark Energy Survey\footnote{\tt
  www.darkenergysurvey.org} (DES),  the Physics of the
Accelerating Universe collaboration\footnote{\tt www.pausurvey.org} (PAU) and the
the Panoramic Survey Telescope and Rapid Response System\footnote{\tt pan-stars.ifa.hawaii.edu} (PanStarrs). But also for upcoming imaging
proposals such as the Large Synoptic Survey
Telescope\footnote{\tt www.lsst.org} (LSST) and the
ESA/Euclid\footnote{\tt www.euclid-imaging.net} survey.

This paper is organized as follows. In Sec.~\ref{sec:model} and
Sec.~\ref{sec:errors} we discuss the models proposed in this
work for the angular correlation function and its full error
matrix respectively. In Sec.~\ref{sec:simulations} we describe the set
of ensembles of mock
redshift bins implemented using a large volume N-body simulation.
In Sec.~\ref{sec:mocks.vs.model.I} and \ref{sec:mocks.vs.model.II} we test the models against the mocks, under
different regimes and assumptions. Section~\ref{sec:conclusions} contains our conclusions and future
lines of research. 
We also include several appendixes. In Appendix~\ref{Appendix:A} we
give a
description of our model for the 3-d nonlinear matter correlation
function. In Appendix~\ref{Appendix:B} we study
the limitations of the widely used Limber
approximation. Finally, Appendix~\ref{Appendix:C} gives a brief note on the
covariance of the angular power spectrum induced by partial sky coverage.

\section{An analytic model for the angular correlation function}
\label{sec:model}

Let us start by considering the projection of the spatial 
galaxy fluctuations $\delta_g(\vx,z)$ along a given direction in the sky ${\vn}$
\beq
\delta(\vn) = \int dz \,\phi(z) \, \delta_g(\vn,z),
\label{eq:projected-density}
\eeq
where $\phi$ is the radial selection function. The angular correlation
function is  then obtained as a simple projection of the 3-d
correlation function $\xi$ \cite{peebles1973},
\begin{eqnarray}
w(\theta) &\equiv& \langle \delta_g(\vn) \delta_g(\vn+\hat{\bf \theta}) \rangle =  \nonumber \\
& = & \int dz_1 \, \phi(z_1) \int dz_2 \, \phi(z_2)
\, \xi_{gg}(r(z_1),r(z_2),\theta)
\label{eq:wtheta1}
\end{eqnarray}
where $\theta$ is the angle between directions $\vn$ and $\vn + \hat{\bf \theta}$, related to
the pair-separation through 
\beq
r_{12}(\theta)=\left\{r(z_1)^2+r(z_2)^2-2 r(z_1)r(z_2)
  \cos(\theta)\right\}^{1/2}
\eeq
and $r(z)$ is the co-moving distance to redshift $z$ given by
\beq
r(z)=\int_0^z \frac{c}{H(u)}du,
\eeq
where $H(z)/H_0=\sqrt{\Omega_m(1+z)^3+\Omega_{\rm DE}(1+z)^{3(1+w)}}$ is
the Hubble parameter,
$\Omega_m$ and $\Omega_{\rm DE}=1-\Omega_m$ are the matter and
dark energy densities respectively, and $w$ is the dark energy equation of
state \footnote{These expressions explicitly assume a flat
  cosmology and constant $w$, for more general cases see \cite{matsubara04}
  and references therein}. 
Since we are interested in redshift bins
and not in extended selections we can neglect the growth evolution
within the bin and
simply evaluate the 3-d correlation in some fiducial redshift $\bar{z}$ (e.g.
the mean redshift of the bin, weighted by $\phi$). 
In addition we will assume a {\it local and linear bias} relation between
fluctuations in the tracer (e.g. galaxies) and matter density field,
$\delta_g=b(z) \delta$ (see Sec.~\ref{sec:biased_tracers} for a justification). Under these assumptions Eq.~(\ref{eq:wtheta1}) is converted to
\beq
w(\theta) = \int dz_1 \, f(z_1) \int dz_2 \, f(z_2) \, \xi(r(z_1),r(z_2),\theta,\bar{z})
\label{eq:wtheta}
\eeq
where $f(z)\equiv \,b(z) \, \phi(z)$ and $\xi$ is the matter 3-d correlation function. 

Hence, in order to predict $w(\theta)$ we need a model for the spatial clustering
accurate in a sufficiently large range of scales to allow the
projection in Eq.~(\ref{eq:wtheta}), in particular when photo-z errors
broadens the extent of the radial distribution (see \cite{sanchez10}
for a empirical parametric fit suited for BAO scales).

In what follows we discuss how to include photo-z effects and the model for
spatial clustering in real and redshift space that we will use throughout this paper.

\subsection{Photo-z}
\label{sec:photo-z}

We incorporate the way uncertainties in the true redshift positions obtained
from photometric estimates affect angular clustering by means of the radial
selection function, following \pcite{budavari03} (see also
\pcite{ma06}).

The radial selection $\phi$ is the probability to include a galaxy
in our redshift bin. If the selection of objects is done according to their
true redshifts, then
$\phi$ is equal to the true number of galaxies per unit
redshift times a window function $W$ encoding selection characteristics (e.g. redshift
cuts),
\beq
\phi(z) = \frac{dN_g}{dz} \, W(z).
\eeq
Instead, if the selection is done according to photometric redshift estimates,
 one must incorporate the probability $P(z|z_p)$ for the true redshift to be
$z$ when the photometric one is $z_p$. The ending result is the
product \cite{budavari03},
\beq
\phi(z) = \frac{dN_g}{dz} \int dz_p P(z|z_p) W(z_p),
\label{eq:phiphotoz}
\eeq
where $W(z_p)$ is the photometric redshift window function. Throughout
this paper we will only consider top-hat window functions both in
true and photometric redshifts (i.e. $W=1$ within a given redshift range,
and $0$ otherwise). In addition we will only consider the idealized case where
the photometric estimate is Gaussianly distributed around the true redshift (e.g. \pcite{ma06}). Although this
might be far from reality, it serves as an interesting starting point
for more realistic scenarios \cite{hearin10,bernstein10}. Lastly, we recall that $\phi$
should be normalized to unity within the redshift range of interest.

\subsection{Spatial clustering and redshift evolution}
\label{sec:nonlinear_matter_clustering}

We now turn into the
discussion of the 3-d matter correlation model accounting for
{\it nonlinear gravitational effects}, {\it redshift space distortions},
and the way we evolve it with redshift.
We postpone to Appendix~\ref{Appendix:A} the testing of this model against 
 measurements of 3-d matter clustering in N-body
simulations.


The linear evolution of the clustering pattern preserves its shape but
increases the overall amplitude. The main effects
due to nonlinear gravitational clustering at large scales are a
smoothing of the BAO wiggles and a rise in clustering amplitude above linear values towards smaller
scales due to mode-coupling effects
\cite{seo05,eisenstein07,crocce08}. Although these processes can be
modeled from first principles
\cite{RPT,matarrese08,matsubara08,taruya09}, it is also possible and desirable to find simpler
parametric approximations. In the correlation function
these two effects can be parameterized as \cite{crocce08},
\beq
\xi(r)= [\xi_{\rm Lin}(r) \otimes {\rm e}^{-(r/s_{bao})^2}](r) +
A_{mc} \,\xi^{(1)}_{\rm Lin}(r) \, \xi^{\prime}_{\rm Lin}(r)
\label{eq:parametric_model}
\eeq
where $s_{bao}$ and $A_{mc}$ are fitting parameters, $\xi_{\rm Lin}$
is the linear correlation function at the given redshift, $\xi_{\rm Lin}^{\prime}$ its derivative and
\beq
\xi^{(1)}_{\rm Lin} \equiv 4 \pi
\int P_{\rm Lin} (k,z) \, j_1(k\,r) k \, dk.
\eeq
In Eq.~(\ref{eq:parametric_model}), the symbol $\otimes$ denotes a convolution. 
This model have been already used in
the analysis of matter, halo and galaxy clustering \cite{sanchez08,sanchez09}.

However the standard approach for analyzing clustering data in a photometric
redshift survey covering from low ($z\sim 0.2$) to high redshift ($z\sim 1.4$) is to divide the data into several redshift bins 
(whose minimum width are ultimately determined by the photo-z
accuracy, e.g. \pcite{padmanabhan07}). If one then performs a joint analysis of all these
bins it is desirable to have the least number of nuisance parameters
possible in order to optimize constraints on derived cosmological
parameters. From this point of view it is interesting to investigate
to what extent a single set of best-fit parameters can be used to describe the
3-d clustering from low to high redshift, and hence the angular clustering
after the projection in Eq.~(\ref{eq:wtheta}).

We implement this as follows. The first term in Eq.~(\ref{eq:parametric_model}) is proportional to
the linear correlation, therefore it scales with the growth factor
squared $\sim D^2(z)$. The second term arises from leading order mode-mode coupling
and thus scales as $\sim D^4(z)$. In turn the damping of BAO is
proportional to the amplitude of large-scale velocity flows,
Eq.~(\ref{eq:sbao}), and so $s_{bao} \sim D(z)$.
Putting these considerations together we scale our parametric model
with redshift as,
\begin{eqnarray}
\xi(r,z)& =& D^2(z) \, [\,\xi_{\rm Lin,0}(r) \otimes {\rm e}^{-(r/D(z)
  s_{bao})^2}]\,(r)  \nonumber \\
 &+& A_{mc} \, D^4(z) \,\xi^{(1)}_{\rm Lin,0}(r) \, \xi^{\prime}_{\rm
   Lin,0}(r) 
\label{eq:xiz}
\end{eqnarray}
where sub-script $0$ means (linear) quantities evaluated at $z=0$. The
values for $s_{bao}$ and $A_{mc}$ can be taken from a best-fit
analysis to $\xi(r)$ at any given redshift (or to $w(\theta)$ in any
given redshift bin, after the projection in Eq.~(\ref{eq:wtheta})). 
In our case will be those from the best-fit at $z=0.3$,
$s_{bao}=6.37\Mpc$ and $A_{mc}=1.55$. This is detailed in 
Appendix~\ref{Appendix:A}, where we present a detailed comparison of our
model against numerical simulations, with particular emphasis on the
scaling introduced in Eq.~(\ref{eq:xiz}).

Lastly, we move to the inclusion of redshift space distortions.
The true distance to a galaxy differs from the one derived from its
redshift through the Hubble law because of the radial peculiar
velocity of the galaxy on top of the Hubble flow \cite{kaiser1987}. 
At large scales, the coherent infall of galaxies into
large overdensities, such as clusters, make their observed radial separation
smaller, squashing the structure along the line-of-sight and boosting
the amplitude of the 3-d two point correlation. 
In this way for separations along the line of sight $\pi \simlt 40 \Mpc$ the correlation (or number of pairs)
increases dramatically, while for larger separations the correlation becomes
negative in such a way that the total number of pairs along the l.o.s
is preserved (e.g. Fig. A1 in \pcite{gaztanaga09}). This implies that, by dividing the data in redshift
bins, one is discarding the leverage of large radial separations
effectively increasing the (angular) correlation within the bin (see
\pcite{nock10} for a recent and detailed discussion of this effect).

The linear redshift distortions discussed above, namely the Kaiser
effect, can be easily described
assuming the plane-parallel approximation. We will incorporate it into our modeling of
the angular correlation function by writing $\xi(r_1,r_2) =
\xi(\sigma,\pi)$ in Eq.~(\ref{eq:wtheta1}),
with \cite{hamilton1992},
\beq
\xi(\sigma,\pi)=\xi_0(s)P_0(\mu) + \xi_2(s)P_2(\mu)+\xi_4(s) P_4(\mu),
\eeq
where $\pi=r_2-r_1$ and $\sigma^2=2 r_1 r_2 (1-\cos \theta)$ (to yield
$s=r_{12}$) are the pair-separation along and transverse to the
line-of-sight, $\mu=\pi/s$ and $P_\ell$ are the Legendre polynomials.
The double integrals in Eq.~(\ref{eq:wtheta1}) are still performed in
the $r_1,\,r_2$, variables leaving the evaluation of the radial selection functions unchanged.
The monopoles of the anisotropic correlation $\xi_i(s)$ are given in
Appendix \ref{Appendix:B2} (Eqs.~\ref{eq:xi0},\ref{eq:xi2},\ref{eq:xi4}) in terms of the 3-d monopole correlation $\xi(s)$, that 
we will take as the one including nonlinear gravitational effects
given by Eq.~(\ref{eq:xiz}). 
In Appendix \ref{Appendix:B2} we discuss this effect in more
detail, also in the context of the Limber approximation.

\section{Analytical Modeling of the Covariance Matrix of $w(\theta)$}
\label{sec:errors}

An equally important aspect to the understanding of the
signal in clustering analysis of galaxy surveys is the capability
to estimate the corresponding errors in the measurements.
This is of particular importance for analysis that use 
correlation functions in Configuration space
because the measurements are highly correlated.

Notably there is scarce work in the literature aiming at
developing analytical estimates of the covariance matrix of
angular correlation functions besides the early work of \pcite{bernstein1994}, who developed an error
estimate for the Landy \& Szalay estimator in terms of higher order correlations.

Most ``data analysis'' papers have relied on sub-sampling techniques of the data itself
, such as jack-knife, bootstrap and field-to-field
variations (e.g. \pcite{ross07,meneux09,sawangwit09}). However as noted in the comprehensive work of
\pcite{norberg09} all these approaches have failings, at least in 3-d
clustering, depending on  the way they are
implemented and the regime of scales of interest. On the other hand,
projection along the line-of-sight alleviates this tension leading
to a good agreement with theoretical estimates, as shown by
\pcite{cabre07} in the context of cross-correlations between galaxy and CMB maps.

In what follows we try to revert the lack of analytical work
provided that we are interested on large angular scales, where nonlinear
(i.e. non-Gaussian) effects are weaker and that we know how to model
the signal itself. We thus concentrate in discussing how to model expected
errors in angular clustering, including the effects of sampling variance,
shot-noise, partial sky coverage, photo-z and redshift distortions. We put
particular emphasis on the description of the full error matrix,
and not only the diagonal components, and leave for further work the
assessment  of possible systematic effects.

Let us start by decomposing the fluctuations in the number of objects ``per pixel''
in the sky into spherical coordinates \cite{peebles1973},
\beq
\delta(\vn)=\sum_{\ell \ge 0} \sum_{m=-\ell}^{\ell}a_{\ell m} Y_{\ell m}(\vn),
\eeq
where $\vn$ is the line-of-sight direction and  $Y_{\ell m}$ the 
spherical harmonics. The coefficients in this expansion form the 
{\it angular power spectrum},
\beq
\langle a_{\ell m} a_{\ell^{\prime} m^{\prime}}\rangle \equiv
\delta_{\ell\ell^{\prime}} \delta_{m m^{\prime}} C_{\ell}
\label{eq:Clspectra}
\eeq
that can be related to the {\it angular correlation function} using the
Addition theorem \footnote{$P_{\ell}(\vn \cdot \vn^{\prime})=\frac{4\pi}{2\ell+1}\sum_{m=-\ell}^{\ell} = Y_{\ell
  m }(\vn) Y^{\star}_{\ell m}(\vn)$}  yielding,
\beq
w(\theta) = \sum_{\ell \ge 0} \left(\frac{2 \ell+1}{4\pi}\right)P_{\ell}(cos\theta) \,C_{\ell}
\label{eq:LegendreT}
\eeq
where $P_{\ell}$ are the Legendre polynomials of degree ${\ell}$.
The covariance in the measurements of $w(\theta)$ can then be related
to those in $C_{\ell}$ as,
\beq
{\rm Cov}_{\theta \theta^{\prime}} = \sum_{\ell,\ell^{\prime} \ge 0}
\left(\frac{2l+1}{4\pi}\right)^2 P_{\ell}(\cos\theta) P_{\ell^{\prime}}(\cos \theta^{\prime})\,{\rm Cov}_{\ell \ell^{\prime}}
\label{eq:CovWorig}
\eeq
where 
\beq
{\rm Cov}_{\theta \theta^{\prime}}\equiv \langle \tilde{w}(\theta) \tilde{w}(\theta^{\prime}) \rangle, \ \ \ \ \
{\rm Cov}_{\ell \ell^{\prime}}\equiv \langle \tilde{C}_{\ell} \tilde{C}_{\ell^{\prime}} \rangle, 
\eeq
and $\tilde{w}(\theta)$ and $\tilde{C}_{\ell}$ denote the estimators used
for $w(\theta)$ and $C_{\ell}$ respectively. In a full sky situation, and assuming the
$a_{\ell m}$ spectra are Gaussianly distributed, the $\tilde{C}_{\ell}$
measurements are uncorrelated, ${\rm Cov}_{\ell \ell^{\prime}}={\rm
  Var}(C_{\ell})\delta_{\ell\ell^{\prime}}$. In addition, one can
estimate each $\ell$ power using the $2\ell +1$ available modes,
\beq
\tilde{C}_{\ell}\equiv \frac{1}{2\ell+1}\sum_{m=-\ell}^{\ell} a_{\ell m}^2
\eeq
thus, ${\rm Var}(C_{\ell}) = 2 \, C_{\ell}^2 / (2\ell+1)$. Replacing
these relations back into Eq.~(\ref{eq:CovWorig}) leads to the final expression for ${\rm Cov}_{\theta \theta^{\prime}}$
in a full sky survey.

However a more realistic and interesting scenario is one in which the sky coverage is partial.
In \pcite{cabre07} it was shown, using Gaussian realizations
of the $a_{\ell m}$ spectra, that errors in configurations space scale
as $1/\sqrt{f_{sky}}$ (which, in turn, is the scaling of the available
number of harmonic modes). In what follows we will assume this
scaling, and compute the covariance matrix as
(\pcite{dodelson2003,cabre07})
\beq
{\rm Cov}_{\theta \theta^{\prime}} =
\frac{2}{f_{sky}}\sum_{\ell\ge 0}
\frac{2\ell+1}{(4\pi)^2}P_{\ell}(\cos\theta) P_{\ell}(\cos
\theta^{\prime})\left(C_{\ell}+1/{\bar n} \right)^2
\label{eq:CovW}
\eeq
where we have also included the standard shot-noise contribution
arising in the variance of the $C_{\ell}$ estimates \cite{peebles1973}
(${\bar n}$ is the number of objects per steradian). We
remark that the assumption leading to Eq.~(\ref{eq:CovW}) is {\it not} that the
$C_\ell$ covariance remains diagonal in a partial sky survey 
but instead that ${\rm Cov}(\theta,\theta^\prime)$
can be obtained from its full sky expression by the scaling $1/f_{sky}$.
We discuss this further in Appendix \ref{Appendix:C}.

To proceed further we thus need a model for the angular spectra.
In {\it real space} the $C_\ell$ spectra are given by (see Appendix~\ref{Appendix:A})
\beq
C_{\ell, {\rm Exact}} = \frac{1}{2\pi^2} \int 4 \pi k^2 dk P(k) \Psi^2_{\ell}(k)
\label{eq:cl}
\eeq
where $P(k)$ is the real space matter power spectrum and,
\beq
\Psi_{\ell}(k) = \int dz \phi(z) D(z) j_{\ell}(k r(z)).
\label{eq:psi}
\eeq
Throughout this paper we will use the {\it linear} theory power
spectrum in Eq.~(\ref{eq:cl}). We have tested that the inclusion of small
scale nonlinear effects (or the damping of the baryon acoustic features)
have no impact in our predictions for the errors at the (large) angular
scales we are interested in.

{\it Redshift space distortions} are accounted for by following the same
procedure that leads to the $C_{\ell}$ expression in Eq.~(\ref{eq:cl})
but starting from a power spectrum that includes the Kaiser
effect discussed in Sec.~\ref{sec:nonlinear_matter_clustering} and Eq.~\ref{eq:xisigpi}.
The final result is simply the following     
additive contribution to the kernel in Eq.~(\ref{eq:psi}) \cite{padmanabhan07},
\begin{eqnarray}
\Psi^r_{\ell}(k) = \beta \int dz \phi(z) D(z)
\left[\frac{(2\ell^2+2\ell-1)}{(2\ell+3)(2\ell-1)} j_{\ell}(k r)
\right. \nonumber \\
\left. - \frac{\ell(\ell-1)}{(2\ell-1)(2\ell+1)}j_{\ell-2}(k r) -
 \frac{(\ell+1)(\ell+2)}{(2\ell+1)(2\ell+3)}j_{\ell+2}(k r)\right] 
\label{eq:psir}
\end{eqnarray}
where $r=r(z)$. In turn, {\it photo-z} effects  are
included through the radial selection function $\phi(z)$, as discussed in Sec.~\ref{sec:photo-z}.

Notice however that the expressions in Eqs.~(\ref{eq:cl},\ref{eq:psi},\ref{eq:psir}) are numerically expensive
to evaluate due to the oscillatory behavior of $j_\ell(x)$ for $x\gg
1$. In Appendix \ref{Appendix:B1} we discuss our own approach to perform
these integrals, valid at large scales and involving the natural
cut-off $s_{bao}$ in Eq.~(\ref{eq:parametric_model}).

\section{N-body simulation and mock survey catalogues}
\label{sec:simulations}

As discussed previously, we aim at developing and testing analytical expressions for
the  signal, variance and co-variance of the angular correlation function
$w(\theta)$ against measurements in simulated upcoming photometric surveys.

Hence, in order to have a robust statistics but at the same time be
representative of such future surveys we used partitions of a
very large N-body simulation, provided by
the MICE collaboration\footnote{\tt http://www.ice.cat/mice}, to build
survey mocks.
The simulation, named MICE7680, tracked the gravitational evolution of $2048^3$
dark-matter particles within an unprecedented comoving volume of $L_{\rm box}=7680\Mpc$
using the {\tt Gadget-2} code \cite{gadget}. Initial conditions were set at
$z_i=150$ using the Zeldovich dynamics and assuming a flat LCDM cosmology with
parameters $\Omega_m=0.25$, $\Omega_{\Lambda}=0.75$, $\Omega_b=0.044$
and $h=0.7$. The spectral tilt was set to $n_s=0.95$ and
the initial amplitude of fluctuations set to yield $\sigma_8=0.8$ at
$z=0$. The resulting particle mass was $3.65\times 10^{12}\Msun$ (see
\pcite{fosalba08} and \pcite{crocce09} for further details).

Without loss of generality we next assumed a survey covering a continuous
$5000\,{\rm deg}^2$ of sky (i.e. a sky fraction $f_{\rm sky}=1/8$), 
and redshift coverage in the range $0.2<z<1.4$. In broad terms, this
matches the specifications for DES. 

In turn, the fact that redshifts are estimated photometrically implies
that much of the ``small-scales'' radial information is lost. In such scenario the best
approach is to study angular clustering in redshift bins 
of width larger or comparable to the mean photo-z error (e.g. \pcite{simpson09}).
We thus built photo-z survey mocks by extracting spherical shells of
varying width from ``comoving'' outputs of MICE7680. Each shell is
restricted to span only one octant in angular size and its radius $\bar{r}(z)$ is
set to match the comoving distance to the redshift of the output.

To place the spherical shells within the simulation boxsize we defined observers in
a regular grid and set the number of grid-points in each direction $N_i$ in order to
have none or minimal volume overlap between different
mocks (while placing the shells in the positive
octant of the observer). Strictly non-overlaping mocks can be achieved by setting the spacing along two cartesian
axis equal to $\bar{r}+\Delta r/2$, while the third to $\sqrt{2 \bar{r} \Delta r}$. 
Thus we defined $N_z$ as the integer part of $L_{box}/\sqrt{2 \bar{r} \Delta}$, while
$N_x$ and $N_y$ as the round-off of $\bar{r}+\Delta r/2$ to its nearest integer \footnote{Except when we include
  redshift distortions and/or photometric error where take $N_x=N_y=N_z$ and round off
  $L_{box}/r(z_{\star})$ instead, with 
$z_{\star}$ the redshift at which the true redshift distribution in Eq.~(\ref{eq:phiphotoz})
is negligible (see Fig.~\ref{fig:dNdz_photoz})}. Notice that rounding-off instead of
taking the integer part could lead
to a sligth volume overlap in those directions. However we have
explicitly checked that volume overlap never exceeds $1\%$
in any of the mocks. 

Note that the very large size of MICE7680 is
a critical point in order to have a robust statistics in the whole
redshift range of the assumed survey.

In order to have a clearer understanding of the different components
of the model we built ensembles of mocks in increasing ``layers of
reality''. We first selected dark-matter particles directly from real
space assuming a radial distribution as expected in DES. We also 
tested, and dismissed, effects due to biased tracers by repeating this exercise 
starting from halo catalogues.
Next we moved particles to redshift space before doing the
selection. Alternatively, we imposed a random radial uncertainty in
the position of each particle before
selection to mimic photometric error. Finally we imposed the radial
distribution in addition to redshift distortions and photometric error
to build mocks which are closest to a real survey.

Once the particles were selected we built angular number density
maps in the {\tt Healpix} format with $N_{side}=256$
\cite{healpix}\footnote{http://healpix.jpl.nasa.gov}. This $N_{side}$ corresponds to $98304$ pixels in $1/8$ of sky with
an angular resolution of $13.75$ arc-min. In this way, and given our mass resolution $m_p=3.65 \times
10^{12}\Msun$, we obtained mocks with $\sim 0.1-0.5$ galaxies per
squared arc-min
(depending on the bin-width) in the low redshift bins ($z=0.3$) and
$\sim 1-4$ galaxies per squared arc-min at high redshift ($z=1.1$). Once the
density field was pixelized we measured the correlation function using the standard estimator
\cite{barriga02,eriksen04}
\beq
\hat{w}(\theta) = \frac{1}{N_{\rm pairs}}\sum_{ij} \delta_i \delta_j , 
\eeq
where $\delta_i = (n_i-\langle n \rangle)/\langle n \rangle $ is the
density fluctuations in the $i$-th pixel and $N_{pair}$ is the number
of pixel pairs.

In what follows we give a more detailed discussion of the different
cases considered, while Table~\ref{Table:mocks}
contains a descriptive summary of the final suite of mock ensembles used
throughout this paper.

\subsection{Real Space Mocks}
\label{sec:real_space_mocks}

Let us start by describing our mocks in real (or configuration) space.
To fairly sample the range $0.2<z<1.4$ we select comoving outputs of
MICE7680 at the
redshifts $\bar{z}\,=\,0.3,\,0.5,\,0.73$ and $1.1$.
In each output we extract all the particles
within spherical shells of radius equal to the comoving distance to
the given redshift. 
In turn, to be representative of typical photo-z
errors we set 4 different bin width for each ${\bar z}$, namely 
$\Delta z/(1+z)=\,0.03,\,0.05,\,0.1,\,0.15$.
The comparison of these $16$ mock configurations with the
approximate $\sigma_z(z)$ expected for DES \cite{banerji08} is shown in
Fig.~\ref{fig:des_photoz}.

Translated to comoving distance these bins range from $100$ to
$500\Mpc$ in width. These spherical shells are
restricted to have right ascension and declination in
$0^{\circ}-90^{\circ}$, therefore covering 1/8 of sky.
Lastly, by randomly selecting particles within each bin we impose the
following radial distribution
\beq
dN/dz \propto (z/0.5)^2 \exp{\left[-(z/0.5)^{1.5}\right]},
\label{eq:dist}
\eeq
which is also what is expected in DES (we thank DES LSS working group
for providing this).

\begin{table}
\begin{center}
\begin{tabular}{p{20pt}p{30pt}p{38pt}p{23pt}p{27pt}p{22pt}} 
\hline 
\multicolumn{6}{l}{\bf Real Space Mocks} \\ \\
$\bar{z}$   &    $\frac{\Delta z}{(1+z)}$  & $\bar{r}$ & $\Delta r $ &
$\bar{n}$ & $Mocks$ \\ \\
0.3  & 0.03 & 845.7  & 102.65  & 0.11 & 1344  \\
0.3  & 0.10 & 843.3  & 342.14  & 0.33 & 441   \\
0.3  & 0.15 & 839.8  & 513.20  & 0.47 & 392   \\ \\
0.5  & 0.05 & 1345.8 & 178.23  & 0.48 & 324   \\
0.5  & 0.10 & 1343.2 & 356.50  & 0.88 & 175 \\
0.5  & 0.15 & 1338.7 & 534.83  & 1.21 & 150 \\ \\ 
0.73 & 0.05 & 1859.7 & 181.50  & 0.92 & 104   \\
0.73 & 0.10 & 1856.5 & 363.10  & 1.68 & 96    \\
0.73 & 0.15 & 1851.1 & 544.88  & 2.30 & 80   \\ \\
1.1  & 0.10 & 2558.2 & 360.07  & 2.60 & 36    \\ 
1.1  & 0.15 & 2551.9 & 540.56  & 3.90 & 36    \\ \\
\hline \multicolumn{5}{l}{\bf Real Space Halo Mocks} \\ \\
$\bar{z}$   &    $\frac{\Delta z}{(1+z)}$  & $M_{halo}$ &
$bias$ & $\bar{n} $ & $Mocks$ 
\\ \\
0.3  & 0.15 & $10^{13}$  & 2.35  & $4.0 \ 10^{-4}$ &  392  \\ 
0.5  & 0.10 & $2 \times 10^{13}$ & 2.95& $3.2 \ 10^{-4}$ &  175  \\
0.5  & 0.10 & $10^{14}$ & 4.40  &  $2.9 \ 10^{-4}$ & 175  \\ \\
\hline \multicolumn{6}{l}{\bf Redshift Space Mocks} \\ \\
$\bar{z}$   &    $\frac{\Delta z}{(1+z)}$  & $f\equiv\frac{\partial \ln D}{\partial \ln
a}$ & $Mocks$ &   \\ \\
0.5  & 0.05 & 0.705 & 125 &   &\\
0.5  & 0.15 & 0.705 & 125 &   & \\ \\
\hline \multicolumn{6}{l}{\bf Photo-z Space Mocks} \\ \\
$\bar{z}$   &    $\frac{\Delta z}{(1+z)}$  & $\sigma_z$ & $Mocks$ &
\\ \\
0.5  & 0.05 & 0.06 & 125  &   & \\
0.5  & 0.15 & 0.06 & 125  &  & \\ \\
\hline \multicolumn{6}{l}{\bf Photo-z + Redshift Space Mocks} \\ \\
$\bar{z}$   &    $\frac{\Delta z}{(1+z)}$  & $f$ & $\sigma_z$ &  $Mocks$
\\ \\
0.5  & 0.05 & 0.705 & 0.06  &  125 &  \\
0.5  & 0.15 & 0.705 & 0.06  &  125 & \\ \\ \hline
\end{tabular}
\end{center}
\caption{{\it Suite of Mock catalogues.}
Each mock consist of a redshift shell subtending $1/8$ of sky at a radial comoving distance and width
  as listed in the top panel. All mocks corresponds to
  dark-matter particles (except real space halo mocks) with
  a radial distribution given in Eq.~(\ref{eq:dist}).
 The mean distance $\bar{r}$ and width $\Delta r$ are in $\Mpc$. The
 surface density $\bar{n}$ in particles per square arc-min (and for a
 given bin is similar for real, redshift and photo-z space). 
 Halo-masses are in $\Msun$ and $D$ is the ``linear'' growth factor.
The photo-z error $\sigma_z$ equals $140\Mpc$ for our cosmology. The
bias reported was obtained from the ratio of angular correlation functions (see
Sec.~\ref{sec:biased_tracers}, also for associated error bars).}
\label{Table:mocks}
\end{table}

The resulting number of mocks depends  
 on $\bar{z}$ and $\Delta z/(1+z)$, but it ranges from tens to
thousands, making these set of ensembles very suitable for error
studies as well as for testing models and systematics.
The top panel of Table~\ref{Table:mocks} includes the main characteristics for $11$ of these bins,
which are the ones that for concreteness we focus on in this paper,
although our conclusions extend to the full set.

\subsection{Real Space Mocks for  biased tracers}
\label{sec:mock-halos}

In order to study differential features in the angular clustering of biased
tracers as compared to that of dark-matter we also built mocks 
starting from halo catalogues. 

We concentrated in two characteristic redshifts and built mocks in
exactly the same manner as described in
Sec.~\ref{sec:real_space_mocks}. We note however that  from the numerical point of view it is very hard
to resolve halos of $10^{12-13}\Msun$ in a volume as large as the one we are considering here, $\sim 450$ cubic
Gpc/h. Thus, to be able to reproduce the mass-scale of LRG halos we
choose poorly resolved halos (or groups) as tracers. Note that,
although poorly resolved, these groups have the same clustering
amplitude and abundance of real LRG galaxies \cite{cabre09}.

At $z=0.3$ we consider groups of $M>10^{13}\Msun$ (i.e. 5 or more particles) and bin width
$\Delta z/(1+z)=0.15$. The spatial abundance of these tracers is ${\bar n}=1.7\times 10^{-4}
\invMpccube$. 

At $z=0.5$ we consider halos with $M>2\times 10^{13}\Msun$ (8 or more
particles) with ${\bar n}=0.49\times 10^{-4} \invMpccube$.
With these selections we try to mimic LRG halos of $M \ge 10^{13}
\Msun$.  At $z=0.5$ we also consider cluster mass-scale halos of $M>
10^{14}\Msun$ (35 particles or more), their abundance given by ${\bar n}=0.64\times 10^{-5}
\invMpccube$. 

These mocks are summarized in the second panel of Table~\ref{Table:mocks}.
We do not consider bins at higher redshifts, as the galaxy bias is expected to be closer to linear and local.
In addition, since the linear local bias model holds valid to the extent we are
able to investigate, see Sec.~\ref{sec:biased_tracers}, we will next concentrate on the effects of photo-z and
redshift distortions on the matter field itself.

\subsection{Redshift Space Mocks}
\label{sec:mock-zspace}

To understand the importance of redshift distortions, and the accuracy
of the modeling, we built mocks where we impose the radial
distribution in Eq.~(\ref{eq:dist}) but displace the  particles to
redshift space prior to the top-hat selection.

We concentrated in the comoving output of MICE7680 at $z=0.5$ since this redshift
is the typical mean $z$ for upcoming photometric surveys
such as DES or PanStarrs. We then identified redshift bins of
width $\Delta/(1+z)=0.05$ (thin) and $\Delta/(1+z)=0.15$ (thick).
In each redshift shell the mapping from real ${\bf r}$ to {\it Redshift Space} positions
${\bf s}$ is given by the transformation
\beq
{\bf s}={\bf r}+v_r (1+z)/ H(z) \,{\bf {\hat r}}
\eeq
 where $H$ is the Hubble parameter and $v_r$ the peculiar
velocity of the object along the line of sight from the observer.
Therefore given the observer at position ${\bf r}_0$ (the center of
the sphericall redshift shell) we first find the
particle's projected velocity along the l.o.s. to the observer,
\beq
v_r=\frac{{\bf v}\cdot({\bf r}-{\bf r}_0)}{|{\bf r}-{\bf r}_0|},
\eeq
then displace it by $\delta r = v_r(1+z)/H(z)$ along the l.o.s.,
\beq
\delta {\bf r}=\delta r\frac{({\bf r}-{\bf r}_0)}{|{\bf r}-{\bf r}_0|}
\label{eq:radial-velocity}
\eeq
and finally do the top-hat selection. It total, we built $125$ mocks
subtending one octant of angular size for each of these two bin widths
(see Table~\ref{Table:mocks}).

\begin{figure}
\begin{center}
\includegraphics[width=0.4\textwidth]{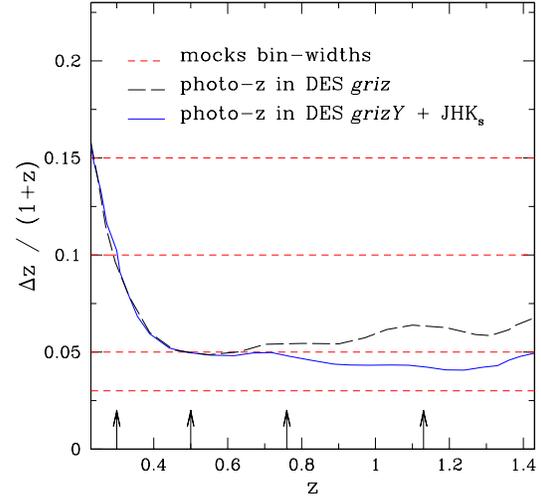}
\caption{{\it Mocks configurations}
  We built several ensembles of mock redshift bins covering $1/8$ of
  sky (5000 sq deg) and with mean redshifts
  ${\bar z}=0.3,\,0.5,\,0.73,\,1.1$ (shown by the inset vertical
  arrows). For each ${\bar z}$ we set four different
  redshift width $\Delta z / (1+z)=\,0.03,\,0.05,\,0.1,\,0.15$ (dashed
  lines, bottom to top respectively). A more detailed description of 
  the ensemble 
  of mocks for each configuration is given in Table I. The aim is to resemble with high statistical accuracy the
  geometry of large area and deep photometry galaxy surveys,
  such as the Dark Energy Survey (DES). The expected photo-z error in DES
  (using {\it griz} bands) is shown by the long dashed line, while the solid line shows the
  resulting photo-z after adding the {\rm JH${\rm K}_s$} filters from the Vista Hemisphere
   Survey (from Banerji et al. 2008).} 
\label{fig:des_photoz}
\end{center}\end{figure}

\subsection{Photo-z Space Mocks}
\label{sec:mock-photoz}

Sets of mocks including (Gaussian) {\it Photo-z} errors were also produced in
almost the same manner, except that the displacement along
the l.o.s was random with a probability
\beq
f(\delta r)=\frac{1}{\sqrt{2\pi}\sigma_r}\exp\left[-\frac{\delta r^2}{2\sigma^2_r}\right]
\label{eq:gaussianP}
\eeq
where $\sigma_r=\sigma_z c/H(z)$, and $\sigma_z$ is the survey photometric
uncertainty at the given $z$. We only considered $\sigma_z=0.06$,
which is the nominal value for DES at $z=0.5$ using the griz photometric
bands (\pcite{banerji08}) as reproduced our in Fig.~\ref{fig:des_photoz}. This is also the expected photo-z accuracy
for the $3\pi$ all-sky survey of the Pan-Starrs collaboration
\cite{cai09} at this redshift using grizy bands alone. In addition, it
is the approximately photo-z precision obtained for the optical sample of LRGs selected
from the SDSS imaging data \cite{padmanabhan05,padmanabhan07}. This is thus a very representative
value for $\sigma_z$. For our cosmology it translates to an
uncertainty in the radial distance of $\sim 140 \Mpc$ (see Table~\ref{Table:mocks}
for details). Once the photometric error was added we selected the
particles using top-hat cuts in {\it photometric} redshift of width
$\Delta z/(1+z)=0.05$ and $0.15$.

\begin{figure}\begin{center}
\includegraphics[width=0.5\textwidth]{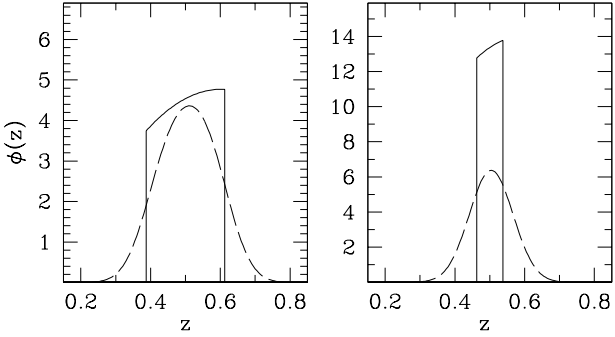} \\
\caption{{\it Radial distribution of galaxies vs. true redshift in our
  redshift distortions and photo-z mocks}.
Solid lines are the radial distribution of objects for a top-hat 
selection in true redshift of width $\Delta z/(1+z)=0.15$ (left
panel) and $\Delta z/(1+z)=0.05$ (right panel), both centered at $z=0.5$
and assuming a constant spatial density. 
Dashed lines shows instead the distribution of objects as a function of
true redshift if the same top-hat selection is now done in photometric redshifts.
In this later case a Gaussian photometric redshift error
of $\sigma_z=0.06$ is assumed.}
\label{fig:dNdz_photoz}
\end{center}\end{figure}

The net effect of selecting a sample according to their photometric
redshift, e.g. with a top-hat criteria, is to have a broader distribution of true 
redshifts of the selected  galaxies.
This is depicted in Fig. ~\ref{fig:dNdz_photoz} where the solid line
corresponds to the standard top-hat selection in true redshift as done
in the previous mocks.
Dashed line shows instead the
true distribution of objects (i.e. as a function of
their true redshifts) that in our photo-z mocks 
entered the (top-hat) photo-z bins of
$\Delta z/(1+z)=0.15$
(left panel) and $\Delta z/(1+z)=0.05$ (right panel).

These radial selection functions were obtained from
Eq.~(\ref{eq:phiphotoz}) using the underlying distribution from Eq.~(\ref{eq:dist}) (shown by the
solid line) and a Gaussian $P(z|z_p)$ of width $\sigma_z=0.06$,
Eq.~(\ref{eq:gaussianP}), and have been normalized to unity when
integrated over redshift. Their importance lies in the fact that this
is all one needs in order to compute the model correlation function,
as discussed in Sec.~\ref{sec:model}.
In addition, notice that we have chosen mock bin widths such that
$\Delta z \sim\sigma_z$ (i.e. a ``narrow bin'' of width comparable to the photo-z)
or $\Delta z \sim 4 \, \sigma_z$ (a ``broad bin''). 


\subsection{ ``Survey'' Mocks : dNdz, RSD and Photo-z}
\label{sec:mock-survey}

Finally we built mocks that include all the aforementioned effects: a
realistic radial distribution Eq.~\ref{eq:dist}, redshift distortions
and photometric redshift with uncertainty $\sigma_z=0.06$. We again
concentrated in the mean redshift $z=0.5$ and two bins of width
$\Delta z/(1+z) = 0.05\sim \sigma_z$ (``narrow'') and $\Delta z/(1+z) = 0.15\sim
4\times \sigma_z$ (``wide'').
Therefore, these are the closest to an actual photo-z survey such as
DES or PanStarrs.

Before moving on we want to stress that our mocks are obtained
from {\it comoving outputs} of an N-body simulation and thus contain all correlations induced by non-linear
gravitational evolution as well as projection effects, partial sky
coverage and realistic radial selection function. In addition some
account for photo-z and/or RSD effects. However,
since we do not use {\it light-cone outputs} they
do not include evolutionary effects of the sample within the redshift
bin. We argue that, for the narrow bins under consideration,
light-cone effects are negligible in front of photo-z and  RSD (in the
same way nonlinear gravitational effects are, see Sec~\ref{sec:mocks.vs.model.I}).
To leading order, light-cone effects will introduce a {\it linear}
evolution, which can be estimated from Eq.~(\ref{eq:wtheta}) by
weighting the selection by the corresponding linear growth
(i.e. defined with respect to the mean redshfit),  $f = D(z,\bar{z})
\,b(z) \phi(z)$. By doing this, we
have found no difference with the same calculation that instead evaluates
the growth a the mean redshift of the bin. Nonetheless, this is certainly an
interesting subjet that needs to be addressed more properly (i.e. with
light-cone
vs comoving outputs). We leave that for further work.

\section{model vs. mocks I : the correlation signal}
\label{sec:mocks.vs.model.I}

In Sec.~\ref{sec:model} we discussed in detail our model for the angular
correlation function. We now proceed to show how it performs against clustering
measurements in the ensembles of mock redshift bins described in
Sec.~\ref{sec:simulations}.

\subsection{Nonlinear Gravity and evolution}

\begin{figure}
\begin{center}
\includegraphics[width=0.4\textwidth]{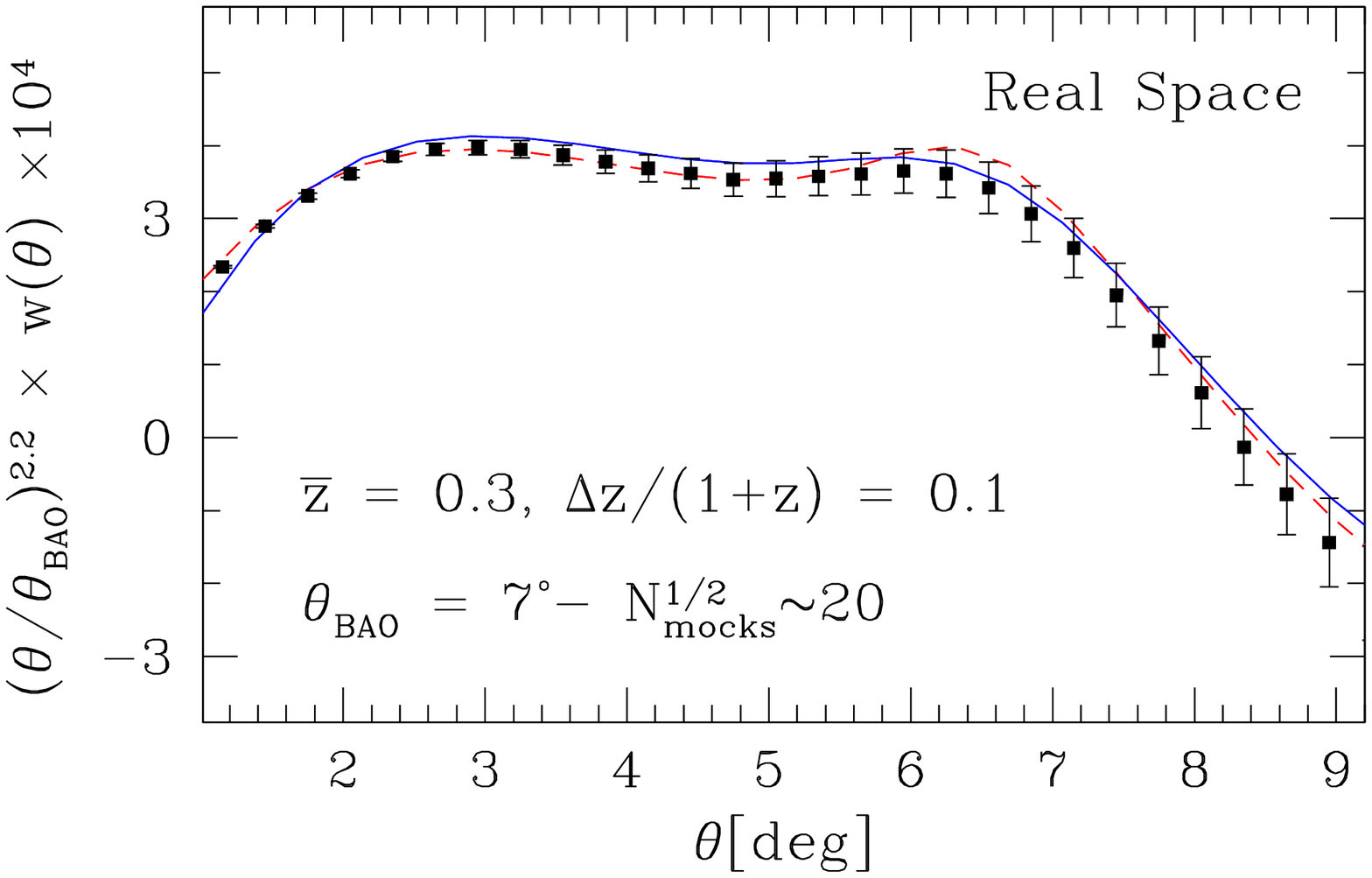}\\
\includegraphics[width=0.4\textwidth]{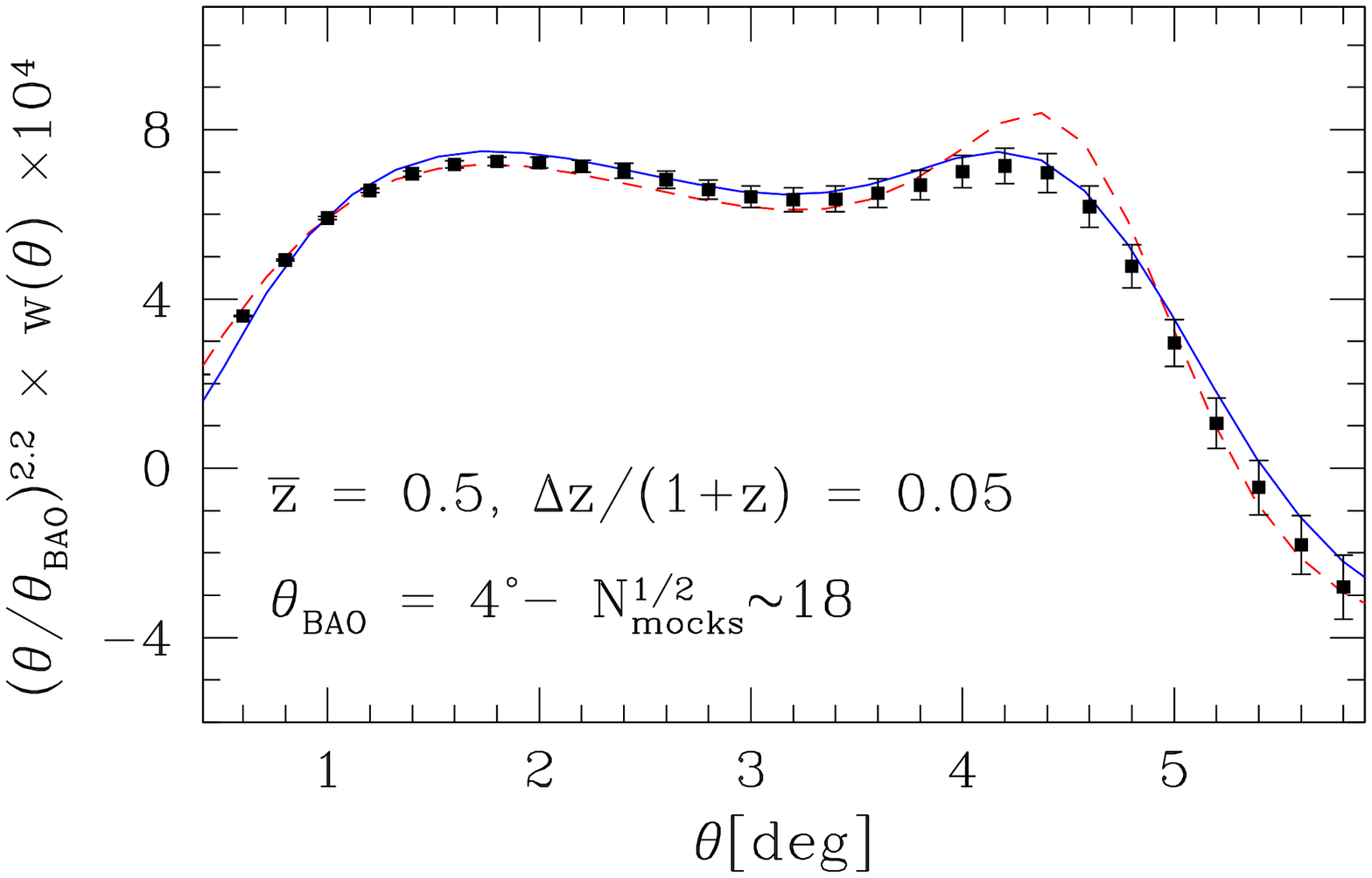}\\
\includegraphics[width=0.4\textwidth]{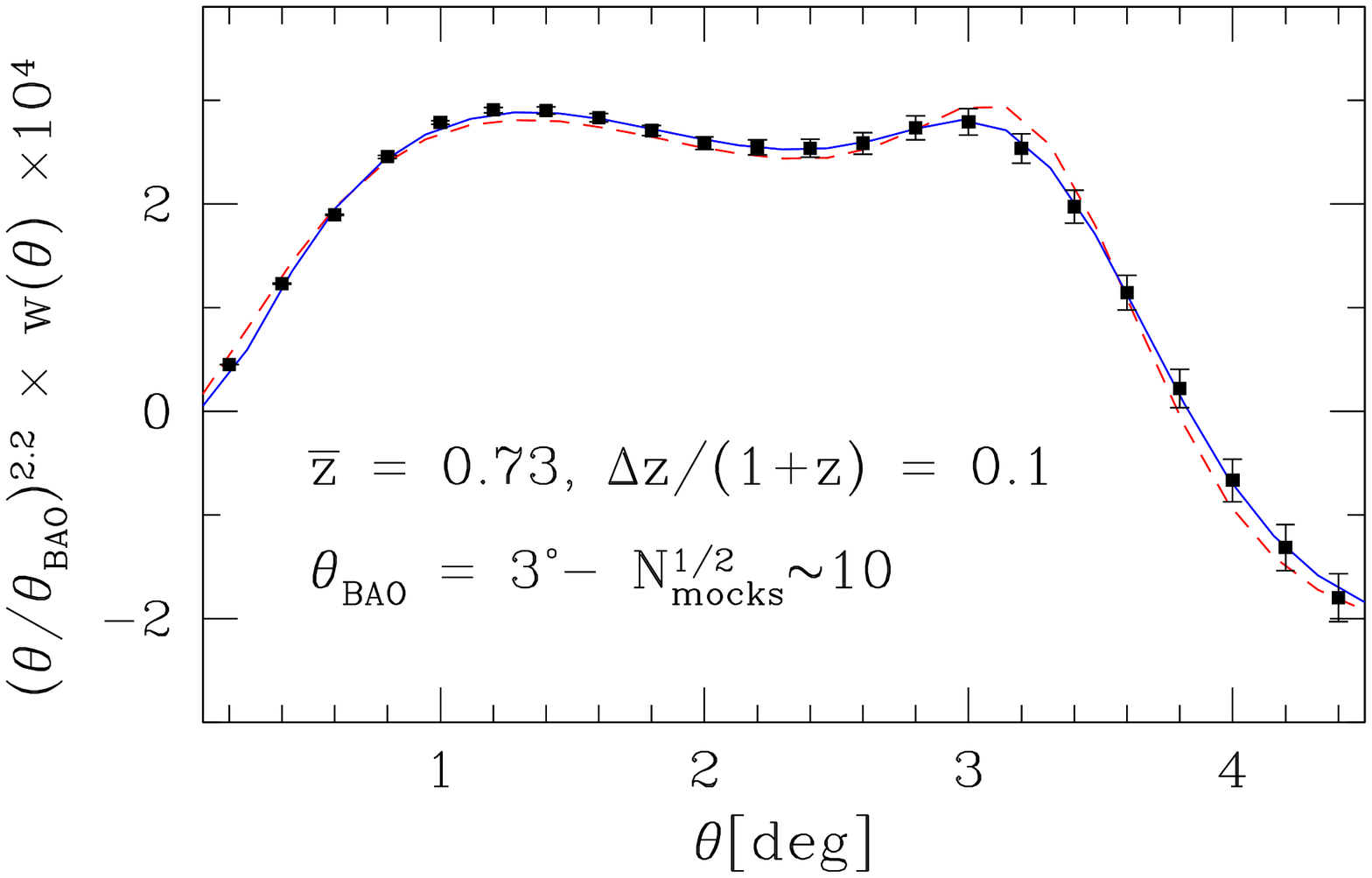}\\
\includegraphics[width=0.4\textwidth]{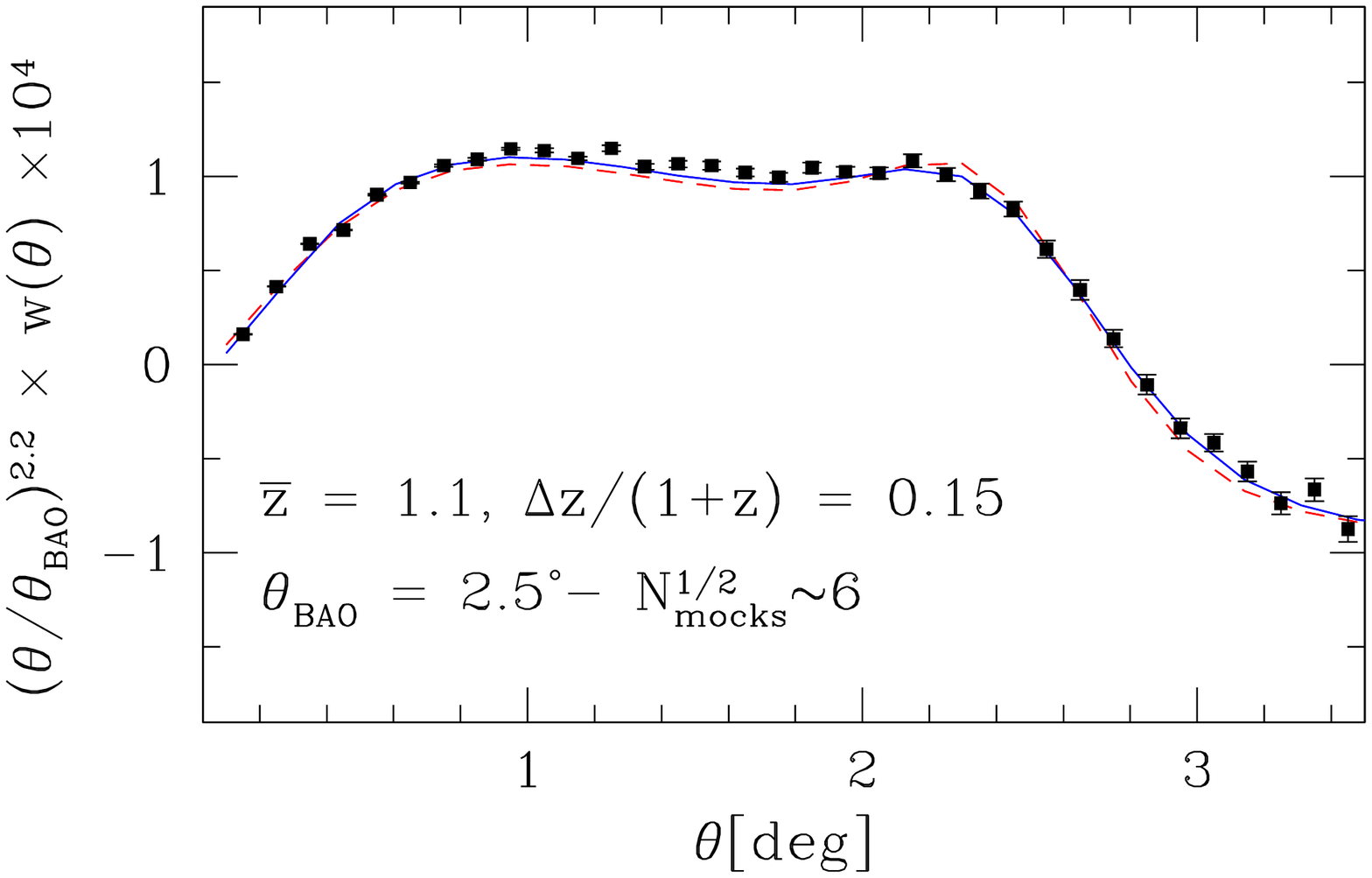}
\caption{{\it Angular Correlation Function} measured in 4 ensembles
  of mocks z-bins in {\it Real Space}
  compared with our nonlinear model (solid blue) and linear theory
  (red dashed line). Error bars correspond to the uncertainty in the
  mean, the actual r.m.s. is $N_{mock}^{1/2}$
  larger (which is specified in the inset label).
  We have explored many more
  configurations than those shown (see details in
  Table~\ref{Table:mocks} and Sec.~\ref{sec:real_space_mocks}) and the
  same level of agreement was found.}
\label{fig:angular_correlation}
\end{center}
\end{figure}

Using the model for $\xi(r,z)$ in Eq.~(\ref{eq:xiz}) we now project into redshift bins
according to Eq.~(\ref{eq:wtheta}) to find the angular correlation
function. The resulting correlations are shown in Fig.~\ref{fig:angular_correlation}
compared with measurements in the mocks at the $4$ different mean
redshifts, $z=0.3,\,0.5,\,0.73,\,1.1$ (top to bottom). In each case
for a bin width of the size of the typical photometric error achievable at the given redshift in a survey like DES \cite{banerji08}.

The agreement between our theoretical model and the mean of the
measurements is excellent for all configurations tested, see Table~\ref{Table:mocks}, in particular
those shown in Fig.~\ref{fig:angular_correlation}. And we recall
that we are using only two parameters obtained from a best-fit
to $\xi$ at $z=0.3$ \footnote{The level of matching does not change if we
  use instead the theoretical expectations for these parameter
  discussed in Sec.~\ref{sec:nonlinear_matter_clustering}}. Hence the evolution with redshift is a component of the
model. In each case displayed error bars correspond to the error on the mean
of the ensemble (i.e. $\sigma/\sqrt{N_{mocks}}$), that
given the large number of ensemble member we count on is remarkably
small.

The importance of nonlinear effects in front of projection ones are
minor if we consider the large error bars achievable in one single
mock measurement. It is however encouraging that one is able to
model accurately a large range of angular scales, given the 
mixing of all the distance scales involved in the redshift bin projection. 
We note that the second term in Eq.~(\ref{eq:parametric_model}) does not impact the shape
and position of the BAO, but it does bring theory and mocks in better
agreement for smaller angular separations.

One interesting result would be to have an estimate of the minimal
angle the model is able to reproduce.
A conservative approach to investigate this is to 
match the smaller comoving scale $r_{s}$ involved in the projection of a
galaxy pair subtending an angle $\theta$ within a redshift bin
(where $r_{min}(z)$ is the lower limit of the redshift range),
\beq
r_s=r_{min}(z)(2-2\cos \theta)^{1/2}
\eeq
with the minimum scale one is capable of modeling in the 3-d
clustering. 

Interestingly we have found that, if we define $r_{nl}$ as the scale where the model
departs from the data by some fixed percentage, then our minimum scale satisfies,
\beq
r_{nl}(z)\sim r_{nl}(z=0) \times D(z)
\eeq
and $r_{nl}(0)\sim 20\Mpc$ for a $15\%$ error in $\xi$. This ensures
the following minimal angles above which the model should perform well for the $4$ cases shown in Fig.~\ref{fig:angular_correlation},
$\theta_{min}=1.3^\circ,\,0.7^\circ,\,0.5^\circ,\,0.4^\circ$, in
agreement with Fig.~\ref{fig:angular_correlation}. But again,
this is a conservative limit because the relative contribution of
scales $\sim r_{s}$ to the full redshift bin projection is minor.

\subsection{Biased tracers}
\label{sec:biased_tracers}

We now revisit to what extent the possible
presence of scale dependent bias in the spatial correlation function
of tracers (\pcite{smith08,manera09,desjacques10,desjacques10b}) translates into the angular correlation function, depending on the
bin-width and mean redshift. Or equivalently, to what extent the assumption of
linear bias holds in the angular clustering of halos. For this
exercise we thus use the suite of mocks described in Sec.~\ref{sec:mock-halos}.

\begin{figure}\begin{center}
\includegraphics[width=0.45\textwidth]{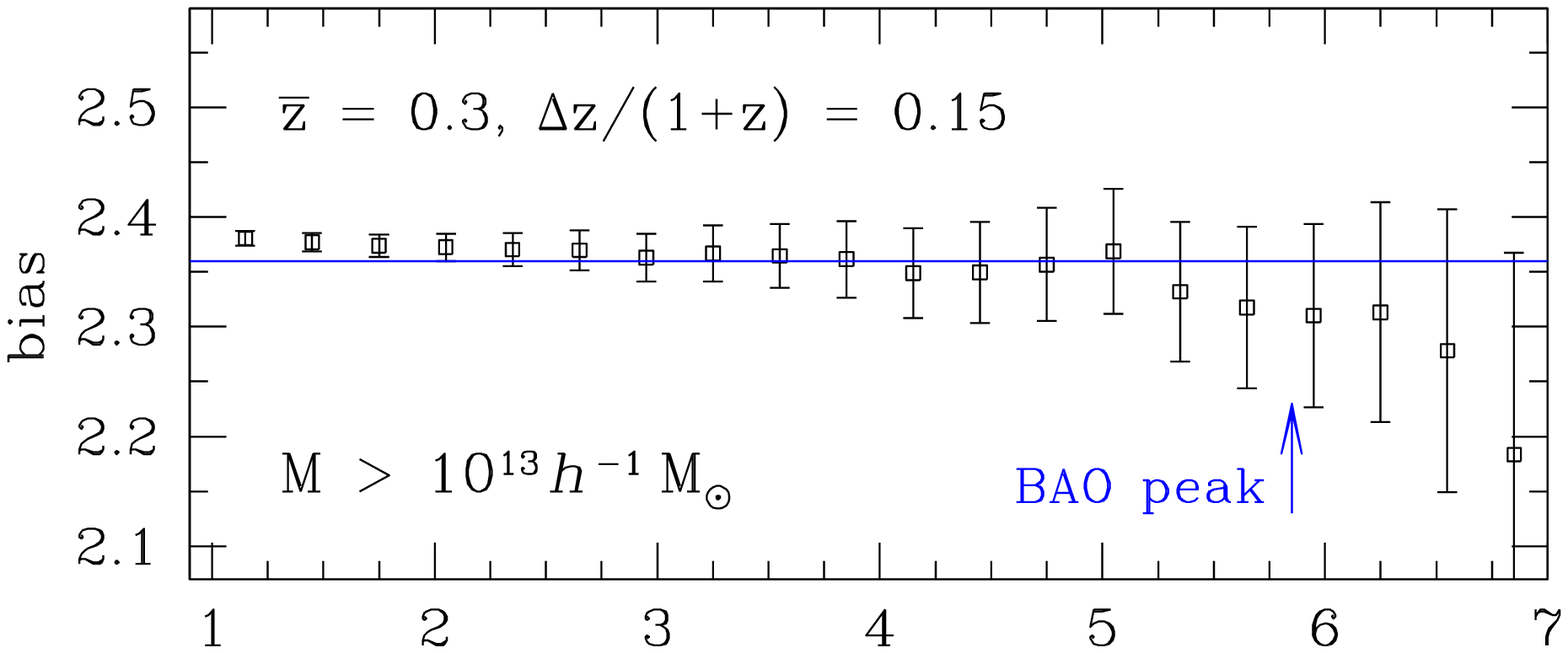} \\
\includegraphics[width=0.45\textwidth]{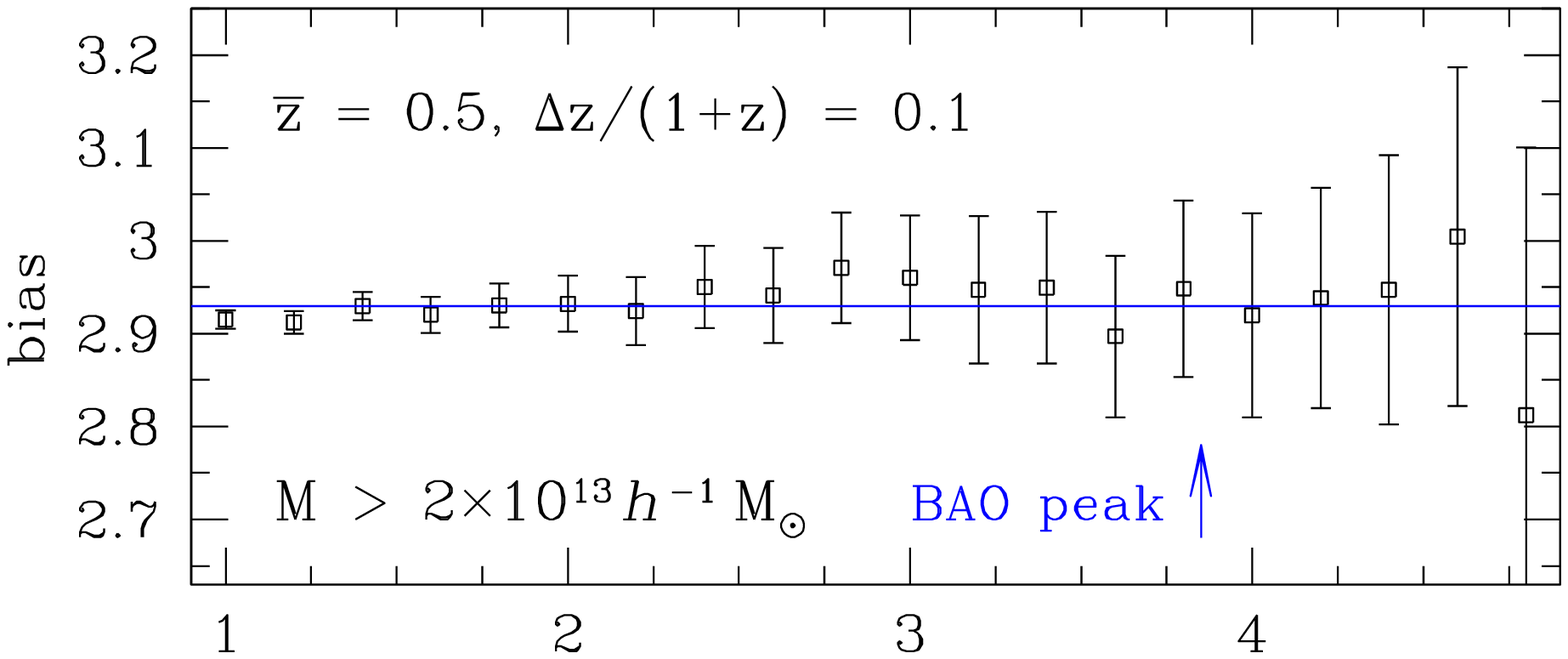} \\
\includegraphics[width=0.45\textwidth]{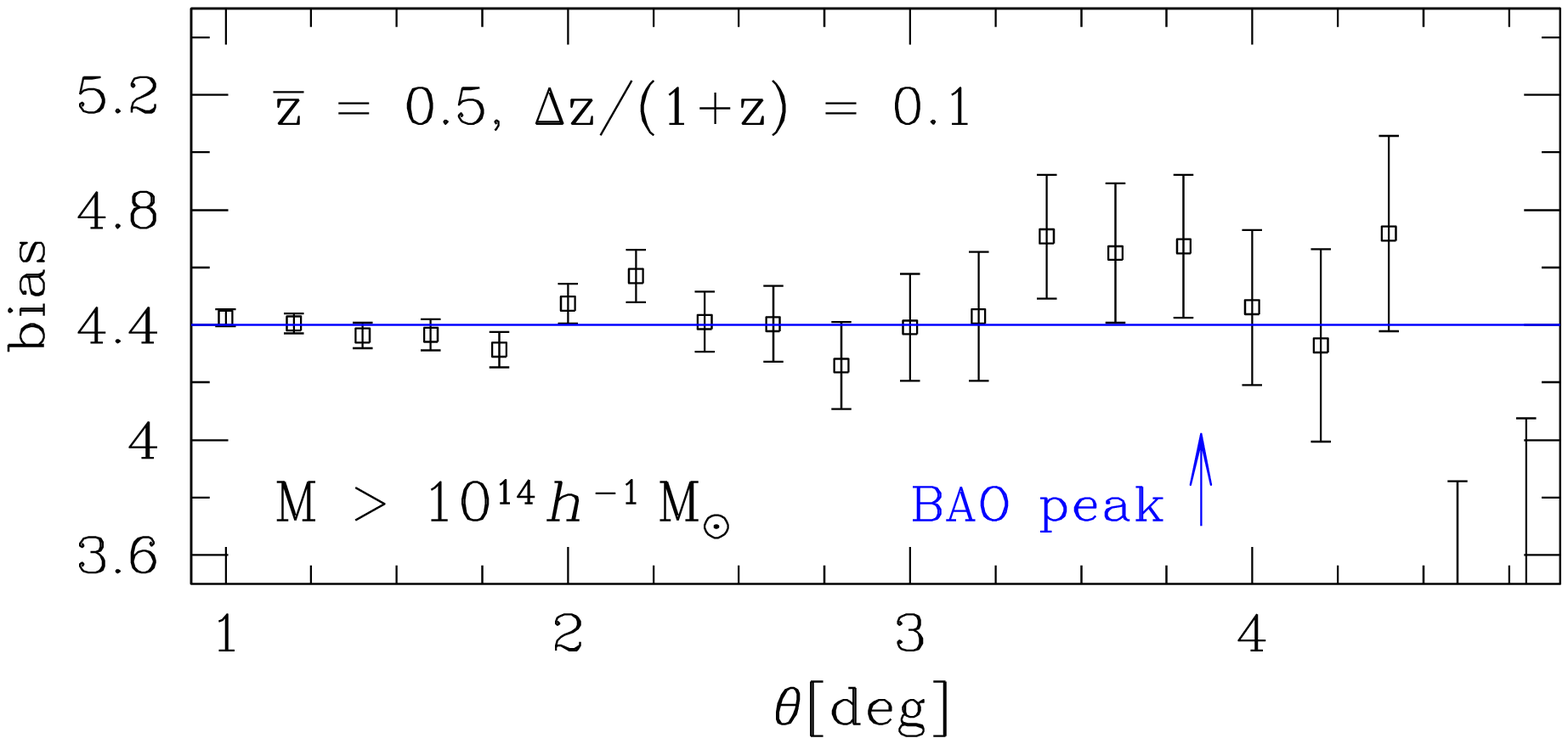} 
\caption{{\it Large Scale Halo Bias (Real Space)}, in the angular correlation function for bins
  at $z=0.3$ (top panel) and $z=0.5$ (middle and bottom). The bias is
  linear within few percent, and well within error bars. 
  Except perhaps in the lower panel, that corresponds to
  cluster-scale mass halos with large errors fully dominated by their
  low abundance. See text for more details.}
\label{fig:bias}
\end{center}\end{figure}

The top panel of Fig.~\ref{fig:bias} shows the ratio of correlation functions
(i.e. the bias) measured in the $392$ mocks of $z=0.3$ and
$\Delta z / (1+z)=0.15$ for halos $M>10^{13}\Msun$. Middle and bottom panels shows
the same ratio but from the $175$ mocks bins at $z=0.5$ and $\Delta z / (1+z)=0.1$
for masses $M>2\times 10^{13}\Msun$ and $M>10^{14}\Msun$
respectively. For reference the vertical blue arrow shows the position of the BAO
peak in each case.

In all cases the bias is scale independent well within error bars
(corresponding to the standard deviation of the mean of the ensemble) and at
the $2-3 \%$ level for the cases mimicking galaxy clustering ($M\sim
10^{13}\Msun$). 
For cluster mass-scale the shot-noise of the sample is much
larger and consequently so the error bars.  Nonetheless, there is not
clear tendency with scale. We recall that our galaxy-type halos are
poorly resolved. However we have found that increasing the resolution
(i.e. the minimum number of particles per halo) and so the typical halo-mass, does not 
alter the scale independence of the bias shown in the top panels of Fig.~\ref{fig:bias}.

The error bars displayed in the top and middle panels of
Fig.~(\ref{fig:bias}) were obtained from the r.m.s of the mocks and
correspond to the mean of the ensemble.

For cluster mass-scale halos (bottom panel) it was not possible to
obtain the bias ratio $(w_{hh}(\theta)/w(\theta))^{1/2}$ for every single mock and
angular separation due to the large shot-noise dominated errors that led to $w_{hh}<0$ in some cases.
We have then estimated the bias as $(\langle w_{hh} \rangle / \langle w  \rangle)^{1/2}$
and the 
errors propagating the
r.m.s. ensemble errors in the halo angular auto-correlation $w_{hh}$ and the
matter angular auto-correlation $w$ as,
\beq
\delta b/b=(1/2)\left[(\delta w_{hh}/w_{hh})^2+(\delta w/w)^2\right]^{1/2}
\eeq
and converting to errors on the mean by $\delta b \rightarrow \delta b
/ \sqrt{N_{mocks}}$. We have tested using the lower mass-scale halos that this
approach of the purely ensemble average leads to the same bias and errors.

In summary, we conclude that there is no evidence of scale dependent bias for
these tracers within the error bars.

\subsection{Redshift Distortions and Photo-z}

\begin{figure}\begin{center}
\includegraphics[width=0.4\textwidth]{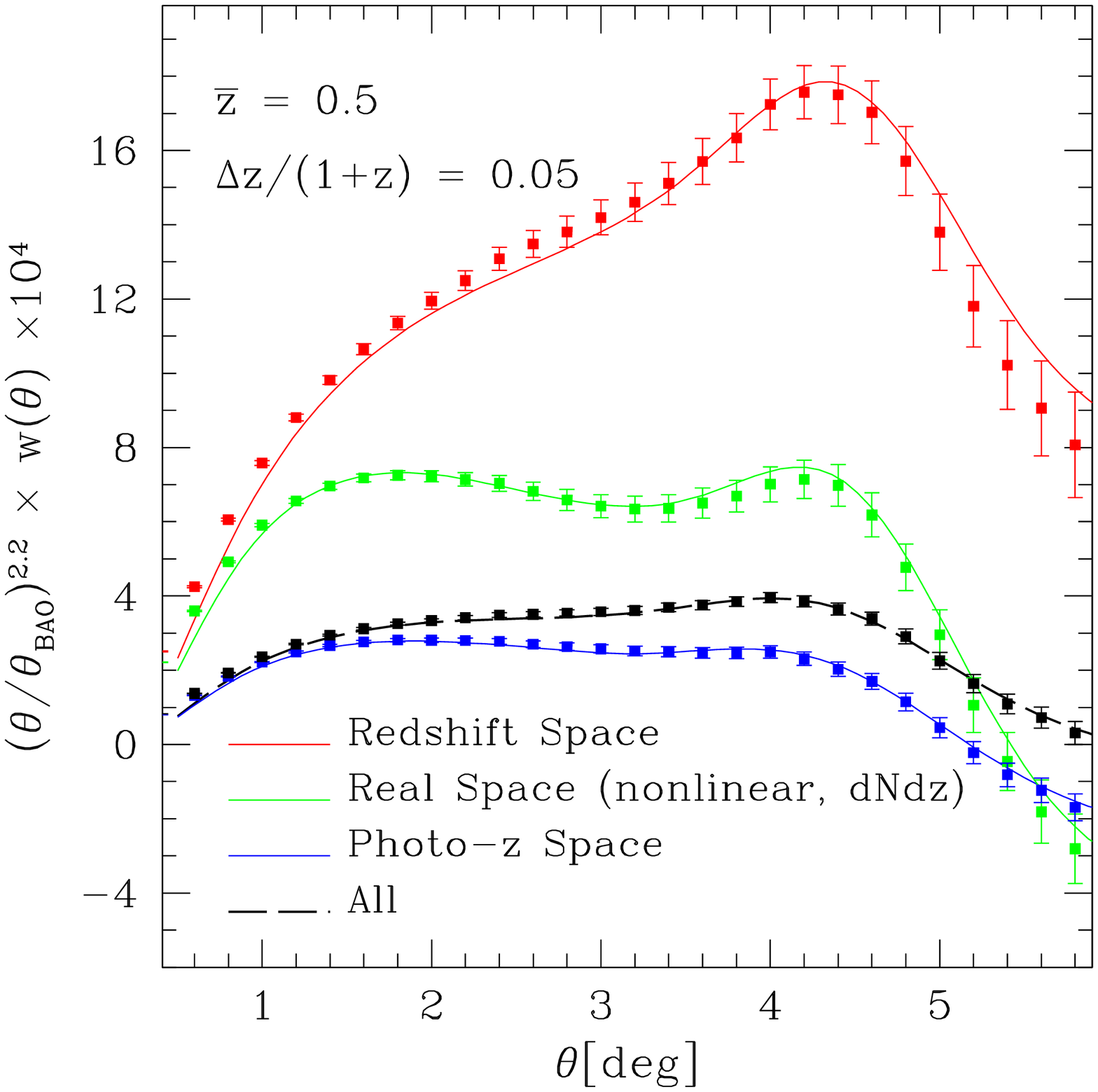}\\
\includegraphics[width=0.4\textwidth]{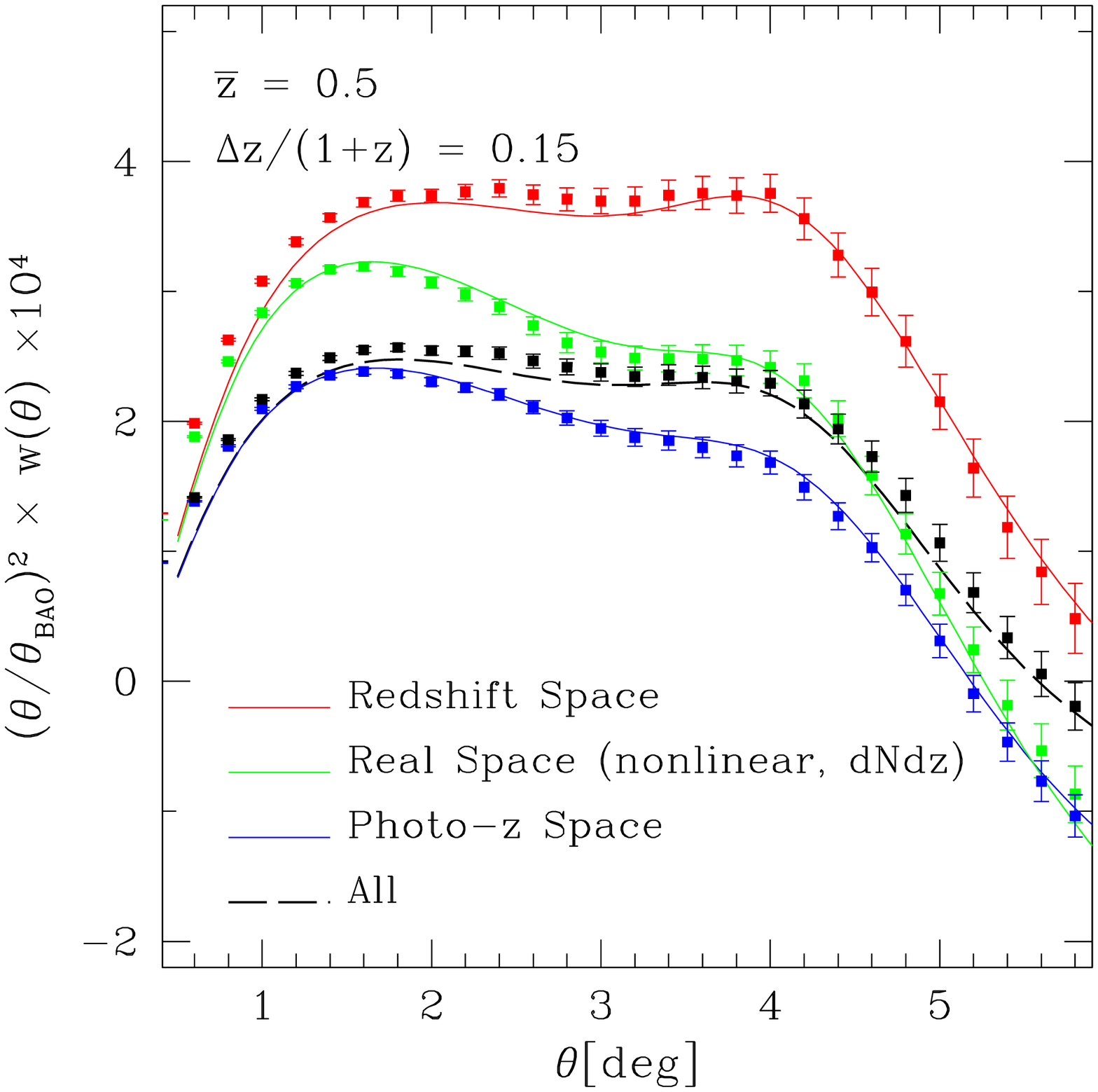}
\caption{{\it Redshift Distortions vs. Photo-z effects}: We show the
  mean correlation over 125 mock measurements, and the corresponding
  predictions, in configuration (middle green line and
symbols), redshift (top red line and symbols) and photo-z space
(low blue line and symbols). Black line and symbols corresponds to
mocks measurements and model when all effects are combined
together. Redshift distortions (RD) depend on the
parameter $\beta=f/b=0.7047$ in our cosmology, while photo-z errors were set to
$\sigma_z=0.06$. Hence, top panel corresponds to a bin of width $\Delta z \sim \sigma_z$
while bottom to $\Delta z \sim 4 \,\sigma_z$.
RD induces a strong and scale
dependent enhancement of the correlation, counteracted by the smearing due
to the photo-z uncertainties. Error bars shown correspond to the 
standard deviation of the mean of the ensemble.}
\label{fig:wtheta_photoz_zspace}
\end{center}\end{figure}

Let us now discuss the impact of redshift distortions and photometric
errors in $w(\theta)$.
Figure~\ref{fig:wtheta_photoz_zspace} corresponds to the angular correlation measured
in the mocks in configuration space from Sec.~\ref{sec:real_space_mocks} (middle green symbols),
redshift space from Sec.~\ref{sec:mock-zspace} (top red symbols) and photo-z space from Sec.~\ref{sec:mock-photoz} (low blue symbols). 
In solid green, red and blue lines we show the corresponding analytical predictions
obtained from
Eqs.~(\ref{eq:wtheta},\ref{eq:phiphotoz},\ref{eq:xiz},\ref{eq:xisigpi}). Black
dashed line and symbols depict the model and measurements when
accounting for all effects at once, with mocks introduced in Sec.~\ref{sec:mock-survey}.
For redshift distortions we used that $\beta=f(z=0.5)=0.7047$ for our
cosmology (since $b=1$) while for photo-z we used the selection functions
shown in Fig.~\ref{fig:dNdz_photoz}.  

Error bars displayed in Fig.~\ref{fig:wtheta_photoz_zspace} are true ensemble errors
corresponding to 
the standard deviation of the mean (i.e. $\sigma/\sqrt{N_{mocks}}$,
where $\sigma$ is the r.m.s variance of $w(\theta)$ measurements).
They increase from $\sim 1\%-2\%$ at $2^{\circ}$ to $\sim 3\%-4 \%$ at
$4^{\circ}$ (the angular BAO scale). Hence, 
Fig. ~\ref{fig:wtheta_photoz_zspace} clearly shows that the model described in this
paper performs remarkably well for both thin and thick bins, in
configuration, redshift, and/or photo-z spaces. 
In redshift space
it tends to underestimate the measurements for $\theta < 3^{\circ}$ at the $\sim 2\%$ level.

Redshift space distortions produce a strong enhancement of the
clustering signal 
(see \cite{nock10} who find similar results for the projected
correlation). This effect is also a strong function
of scale, becoming more important for larger separations.
Indeed, it increases the amplitude of the BAO bump by as much as a factor of $2.3$ for
the thin bin of $\Delta = 0.05 (1+z) \sim 180\Mpc$ (top panel of Fig.~\ref{fig:wtheta_photoz_zspace}) and $1.5$ for the thick one
of $\Delta = 0.15 (1+z) \sim 530\Mpc$ (bottom panel of Fig.~\ref{fig:wtheta_photoz_zspace}).

In turn, photo-z errors have the opposite effect to that of redshift distortions,
decreasing the overall amplitude of angular correlations. This is simply because the  projection in
Eq.~(\ref{eq:wtheta})  extends over a much larger range of scales (e.g. see Fig.~\ref{fig:dNdz_photoz}). Notably, this effect is not far from being scale independent. For
the thin bin the impact is a bit stronger (top panel of
Fig.~\ref{fig:wtheta_photoz_zspace}). If we only consider photo-z
effects, the amplitude of $w(\theta)$ decreases by $\sim 60\%$ when
going from true to photometric redshifts (green to blue lines in the figure).
In the case of a wide bin (bottom panel in
Fig.~\ref{fig:wtheta_photoz_zspace}), although its width is
approximately $4$ times the photo-z 
there is still a reduction in amplitude of $\sim 30\%$.

The total angular correlation, including
redshift distortions, photo-z, nonlinear evolution and bin projection effects
is depicted by the dashed line in Fig.~\ref{fig:wtheta_photoz_zspace}. In the case of bin width comparable to the
photo-z error the effect of photo-z dominates over redshift
distortions (even though the impact of redshift distortions is larger
for thinner bins). For a wider bin (bottom panel) the conjunction of
effects leaves the amplitude of correlation at the BAO peak almost
unchanged, but it does introduce a strong scale dependent bias with
respect to the real space clustering. Notably, the inclusion of
redshift distortions to the $w(\theta)$ in photo-z space enhances the
amplitude by up to $50\%$ at the peak position.  Clearly,
the appropriate inclusion of these contributions can be crucial in the analysis of real
data. 

We note that the above conclusions assume an unbiased
tracer. For biased ones the impact of redshift distortions, sensitive to $f/b$, is smaller. For
instance, if we take $b=1.7$ at $z=0.5$, that is a characteristic value
for optically selected LRGs \cite{padmanabhan07,ross07}, but keep the
same $\sigma_z$ we find the amplitude of $w(\theta)$ at the BAO
peak position ($\theta\sim 4^{\circ}$) reduced by $\sim 10\%$ with
respect to the $b=1$ case, for a bin $\Delta \sim 4\times
\sigma_z$ (but still gives $30\%$ boost with respect to the case where redshift
distortions is neglected all together). This, of course, might or not
be accompanied by a corresponding change in the amplitude of errors
depending on whether the galaxy sample is shot-noise dominated or not.

\subsection{Dependence on redshift distribution}

\begin{figure}\begin{center}
\includegraphics[width=0.4\textwidth]{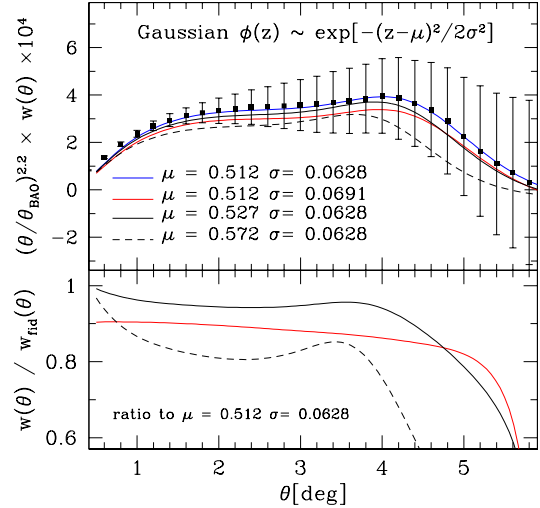} \\
\caption{{\it Dependence of $w(\theta)$ on the redshift
    distribution}. Top panel: Solid blue line is the correlation from the
  ``true'' distribution modeled as a gaussian of mean $\mu=0.512$  and 
  variance $\sigma=0.0628$. Other lines are the model correlation from
  distributions slightly off this one, as
  indicated in the labels. Also
  shown are mock measurements with  errors corresponding to a $5000 \,
  {\rm deg}^2$ photometric survey with nominal photo-z
  $\sigma_z=0.06$. 
  Bottom panel: ratio of the different $w(\theta)$ to
  the ``true'' case with $\mu=0.512$ and $\sigma=0.0628$.
  Varying the mean by $3\%$ and $10\%$ leads
  to shifts in the location of the BAO feature by approximately the same relative
  amounts (solid and dashed black lines respectively), in addition to
  changes in the amplitude. Increasing the variance by $10\%$ lowers the
  correlation by $\sim 10\%$ and introduces a scale dependent modulation of the amplitude but does not affect the peak (solid red
line).}
\label{fig:dndz-dependence}
\end{center}\end{figure}

The very good match shown so far between the model correlation and the mock
measurements relies in the fact that we have a perfect knowledge of the true
redshift distribution. However when analyzing real data this
can only be true to a certain extent. This is particularly so for
photometric data, where different photo-z codes assign different
redshift estimates to the same galaxies and thus could lead to slightly distinct
redshift distributions in each redshift bin (e.g. \pcite{abdalla08} and references
therein). To understand the extent by which the angular
correlation is affected by this potential unknown we first fitted
the redshift distribution for our narrow top-hat bin in photo-z space
(assuming $\sigma_z=0.06$) with a normal distribution,
\beq
\phi(z) = \frac{1}{\sqrt{2\pi}\sigma}\exp[{-(z-\mu)^2/2\sigma^2}].
\label{eq:gaussian}
\eeq
We found that $\mu=0.512$ and $\sigma=0.0628$ leads to an excellent agreement between the
analytic expression in Eq.~(\ref{eq:gaussian}) and the exact distribution shown by a
dashed line the the right panel of
Fig.~\ref{fig:dNdz_photoz}. We 
next increased $\mu$ by $3\%$ and separately $\sigma$ by $10\%$ and
re-computed the angular correlation.
These off-sets are similar to the variations recently
found by \pcite{thomas10} when analyzing the photometric sample of LRGs
in SDSS with 6 different photo-z codes, and selecting top-hat redshift bins
comparable to the one discussed here ($\bar{z}=0.5, \Delta z /(1+z) =
0.05$). As a third example we allowed the center of the redshfit
distribution to vary as much as the assumed photo-z, to
$\mu=0.572$. The resulting correlations are shown in Fig.~\ref{fig:dndz-dependence}. 
We include also the measurements in the corresponding photo-z mocks
(already presented by black symbols in top panel of
Fig.~\ref{fig:wtheta_photoz_zspace}). Error bars shown are the r.m.s variance over the
ensemble (rather than variance on the mean) to be representative of a real survey situation.

As noticeable in the figure variations in $\mu$ and $\sigma$ lead to analogous variations in
$w(\theta)$. Incrementing the width of the distribution by $\sim 10\%$ reduces the
amplitude of the correlation function by about $10\%$ and introduces a
scale dependent bias across all the angular range (see bottom panel). In turn,
displacements in the mean redshift of the radial distribution shifts the angular
position of the BAO peak by comparable percentage amounts (this is
more evident when shifting $\mu$ by $10\%$, see dashed line in
Fig~\ref{fig:dndz-dependence}), in addition to the amplitude. If we
instead lower $\mu$ or $\sigma$ the effects persist but are reversed.

The main conclusion is that a poor knowledge of the true
distribution (e.g. from uncertain photometric estimates) can lead to
serious systematic effects depending on the statistical errors of the survey under consideration.

\section{model vs. mocks II : the error matrix}
\label{sec:mocks.vs.model.II}

We now move to test the performance of Eq.~(\ref{eq:CovW}) in
evaluating the full covariance matrix in $w(\theta)$ measurements. We
first discuss the diagonal component, or variance, in Sec.~\ref{sec:variance} and then
the reduced covariance in Sec.~\ref{sec:covariance}.

\subsection{Comparing the  $w(\theta)$ variance}
\label{sec:variance}

\begin{figure*}
\begin{center}
\includegraphics[width=0.24\textwidth]{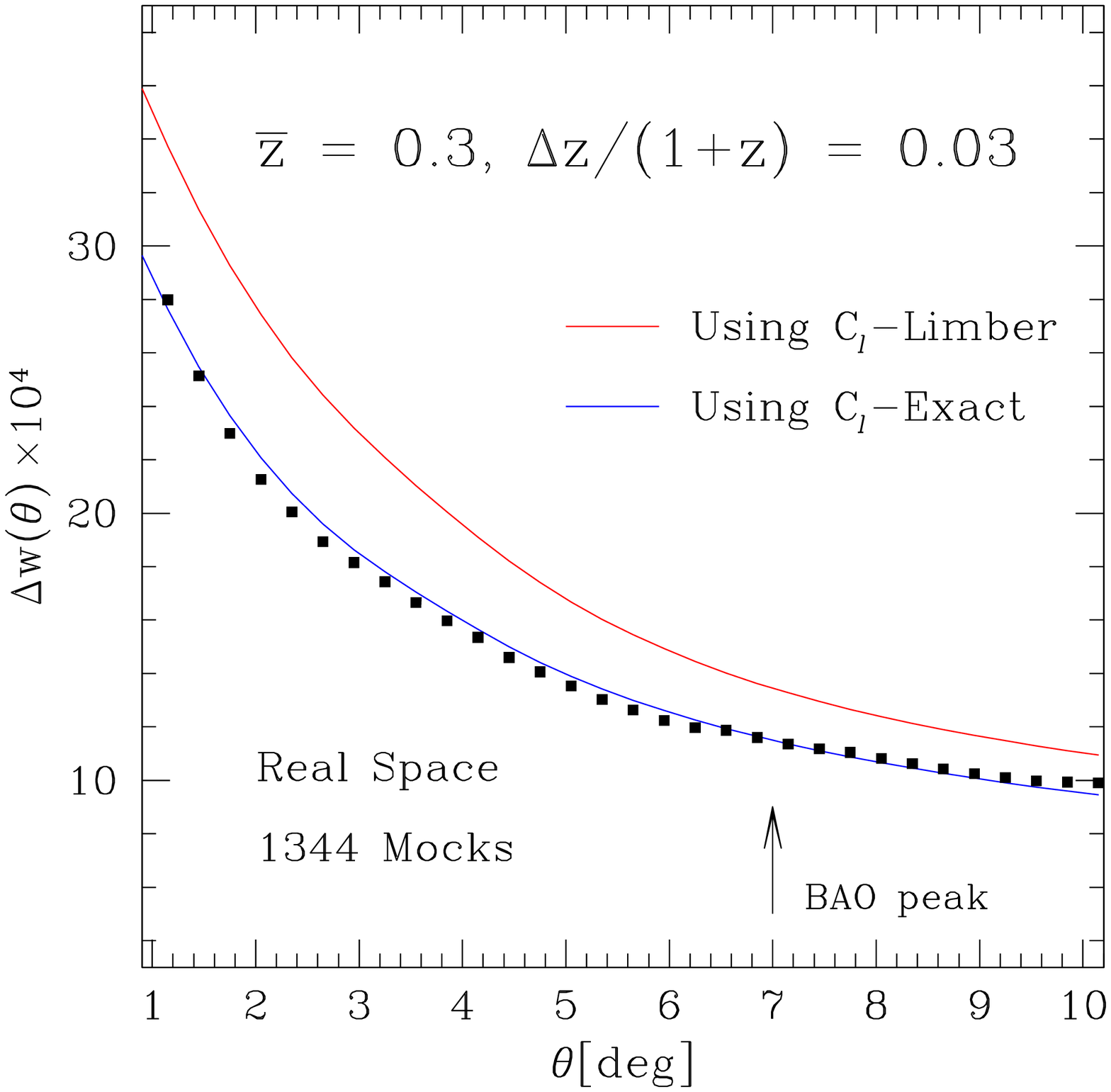}
\includegraphics[width=0.24\textwidth]{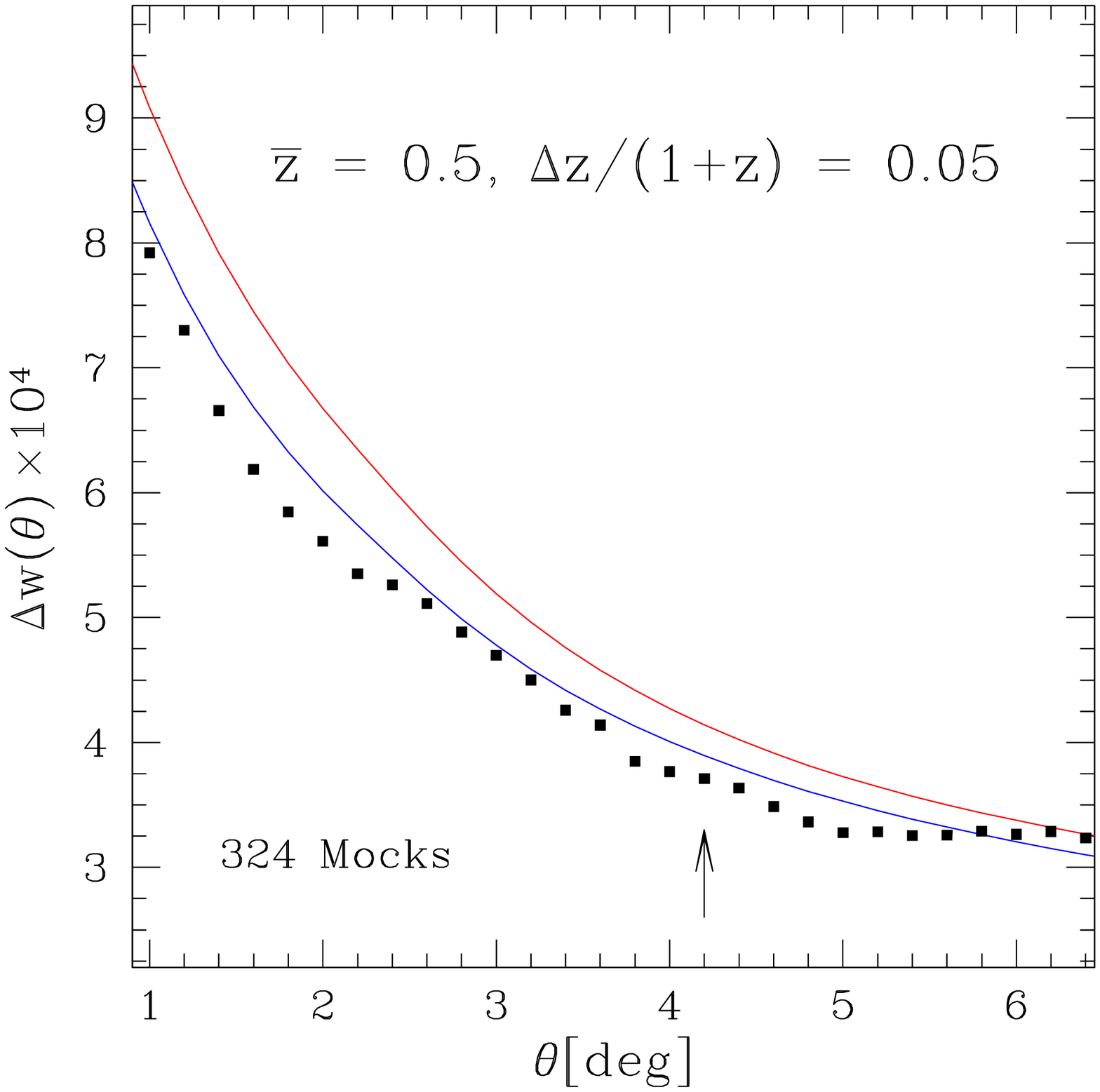}
\includegraphics[width=0.24\textwidth]{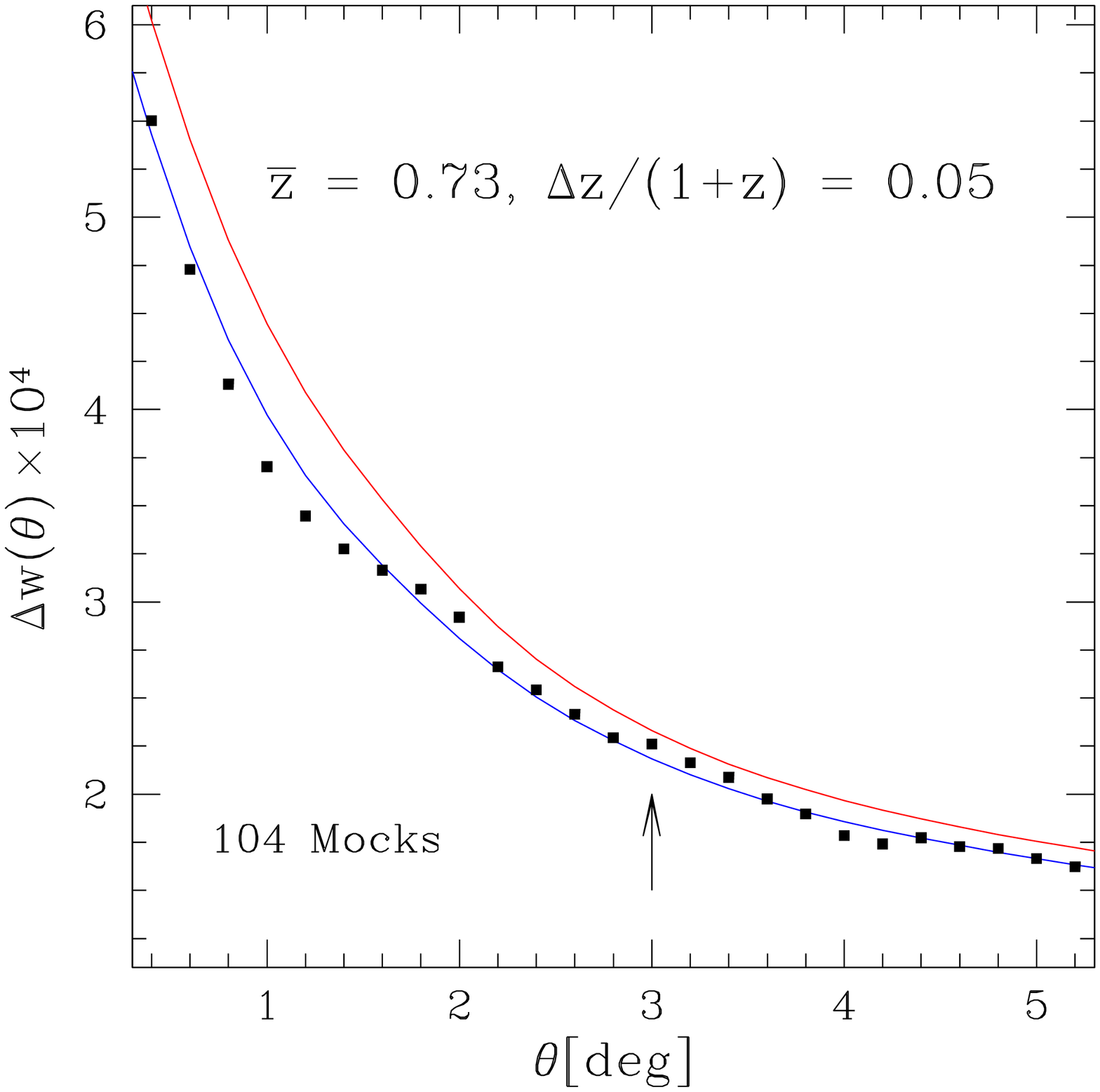} 
\includegraphics[width=0.24\textwidth]{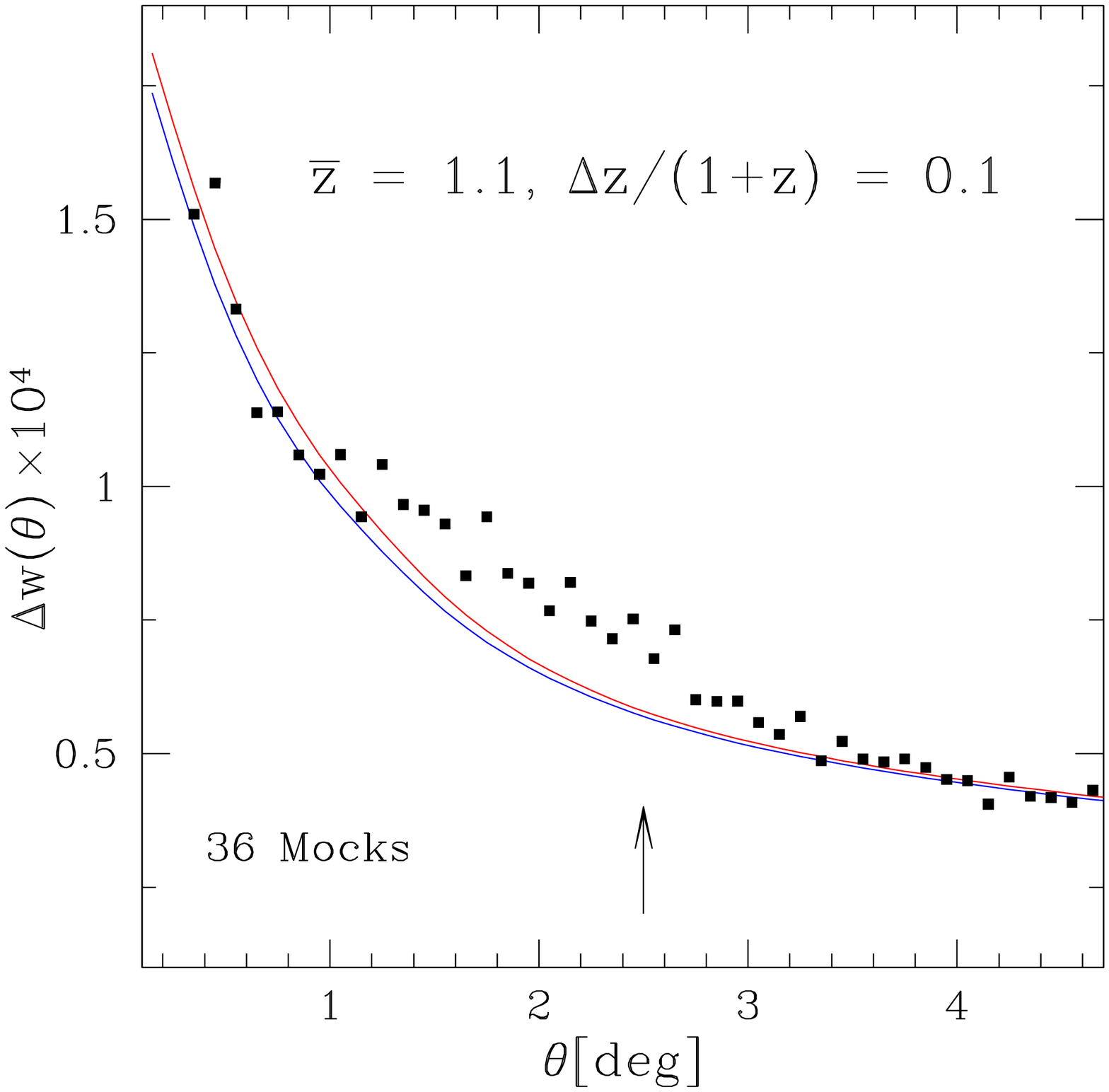}\\
\includegraphics[width=0.24\textwidth]{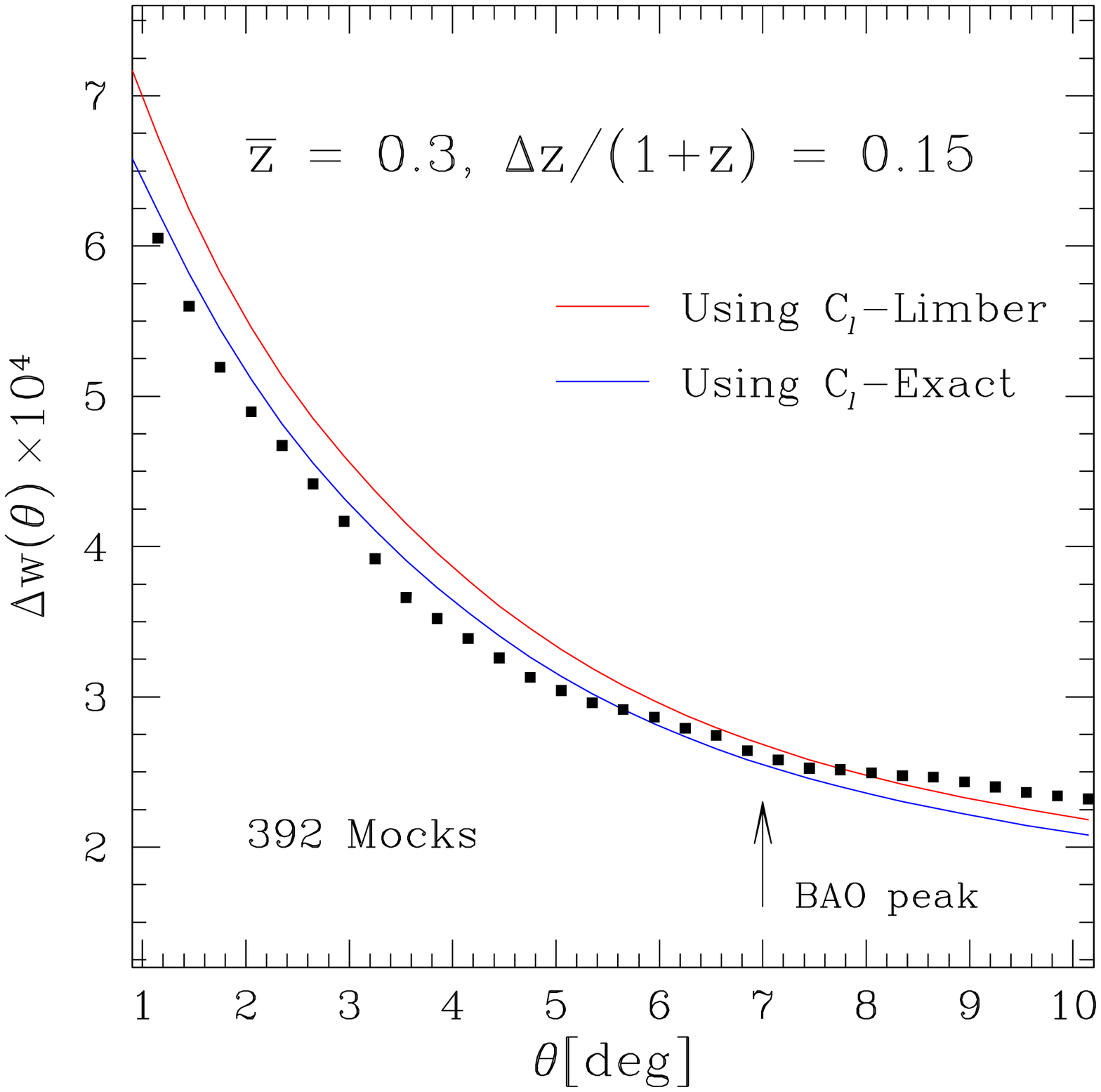}
\includegraphics[width=0.24\textwidth]{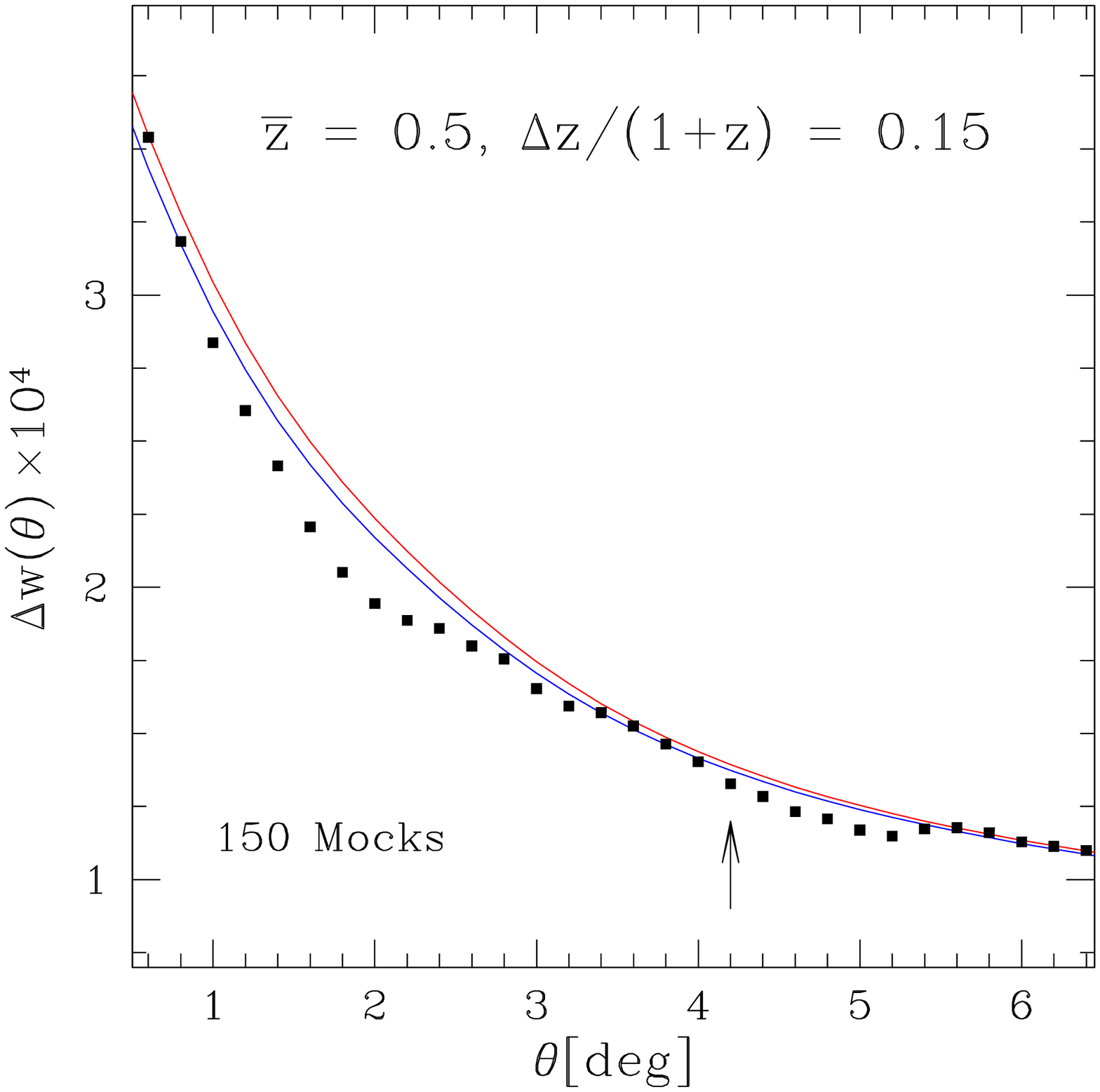} 
\includegraphics[width=0.24\textwidth]{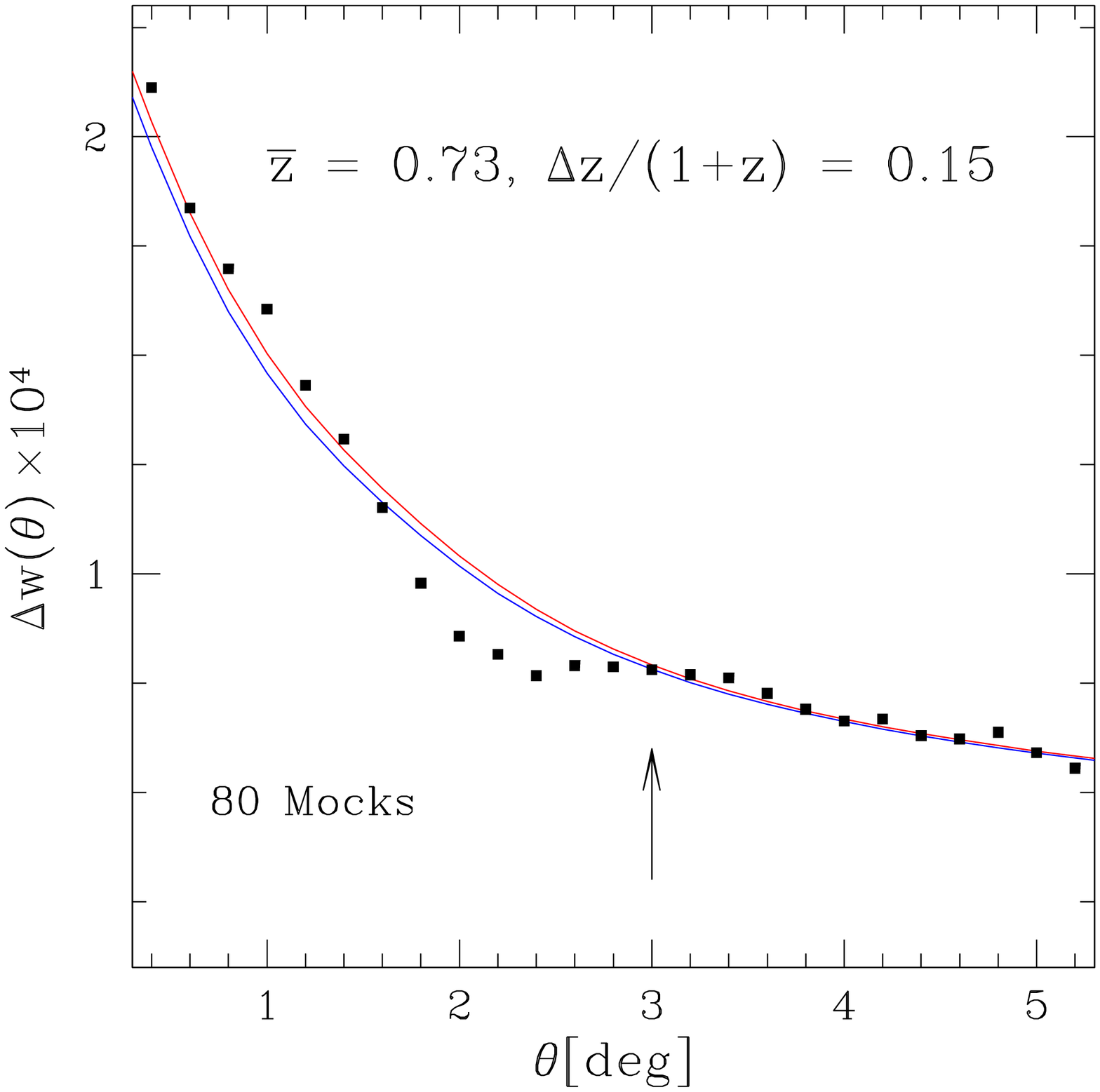}
\includegraphics[width=0.24\textwidth]{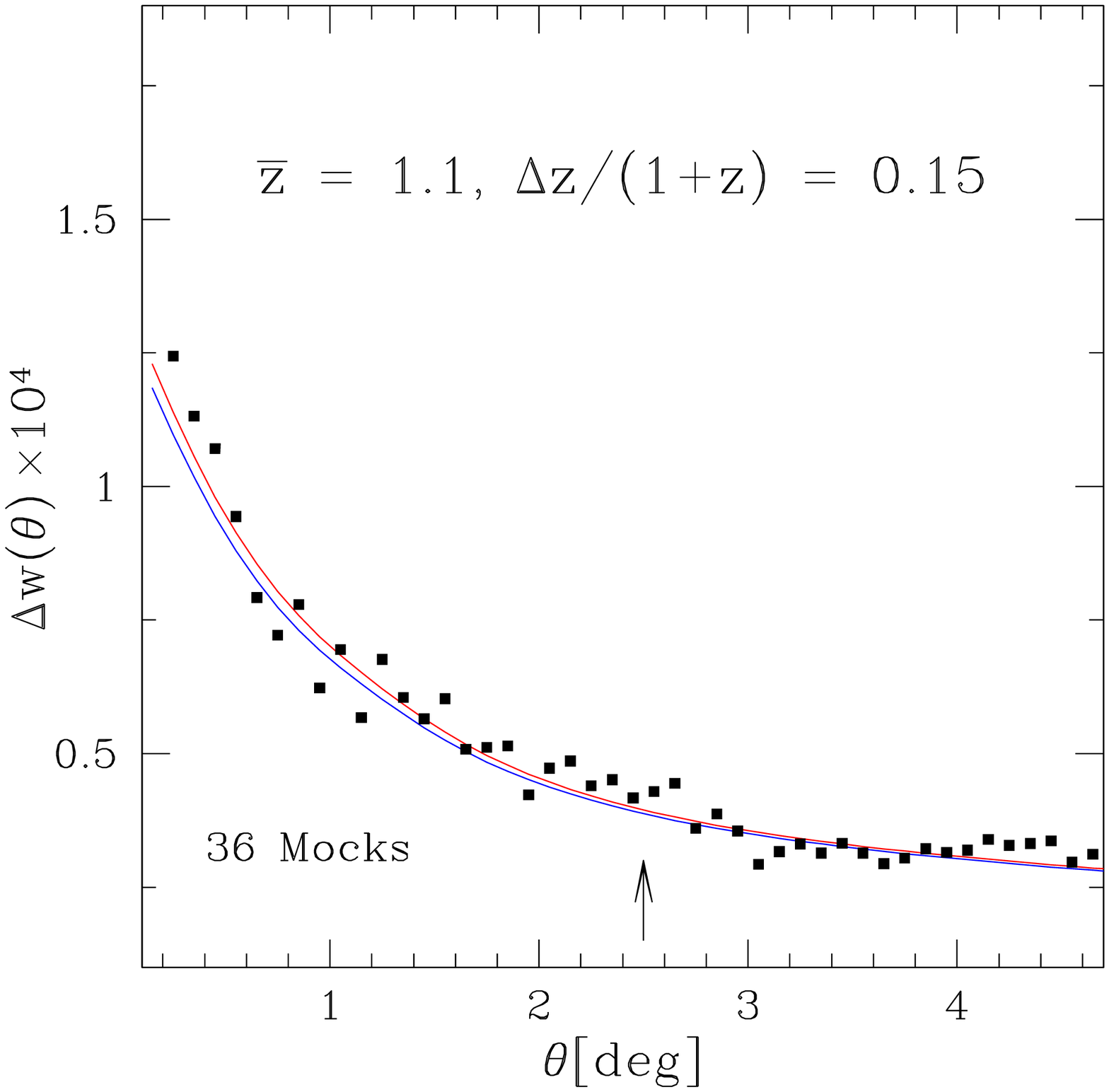}
\caption{{\it Error in the Angular Correlation Function in Real Space}. Predictions for
  $\Delta w(\theta)\equiv{\rm Cov}(\theta,\theta)^{1/2}$  from
  Eq.~(\ref{eq:CovW})  using either the exact integration for
  $C_{\ell}$ (blue solid line)  or the Limber approximation (red solid
  line). Symbols are the r.m.s dispersion in measurements of
  $w(\theta)$ in our ensembles of mock redshift bins in real space
  (see Sec.~\ref{sec:real_space_mocks} and top panel of Table~\ref{Table:mocks}). Top panels show
  thin bins and bottom their wide counter-case. Notice that even as
  many as 100 mocks can still show sample variance fluctuations in the
determination of the error.}
\label{fig:ewtheta}
\end{center}
\end{figure*}

In Fig.~\ref{fig:ewtheta} we show the r.m.s variance resulting from $w(\theta)$
measurements in several ensembles of mock redshift bins in {\it Real Space}
(top panels corresponds to {\it narrow} bin cases, bottom to their
{\it wide} counterpart). The bin selection was top-hat in true
redshift. The total
number of mocks in each ensemble depends on the particular
value of mean redshift and width of the bin (as detailed in Table~\ref{Table:mocks}),
but they are mostly over few hundreds, thus giving a unique statistical
framework for our analysis. 
Solid lines in Fig.~\ref{fig:ewtheta} corresponds to the prediction for the error,
\beq
\Delta w(\theta) \equiv {\rm Cov}(\theta,\theta)^{1/2},
\eeq
from Eq.~(\ref{eq:LegendreT}) using either $C_{\ell, {\rm Exact}}$ (solid blue) or $C_{\ell, {\rm Limber}}$ (solid red).
The agreement between the theory and mocks is remarkably good for all
the range in $\theta$ of interest for large scale structure studies,
in particular BAO, and all bin configurations. As we move to higher
redshifts / wider bins the statistics becomes slightly poorer, but the
agreement is still evident.  In turn,  the
Limber approximation over-estimates the error by as much $30\%$ for
thin bins but it rapidly converges to the exact result for bins wider than
$\sim 200\Mpc$ (e.g. the cases ${\bar z}=0.3-\Delta z/(1+z)=0.15$
or ${\bar z}=0.5-\Delta z/(1+z)=0.15$, at the bottom left panels).

\begin{figure*}
\begin{center}
\includegraphics[width=0.32\textwidth]{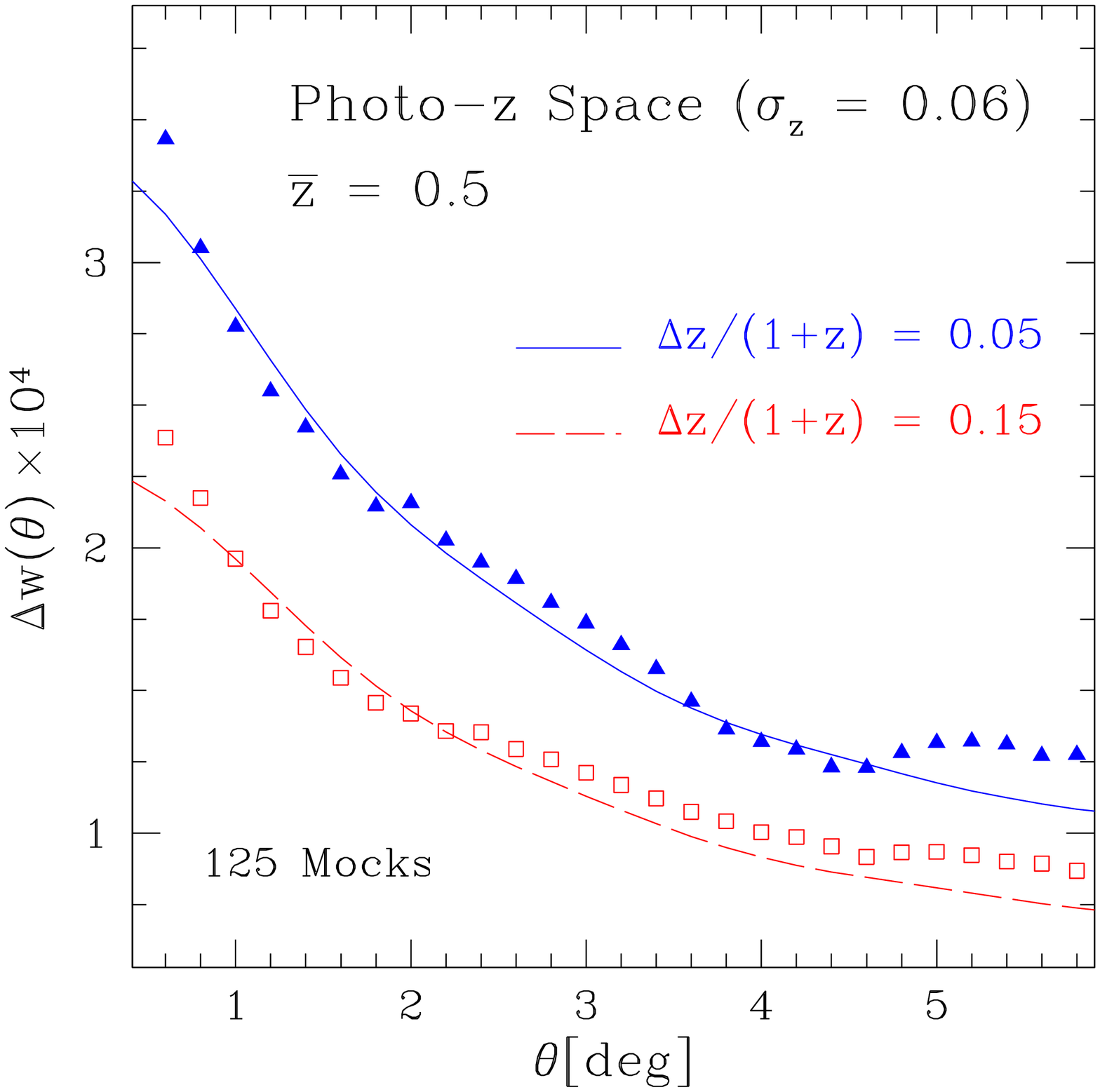}
\includegraphics[width=0.32\textwidth]{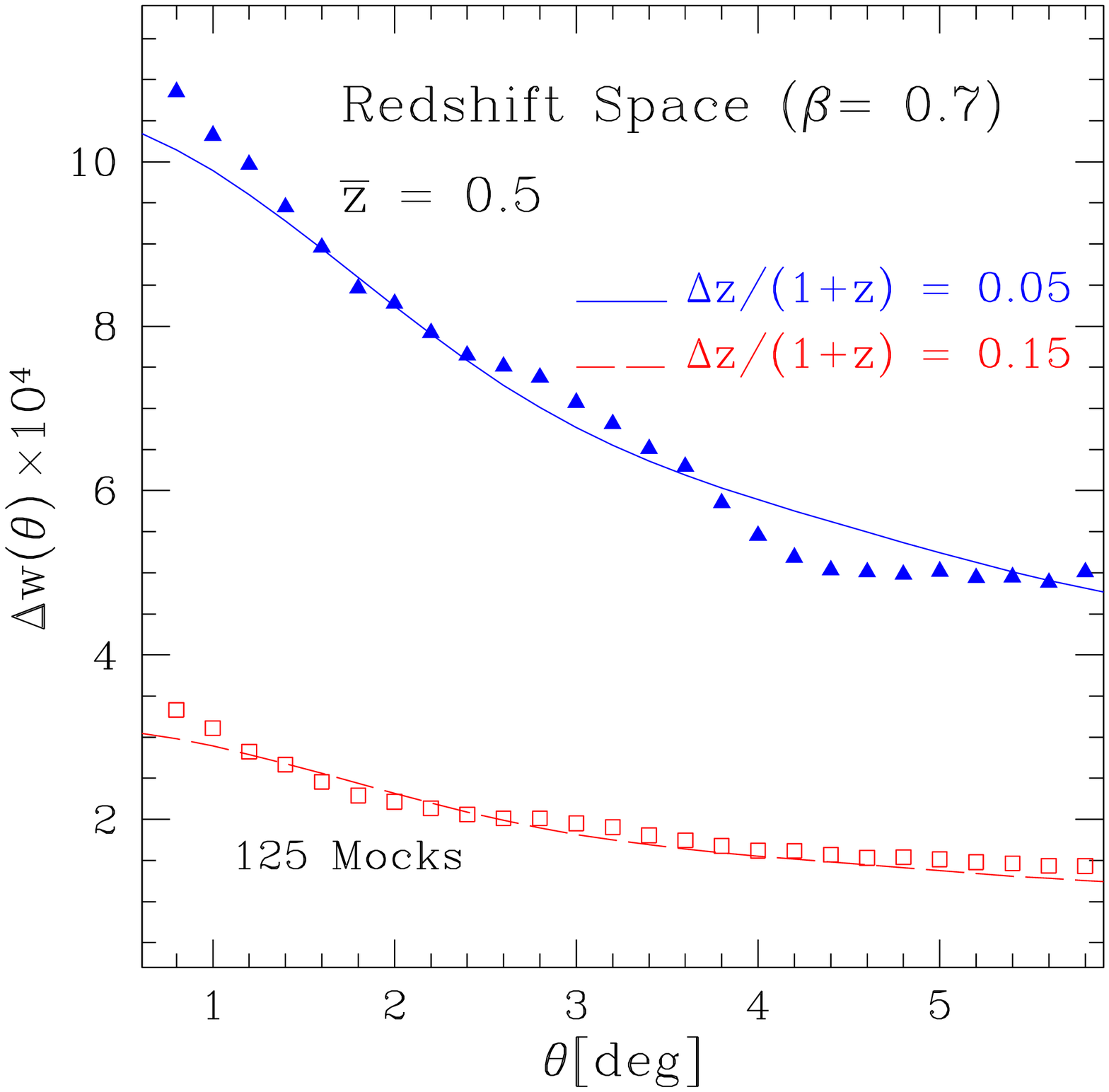}
\includegraphics[width=0.32\textwidth]{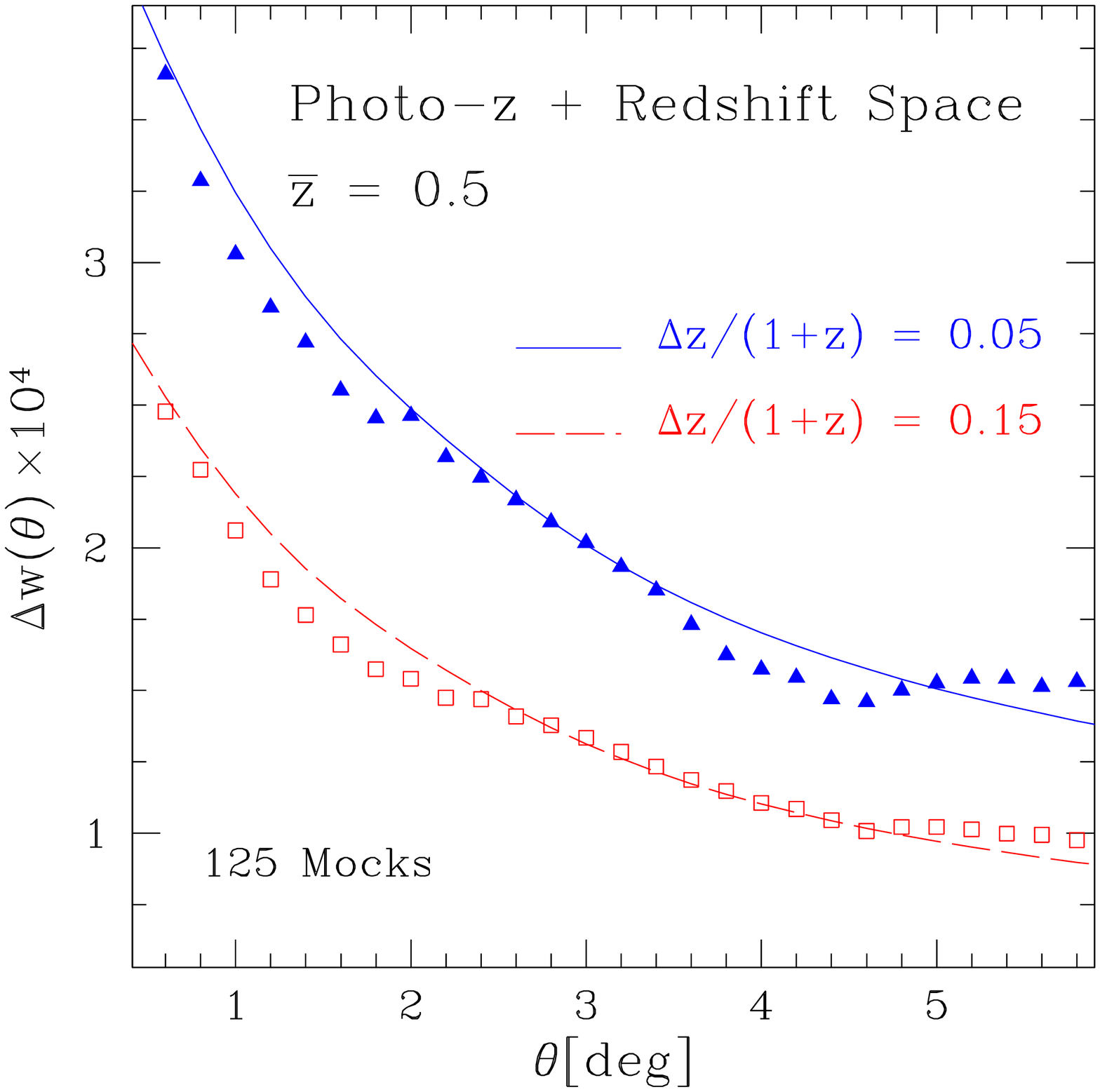}
\caption{{\it Errors in $w(\theta)$ including photo-z and redshift
distortions effects}. Left panel shows the r.m.s variance in
measurements of $w(\theta)$ in two top-hat redshift bins in photo-z space,
  both centered at (photometric) ${\bar z} = 0.5$. The assumed photometric
  error is $\sigma_z = 0.06$. Central panel shows the same
  quantities measured in redshift space (i.e. top-hat selection is
  done in true redshift). Right panel incorporates both
  effects, redshift distortions and photometric errors. The corresponding analytic predictions are given
  by the solid and dashed lines. In each panel, empty squares
  corresponds to a ``narrow'' bin ($ \Delta z \sim \sigma_z$) and filled
  triangles to a ``broad'' one ($\Delta z \sim 4\sigma_z$).}
\label{fig:ewtheta_photoz}
\end{center}
\end{figure*}

Let us now discuss the prediction of errors including photo-z effects
and redshift space distortions. 
The left panel of Fig.~\ref{fig:ewtheta_photoz} shows the r.m.s variance
in the $125$ mocks that incorporate photometric redshift
uncertainties as described in
Sec.~\ref{sec:photo-z}. The mean redshift was ${\bar z}=0.5$ and the photo-z
error assumed was $\sigma_z=0.06$ (Gaussianly distributed). 
Note that in this case, the selection was top-hat in {\it photo-z} space.
The true redshift distribution of objects used in modeling of
$C_\ell$ (as detailed in Eqs.~(\ref{eq:cl}-\ref{eq:psi}) and Appendix~\ref{Appendix:B1}) is given in Fig.~\ref{fig:dNdz_photoz}.

We considered two characteristic {\it photometric} redshift bin widths, one comparable to the
photo-z error ($\Delta = 0.075$), and one almost $4$ times larger ($\Delta =
0.225$). In both cases the prediction in Eqs.~(\ref{eq:cl}-\ref{eq:psi}), shown by solid
blue and dashed red lines, is in very good
agreement with the mock measurements (in empty and filled symbols).

The middle panel of Fig.~\ref{fig:ewtheta_photoz} shows instead the
r.m.s error obtained from the same bins in redshift space. The
prediction, in Eqs.~(\ref{eq:cl}-\ref{eq:psi}-\ref{eq:psir}), works to good accuracy in this case as
well, for both wide and thin redshift widths. Notice that in the cases
shown in this section the error is ``sampling variance'' dominated
(i.e. negligible shot-noise),
and therefore is expected to scale with the signal itself. This is
reflected in Figs.~\ref{fig:ewtheta} and \ref{fig:ewtheta_photoz} where in photo-z space the r.m.s error is smaller
than in real space, and viceversa for redshift space. In addition,
there are some wiggly features in the measurements not reproduced by
the model. These, we believe, are due to an insufficient number of mocks
(as they also show up in Fig.~\ref{fig:ewtheta} for ensembles with similar number of mocks).

The right panel of Fig.~\ref{fig:ewtheta_photoz} show the error
measured in our most realistic 125 mocks including both photo-z and redshift
distortions. They resemble (in amplitude) those in photo-z space
indicating that the spreading of the galaxy distribution due to
photometric uncertainties is the dominant effect in front of redshift
distortions for these redshift bins.

\subsection{Comparing the $w(\theta)$ reduced covariance}
\label{sec:covariance}

Once the diagonal error is well modeled we can turn to the prediction
of the full (reduced) covariance matrix. We will concentrate in
mocks that are ``sampling variance'' dominated, in real and redshift plus
photo-z space and leave the case of ``shot-noise''
dominated error to Sec.~\ref{sec:impact_shot_noise}.

\begin{figure}
\begin{center}
\includegraphics[width=0.214\textwidth]{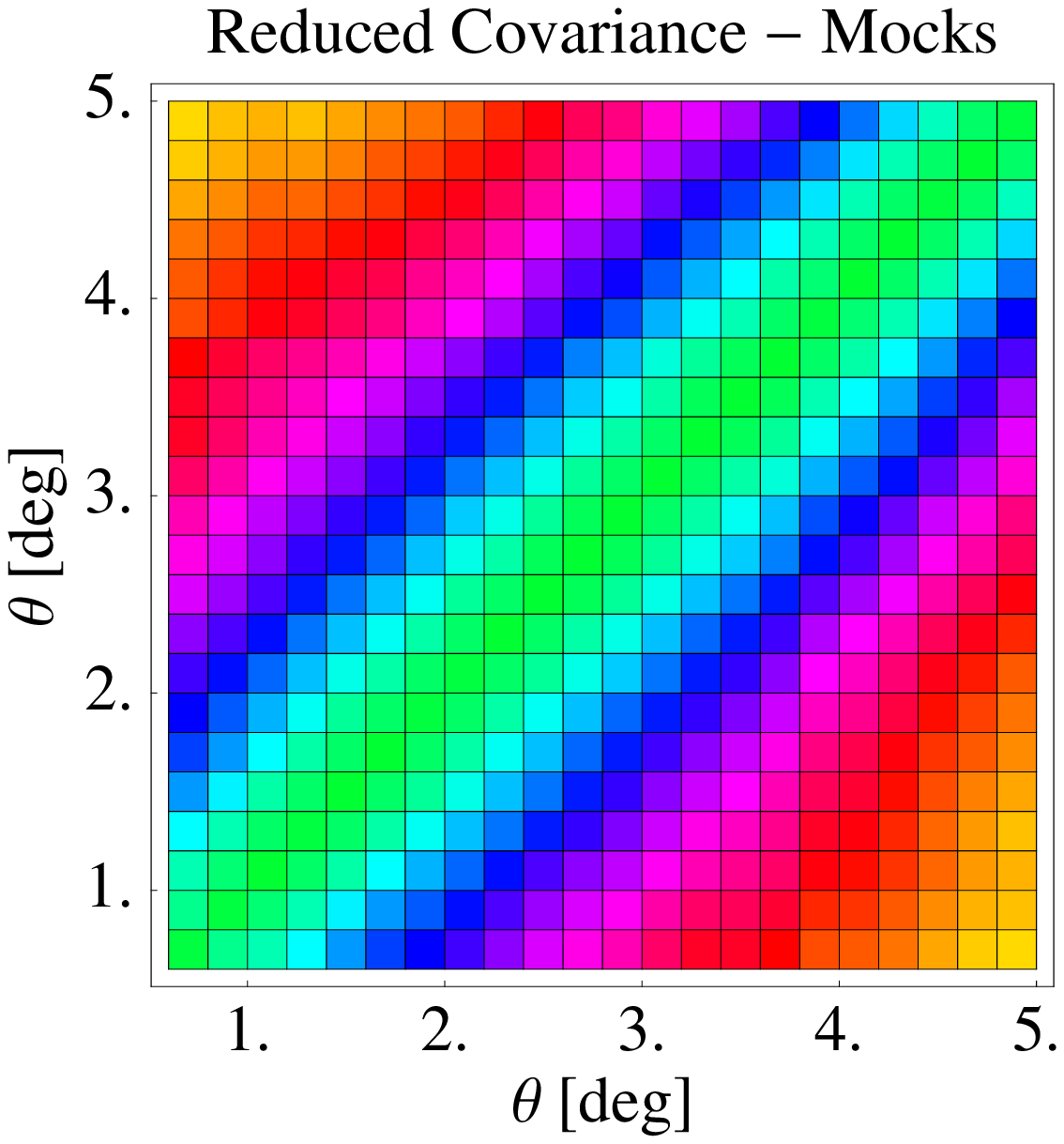}
\includegraphics[width=0.194\textwidth]{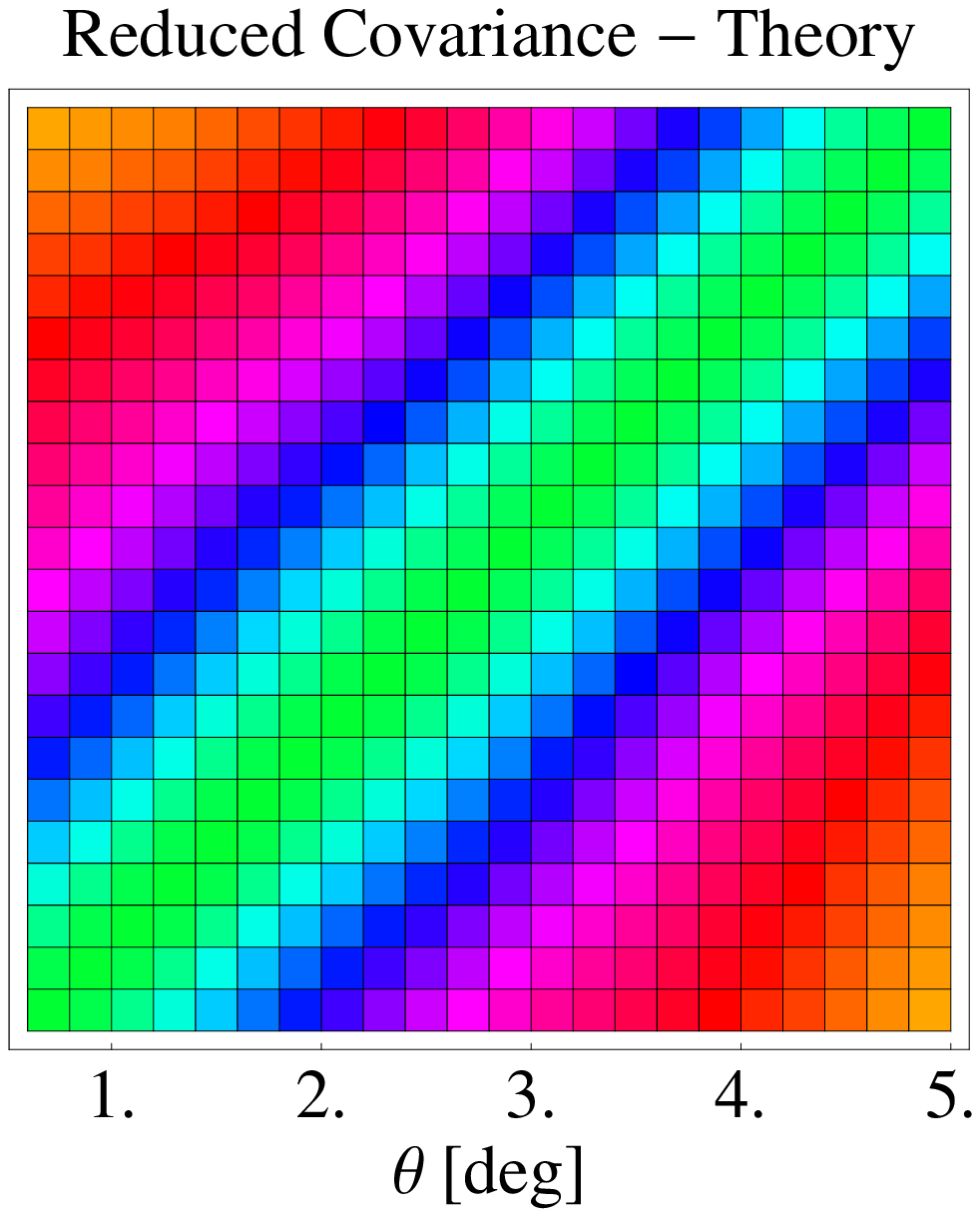}
\includegraphics[width=0.058\textwidth]{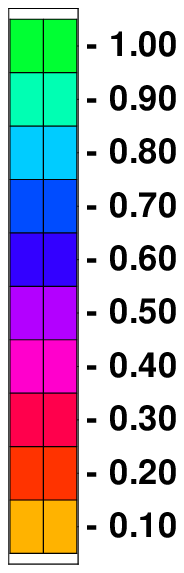}
\newline \\ 
\includegraphics[width=0.37\textwidth]{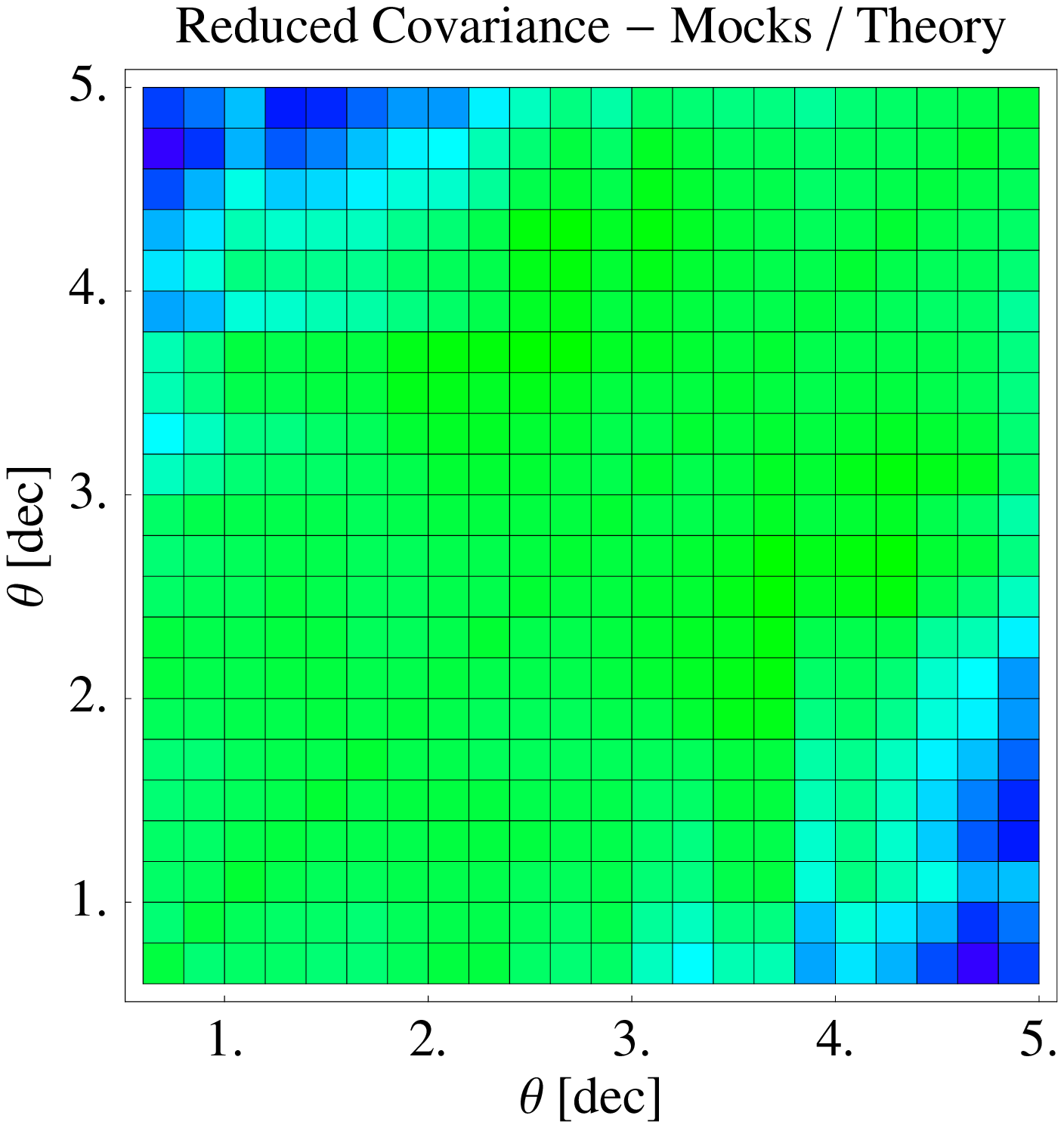}
\includegraphics[width=0.07\textwidth]{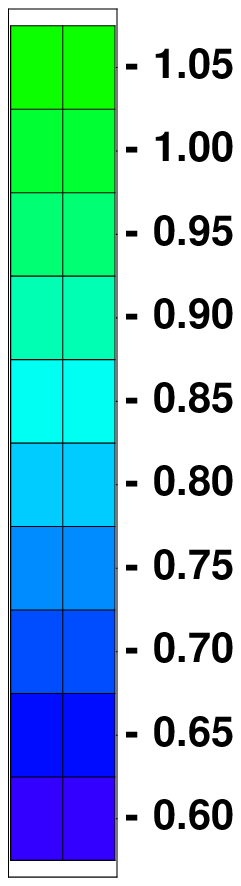}
\caption{{\it Measured Reduced Covariance Matrix (top left) vs. the
    Model in Eq.~(\ref{eq:CovW}) (top right)} for the bin centered at
  ${\bar z} = 0.5$ and width $\Delta z/(1+z) = 0.05$ in Real
  Space. Bottom panel shows their ratio.
Notably, the agreement is within $5\%$ for most of the
  reduced covariance. The largest differences arise in areas far from
  the diagonal where the
  reduced covariance is close to zero anyway.}
\label{fig:cov}
\end{center}
\end{figure}

Figure~\ref{fig:cov} shows the reduced covariance matrix, ${\rm Cov}_{\rm
Reduced}(\theta,\theta^{\prime})\equiv{\rm Cov}(\theta,\theta^{\prime})/\Delta 
w(\theta) \Delta w(\theta^{\prime})$, measured in $324$ mocks in Real Space
 at ${\bar z}=0.5-\Delta z/(1+z)=0.05$ (top left panel). The
 corresponding model from Eq.~(\ref{eq:CovW}) is depicted in the top right
 panel. Bottom panel shows the ratio of mock measurements to model. Remarkably they agree with each other
 within $\sim 5\%$ for the majority of elements, with the largest
 differences arising in areas far from the diagonal where the
 covariance is small anyway.

\begin{figure}
\begin{center}
\includegraphics[width=0.45\textwidth]{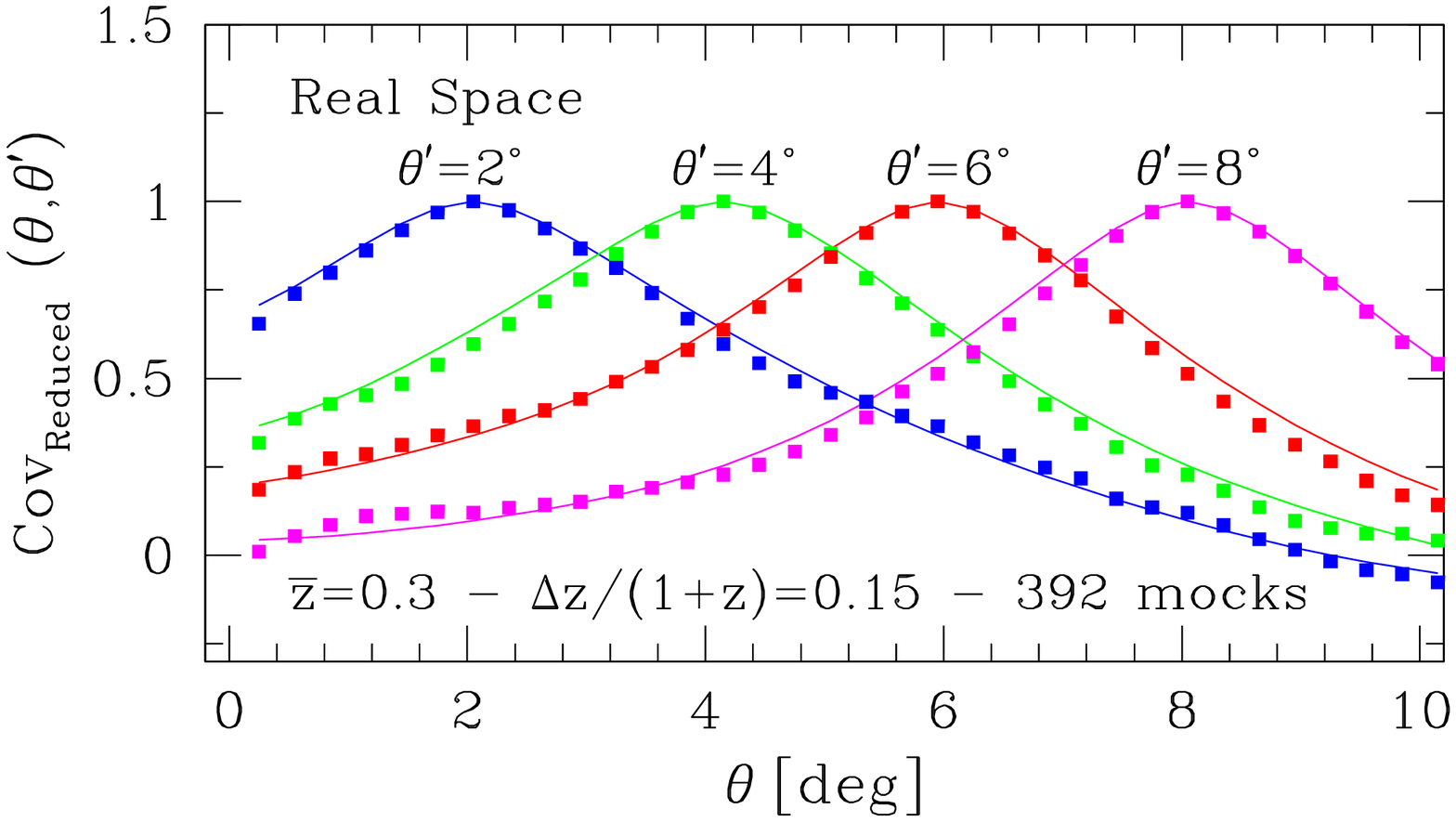}\\
\includegraphics[width=0.45\textwidth]{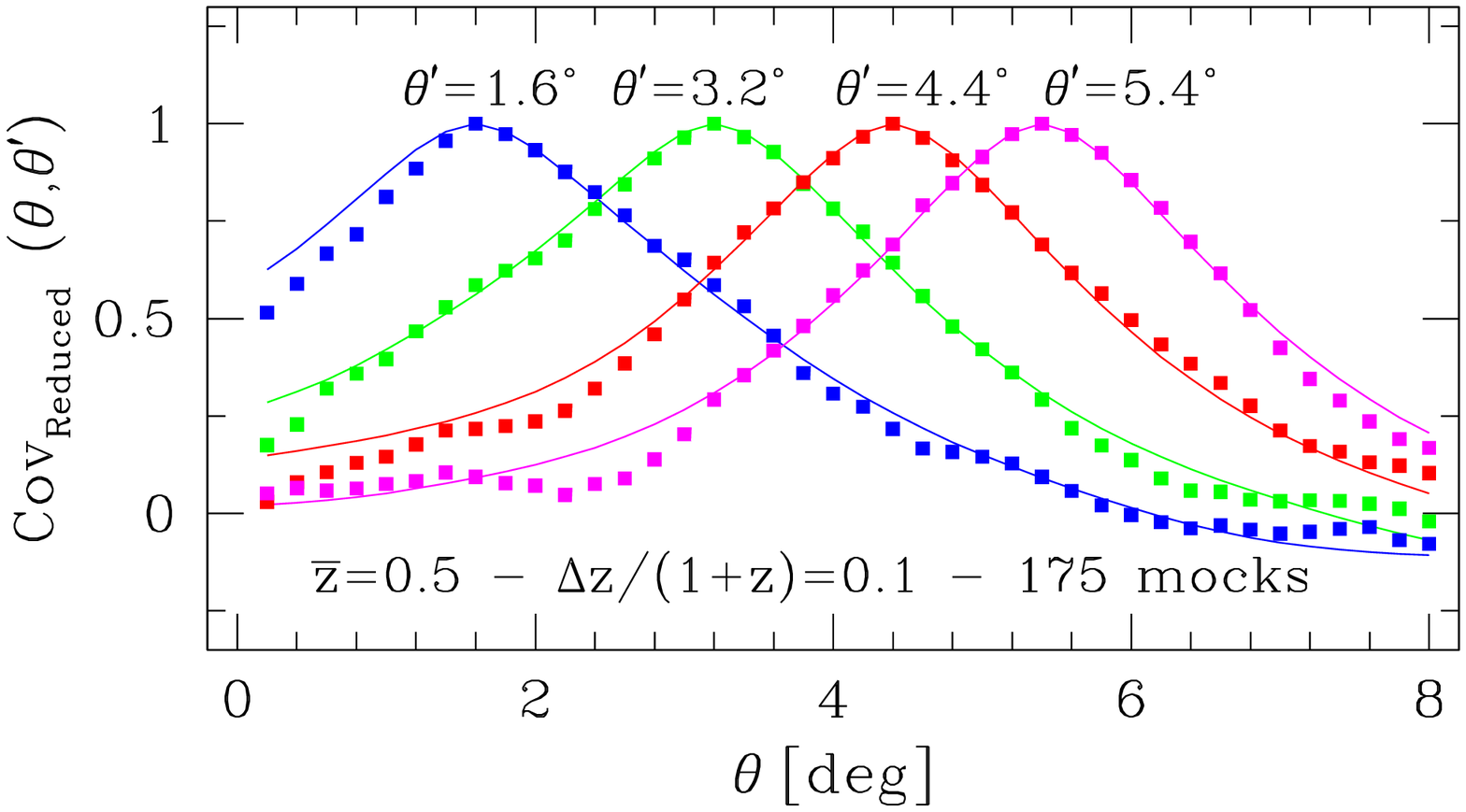}\\
\includegraphics[width=0.45\textwidth]{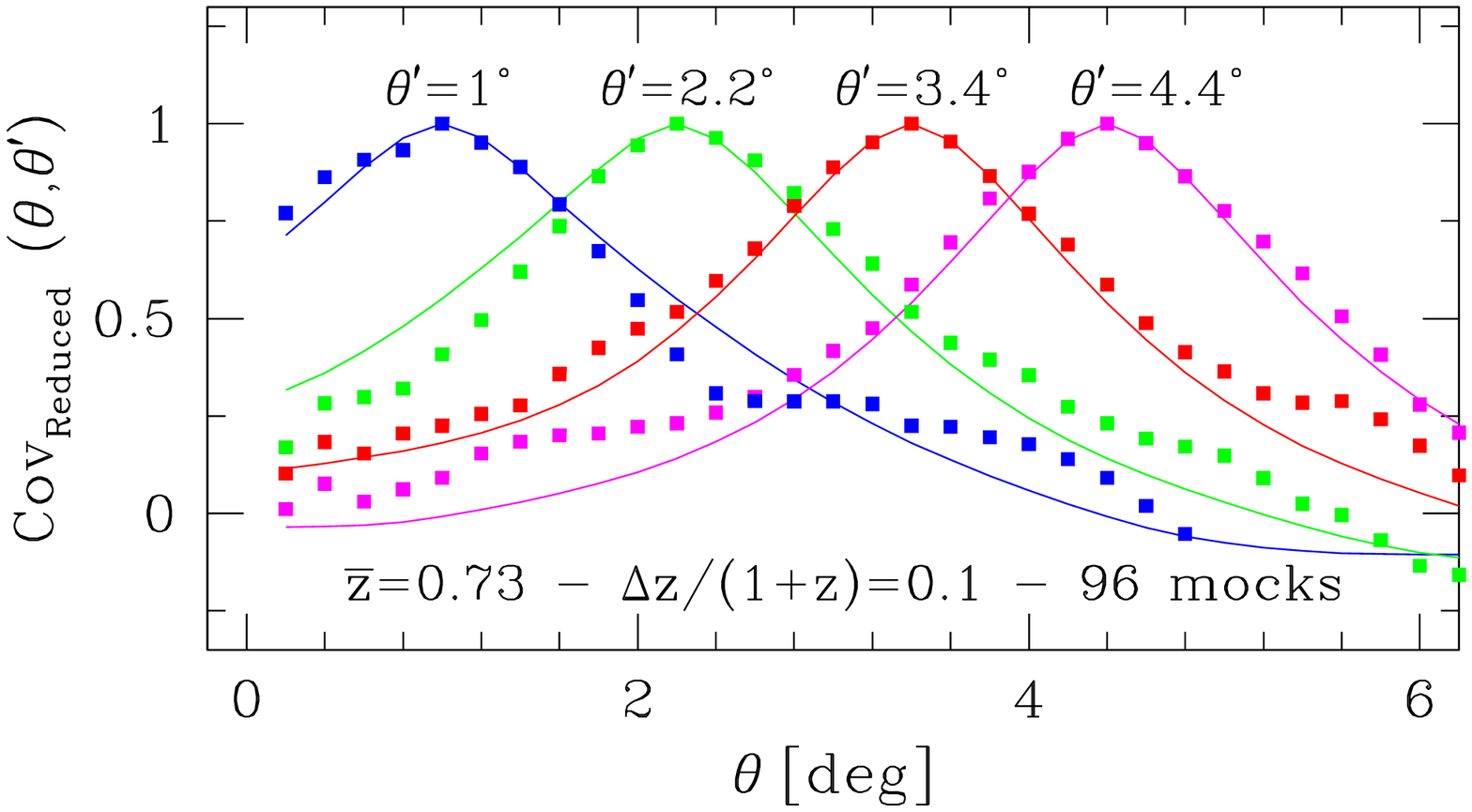}\\
\includegraphics[width=0.45\textwidth]{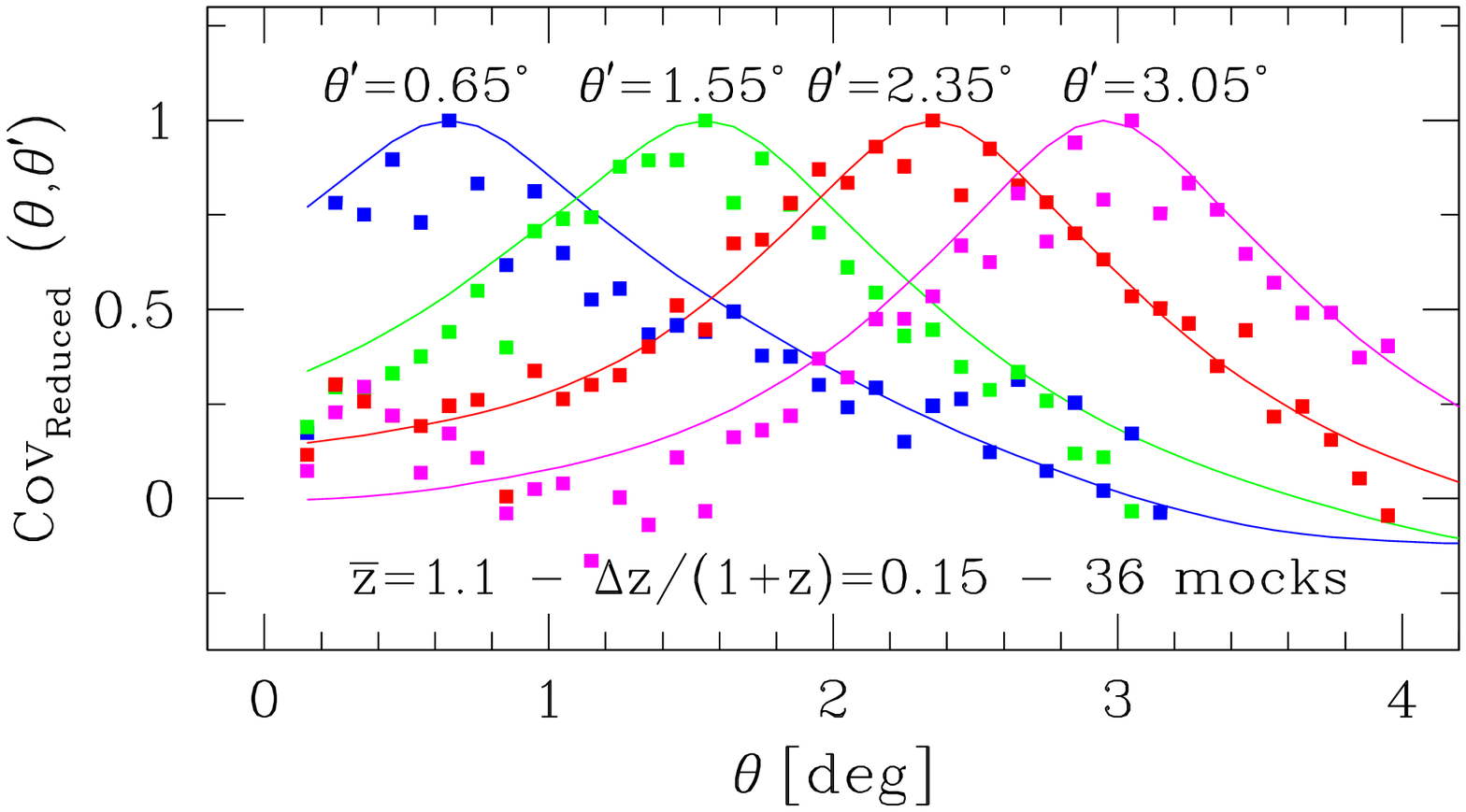}
\caption{{\it Rows of the Reduced Covariance Matrix in Real Space.} Reduced
  covariance between $\hat{w}(\theta)$ and ${\hat w}(\theta^{\prime})$
  for a fixed value
  of $\theta^{\prime}$. From top to bottom we show bin
  configurations with larger width at low and high redshifts and
  smaller at intermediate values (resembling the characteristic
  performance of photo-z estimates). Solid lines are the predictions
  from Eq.~(\ref{eq:CovW}). Notice that close to 200 mocks are
  necessary for a robust estimation of the covariance. In each panel the third from
  left value of $\theta^{\prime}$ corresponds to the angular BAO scale.}
\label{fig:cov_row}
\end{center}\end{figure}

In order to make a more quantitative evaluation of our analytical
expressions we instead plot in Fig.~\ref{fig:cov_row} four
rows of the reduced covariance matrix, that is, the correlation
between $w(\theta)$ and $w(\theta^{\prime})$ as a function of
$\theta^{\prime}$ (at fixed values of $\theta$, as labeled in the
plot). We have chosen to do this at our four mean redshits,
$z=\,0.3,\,0.5,\,0.73,\,1.1$ and characteristics  widths $0.15,\,0.1,\,0.1,\,0.15$ respectively
(from top to bottom). This election reflects the fact that the
calibration of photometric redshifts (mimicked here by the bin width) are better at intermediate
redshift, and worse towards low and high $z$.  Solid lines in Fig. ~\ref{fig:cov_row} are the predictions from Eq.~(\ref{eq:CovW}) using $C_{\ell}$ from Eq.~(\ref{eq:clsplit}).

In all cases, the values of $\theta^{\prime}$ were selected in such
a way to span the whole angular regime where the analytical model for
$w(\theta)$ is accurate.
In particular, the third from the left value of $\theta^{\prime}$
corresponds approximately to the angular
position of the BAO peak

This figure reflects the high degree of correlation between
$\theta$-bins even when widely separated, which is characteristics of
Configuration space. Nonetheless the theoretical estimates accounts
for the whole shape, particularly close to the diagonal. Away from it
there are wiggly features in the measurements that we identify with an insufficient number of mocks
(e.g. only the upper panel with $\sim 400$ mocks does not show these
features). The bottom panel shows an extreme case of using only $\sim
40$ mocks, and the associated level of noise.

\begin{figure}
\begin{center}
\includegraphics[width=0.45\textwidth]{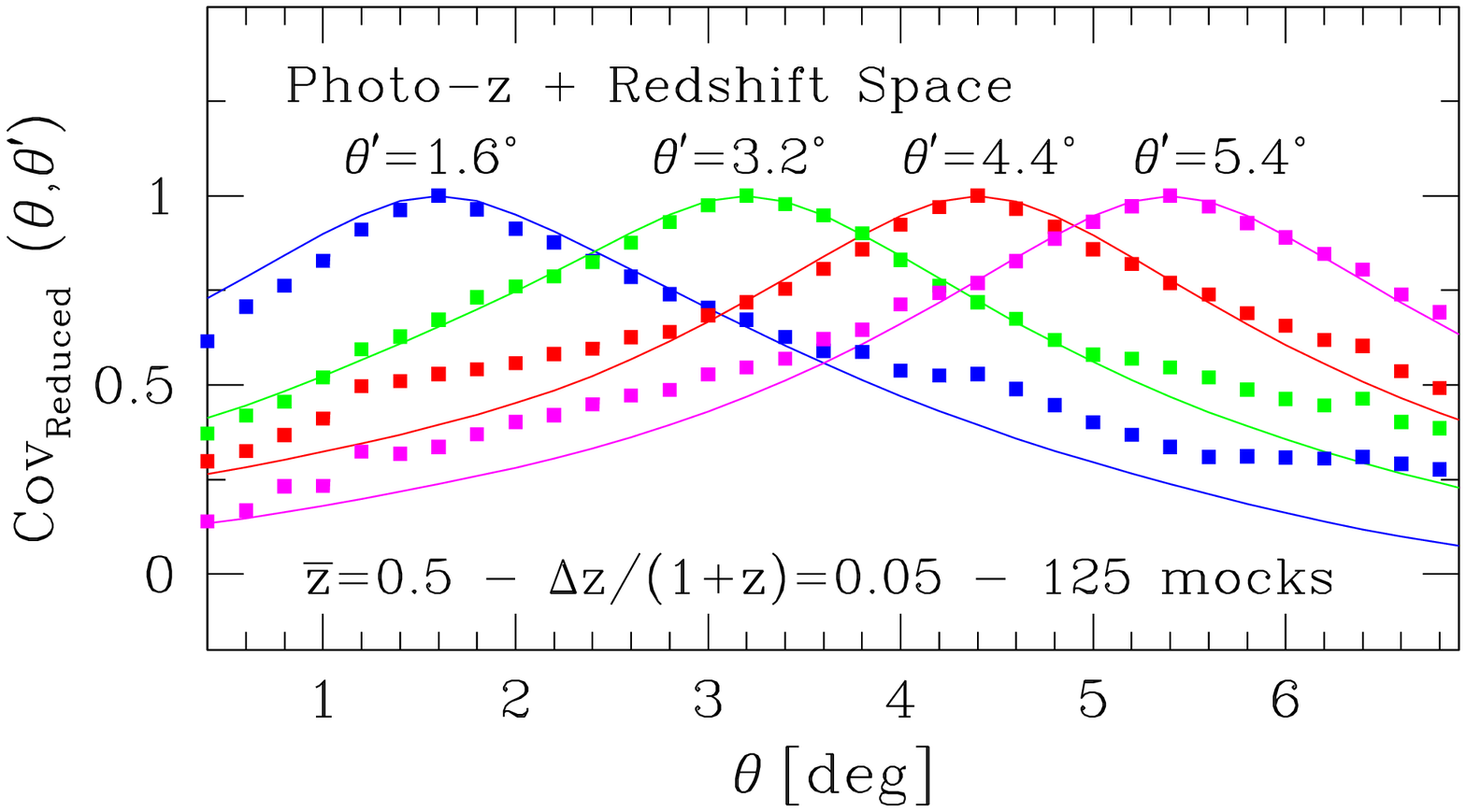}\\
\includegraphics[width=0.45\textwidth]{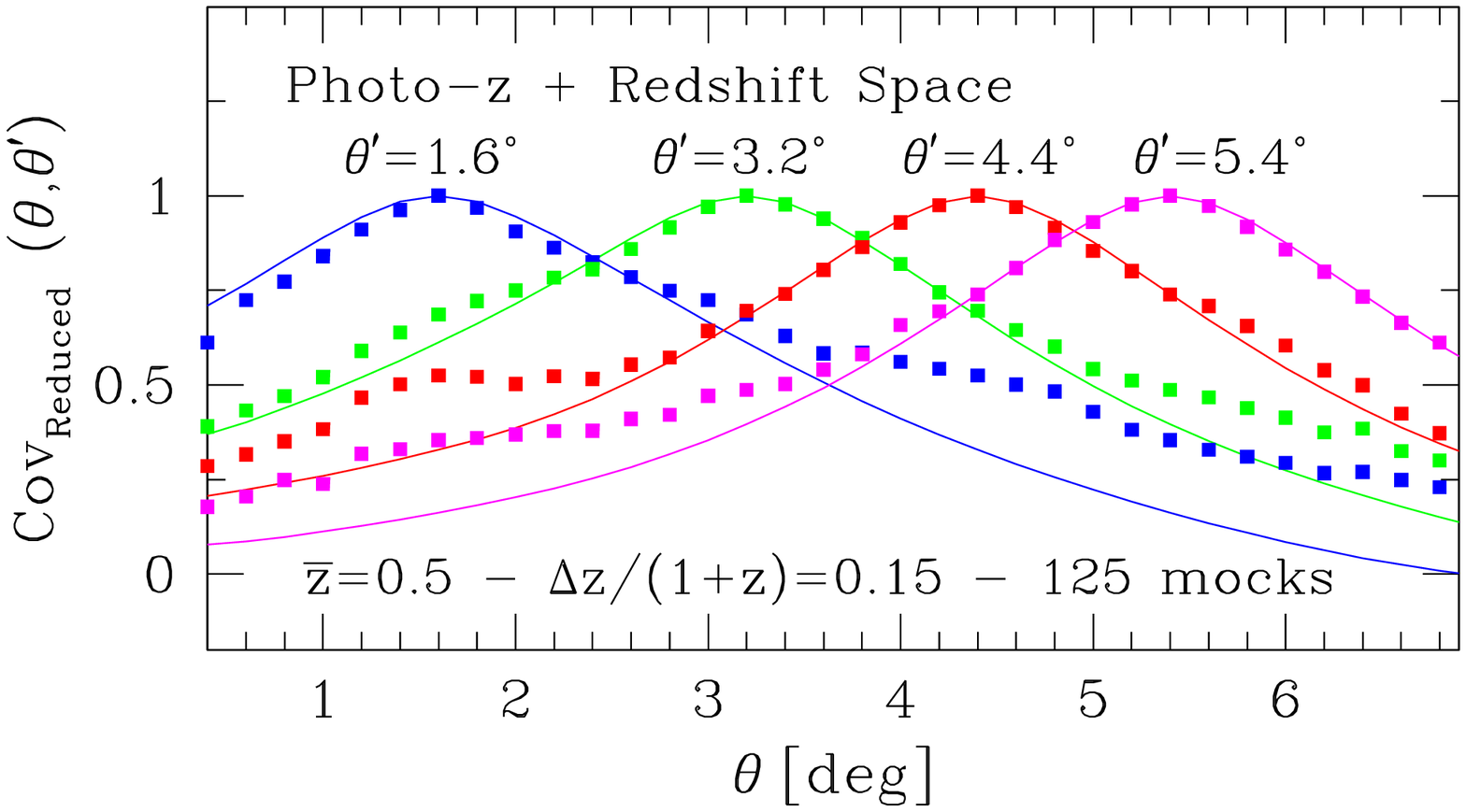}
\caption{{\it Same as Fig. 9,} but for $w(\theta)$ measured
in mocks including photo-z and redshift distortions effects. Top panel
corresponds to a top-hat bin of width $\Delta z / (1+z) = 0.05$ centered at (photometric) $\bar{z}=0.5$,
while bottom panel to $\Delta z / (1+z) = 0.15$.}
\label{fig:cov_row_photoz_zspace}
\end{center}\end{figure}

Figure \ref{fig:cov_row_photoz_zspace} shows the same quantity now
measured in our mocks including photo-z ($\sigma_z=0.06$) and redshift distortions
effects ($\beta = 0.70$). Here the modeling, which follows from
Eqs.~(\ref{eq:cl},\ref{eq:psi},\ref{eq:psir},\ref{eq:clsplit}) 
Eq, works remarkably well close to the diagonal. Away from it, at
angular separations of $\theta \sim 2-1.5^{\circ}$, we see that the
model under-estimates the reduced covariance. This is more noticeable
when the reduce covariance drops $0.5$. This could be due to
shortcomes of the theoretical model or to an insufficient number of
mocks. Nonetheless, we have checked that this difference does not
translate into different estimates of cosmological parameters when
using the theoretical or ensemble covariance matrix. This will be
presented in a forthcoming paper \cite{ross10}.

\begin{figure}
\begin{center}
\includegraphics[width=0.235\textwidth]{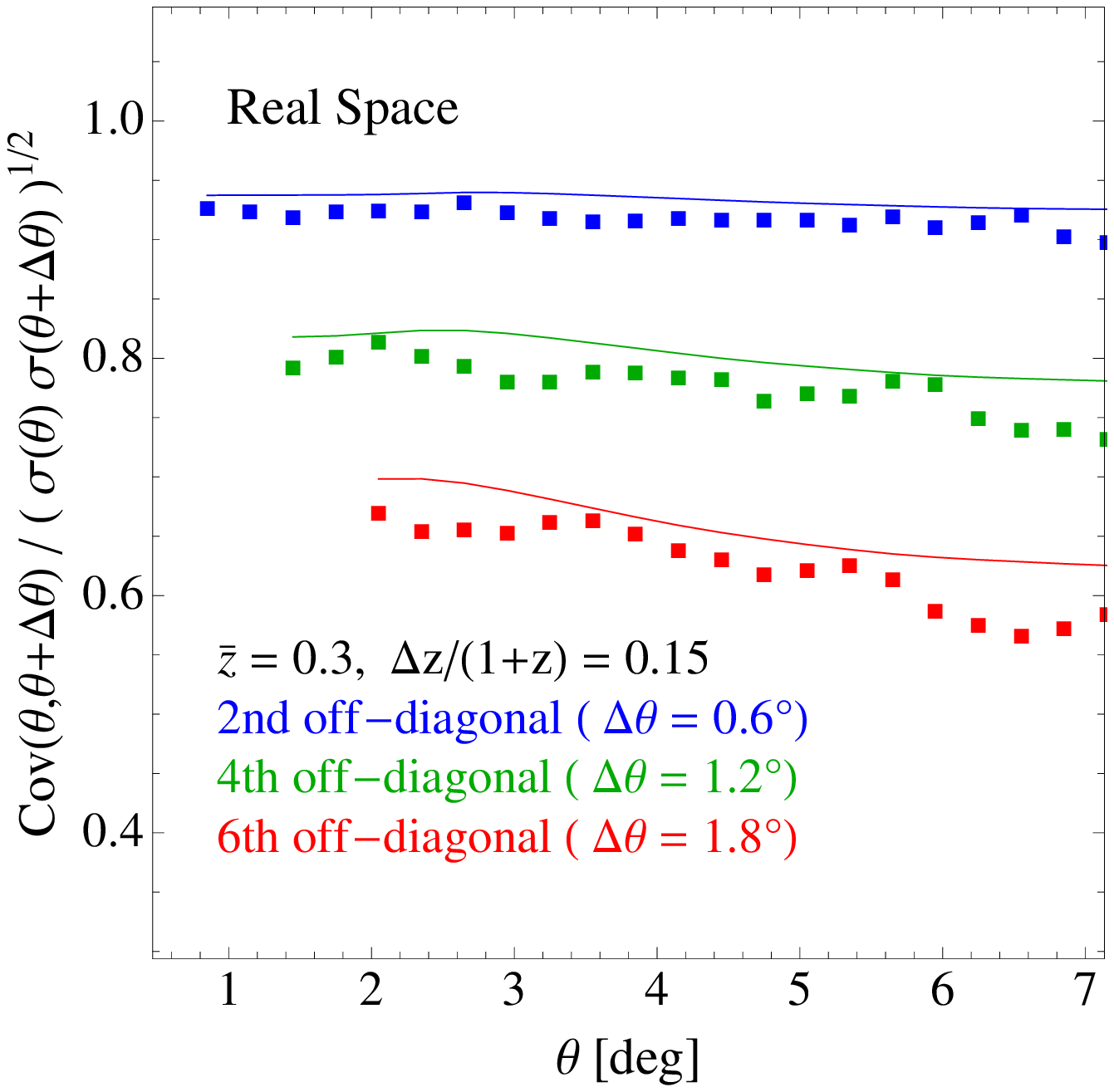} 
\includegraphics[width=0.235\textwidth]{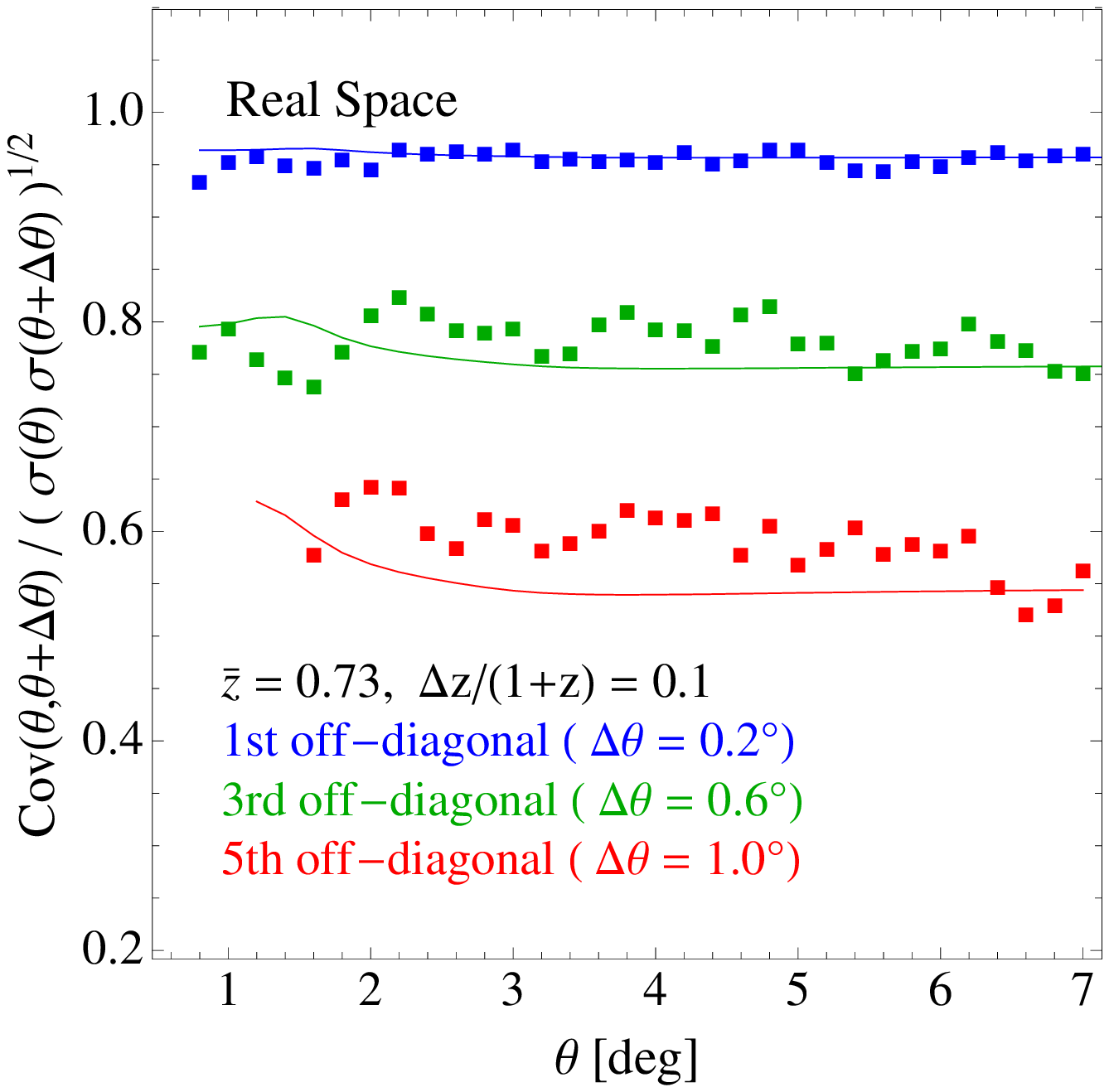} \\
\includegraphics[width=0.235\textwidth]{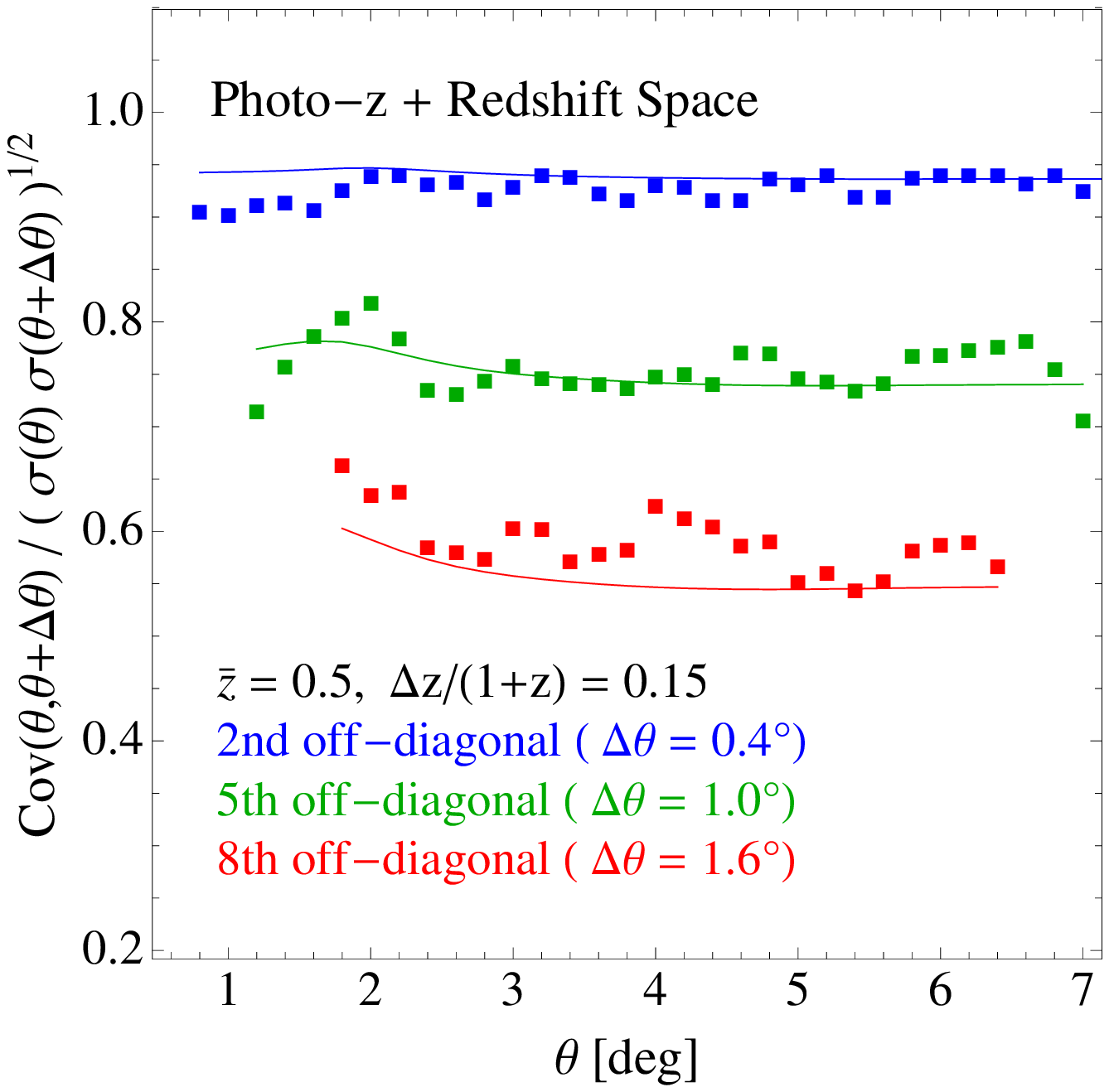} 
\includegraphics[width=0.235\textwidth]{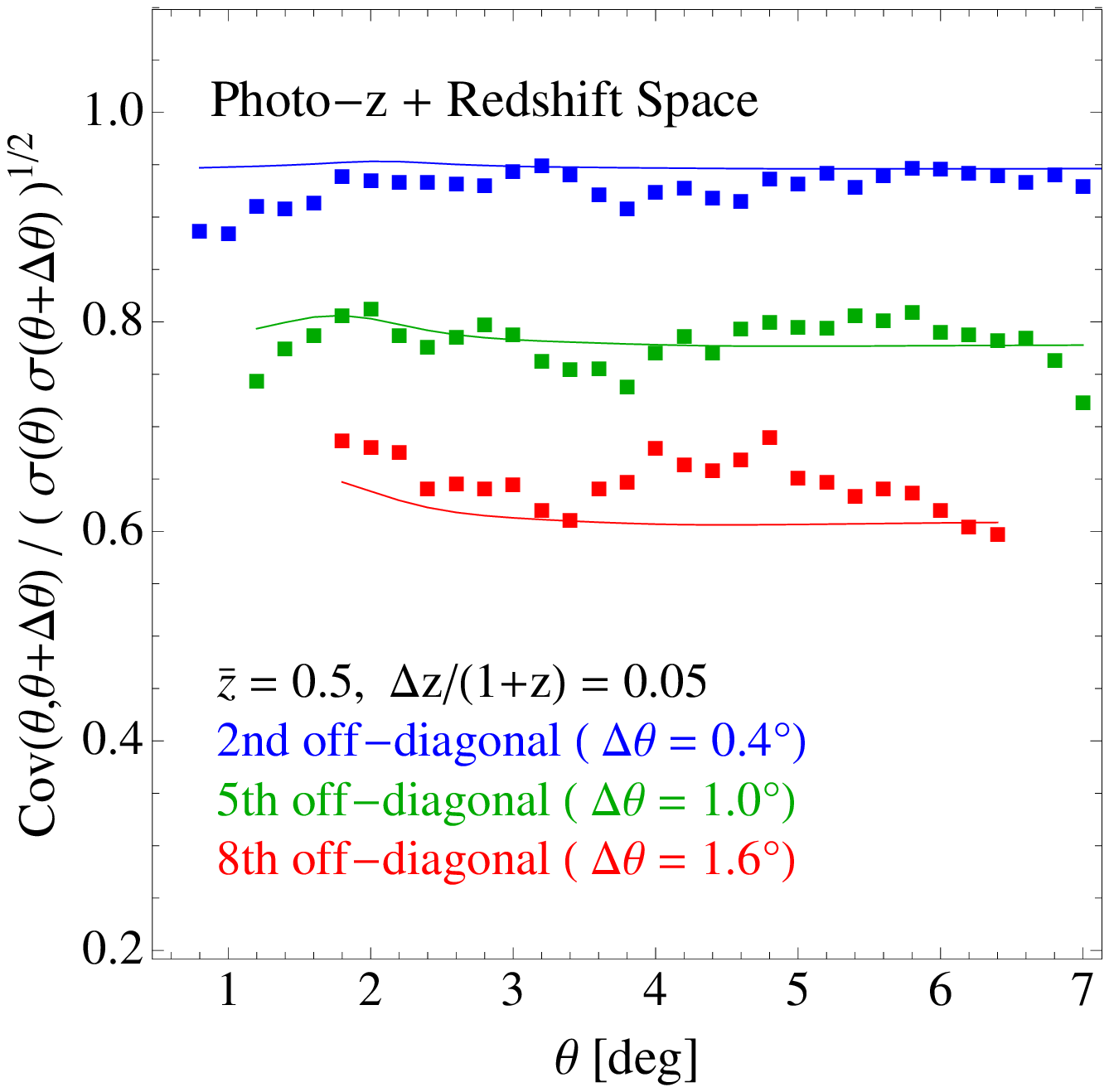}
\caption{{\it Off-diagonal Covariance in Real Space
    (Top Panels) and in Redshift + Photo-z Space (Bottom Panels).}
  Panels show the reduced covariance matrix between $w(\theta)$ and
  $w(\theta+\Delta \theta)$ for increasing values of $\theta$ and
  fixed $\Delta \theta$. Mean red-shift and width of each bin
  configuration are labeled in the plot. Solid lines
  are the predictions from Eq.~(\ref{eq:CovW}) while symbols denote
  measurements in the corresponding mocks.}
\label{fig:cov_offdiagonal}
\end{center}\end{figure}

Figure.~\ref{fig:cov_offdiagonal} shows instead the off-diagonal
elements of the error matrix ${\rm Cov}(\theta, \theta+\Delta
\theta)$ for fixed value of $\Delta \theta$, and as a function of $\theta$.
Top panels correspond to two bins in real space (as labeled) while
bottom panels to our mocks in photo-z $+$ redshift space at
$\bar{z}=0.5$ and $\Delta z/(1+z)=0.015$ (left) or $0.05$ (right). 
The value of $\Delta \theta$ was chosen to sample the
full angular width of the BAO feature. In all, the model performs quite well,
with no general or obvious systematic deviations. It gives accurate results
close to the diagonal (i.e. smallest $\Delta \theta$) and can deviate
by  $\sim 10\%-20\%$ for large $\Delta \theta$ (in a regime where
covariance is noisy nonetheless).

\begin{figure}
\begin{center}
\includegraphics[width=0.45\textwidth]{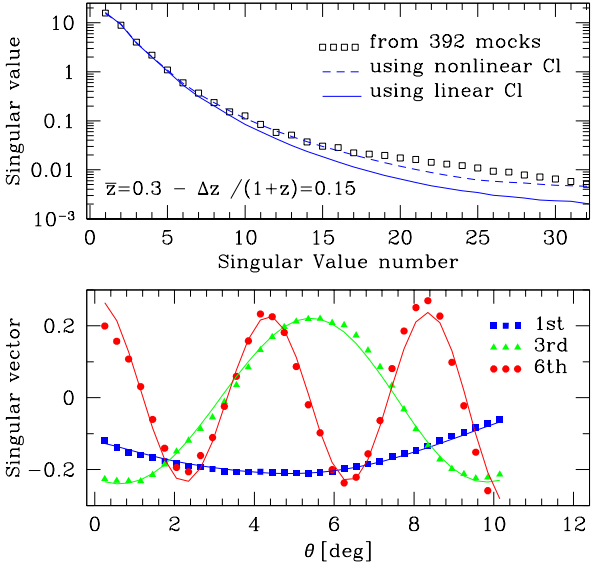}
\caption{{\it Singular Value Decomposition} of the Error Matrix
  measured in 392 mocks at $z=0.5 - \Delta z/(1+z)=0.15$. Using the
  linear power spectrum is possible to reproduce very accurately $\sim 15$ singular
  values (top panel) and their singular vectors (bottom
  panel). Including nonlinear clustering effects into $C_{\ell}$
  leads to a good match for all singular values. The case shown corresponds to unbiased tracers and negligible shot-noise.}
\label{fig:svd_DM}
\end{center}\end{figure}

We finish this section comparing theory and mocks at the level of
the Singular Value Decomposition (SVD) to illustrate how well we can model
the covariance matrix in a different basis, one which is sometimes
used to analyze real data.
The top panel of Fig.~\ref{fig:svd_DM} shows
the Singular values of the covariance matrix measured in the ensemble of 392 mocks with
$\bar{z}=0.5-\Delta z/(1+z)=0.15$. For the SVD we used a range of
 scales $[0.25^{\circ}-8^{\circ}]$, divided in $34$ bins of
width $0.3^{\circ}$ (but results are robust for other ranges and binnings).
The analytical model, using the linear power spectrum (solid blue line),
describe accurately 10 to 15 singular values. We have tested that using more than 10 or
15  singular values the estimate of cosmological parameters
have negligible impact in the outcome (see also \cite{eisenstein01}), as they
describe very short-range correlations (see bottom panel). These results will be presented in
a forthcoming paper \cite{cabre10}. There is however, a simple way to
improve on the agreement. In dashed line we show the results when
using the ``measured'' $C_{\ell}$ spectra in Eq.~(\ref{eq:CovW})
\footnote{We used the publicly available {\tt SpICE} code
  \cite{szapudi01} to measure the $C_\ell$ spectra, which is particularly
  suitable to include mask effects due to partial sky coverage}. This
indicates that the recovery of these high singular values is affected
by the high ${\ell}$ tail of $C_{\ell}$ where non-linear effects increase the
power over their linear value. On the one hand this will have little
importance in practical situations where this regime will be most
probably dominated by shot-noise. On the other hand it can be easily
modeled using fits to the nonlinear power spectrum, such as {\tt
  halofit}, into the Limber formula. In the bottom panel of Fig.~\ref{fig:svd_DM}
we show the singular vectors corresponding to singular values $1$, $3$
and $6$.

So far we have not tested our model against measurements of
$w(\theta)$ covariance between different redshift bins. 
The full covariance matrix in this case can be easily obtained
with the same formalism described in Sec.~\ref{sec:errors} with
 the full sky variance of $C^{ij}_{\ell}$ spectra between bins $i$ and
$j$ given by,
\beq
{\rm Var}(C^{ij}_{\ell})=\frac{1}{2\ell+1}[(C^i_{\ell}+n_i)(C^j_{\ell}+n_j) + (C^{ij}_{\ell})^2]
\eeq
what leads to (for partial sky coverage),
\begin{eqnarray}
{\rm Cov}^{ij}_{\theta \theta^{\prime}} &=&
\frac{1}{f_{sky}}\sum_{\ell\ge 0} \frac{2\ell+1}{(4\pi)^2}
P_{\ell}(\cos\theta) P_{\ell}(\cos \theta^{\prime}) \left[(C^i_{\ell}+n_i) \right. \nonumber \\
 &  & \left. (C^j_{\ell}+n_j) + (C^{ij}_{\ell})^2\right]
\end{eqnarray}
Here $C^{ij}_{\ell}$ is obtained by simply replacing $\Psi^2_\ell\rightarrow \Psi_{i,\ell}
\Psi_{j,\ell}$ in Eq.~(\ref{eq:cl}), where $\Psi_i$ is given by
Eq.~(\ref{eq:psi}) with the radial selection corresponding
to the $i-th$ redshift bin.

\subsection{The impact of shot-noise}
\label{sec:impact_shot_noise}

Previous sections showed that the expression in Eq.~(\ref{eq:CovW})
can describe remarkably well the error due to {\it sampling
variance} in the measurement of the angular correlation function in
redshift bins. 

We now turn into the problem of describing the error due to
shot-noise. To this end we will concentrate on different halo samples
as tracers (from Sec.~\ref{sec:biased_tracers}) and 
test whether the standard Poisson shot-noise term in Eq.~(\ref{eq:CovW}) can account for the
increase in errors due to their low number density.

\begin{figure}
\begin{center}
\includegraphics[width=0.4\textwidth]{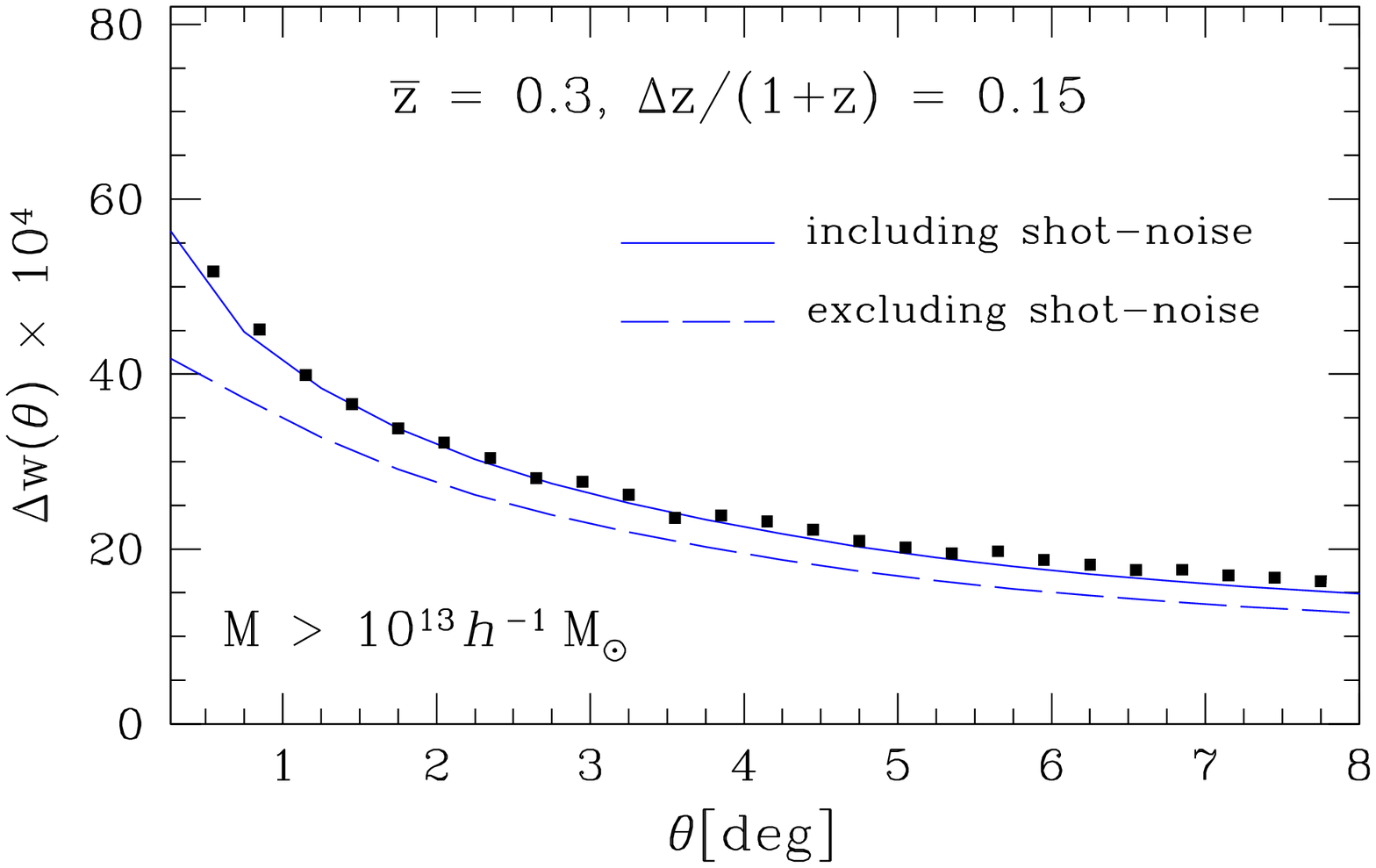} \\
\includegraphics[width=0.4\textwidth]{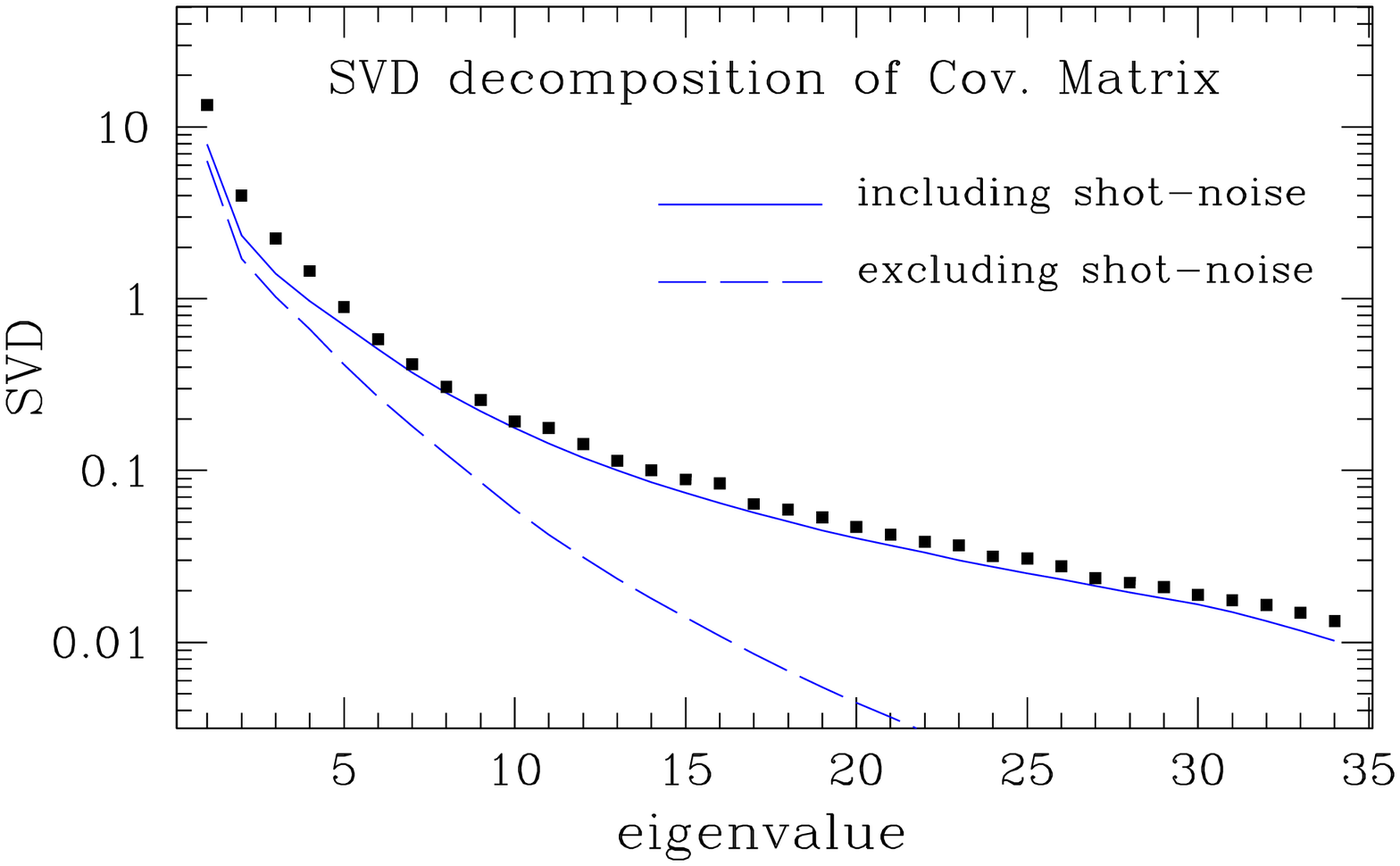}
\caption{{\it The effect of shot-noise at $z=0.3$.} Diagonal error (top panel)
  and Singular Value Decomposition of the Covariance Matrix (bottom
  panel) for the bin $z=0.3$ and $\Delta z/(1+z)=0.15$ for a sample of halos with
  $M> \times 10^{13}\Msun$. The (Poisson) shot-noise term increases the error by $\sim
  20\%$ on top of sampling variance (solid and dashed lines are the
  model with and without the $1/{\bar n}$ term) and is critical to recover the correct singular eigenvalues. We used 392 mocks for this case.}
\label{fig:shotnoise-1}
\end{center}\end{figure}

For concreteness we focus on two characteristic z-bin configurations. 
Figure~\ref{fig:shotnoise-1} corresponds to halos with $M \ge 10^{13}
\Msun$ in the bin at ${\bar z}=0.3$ and width $\Delta z/(1+z)=0.15$.
This sample has a bias $b(z=0.3)=2.36$ (see Fig.~\ref{fig:bias}), what
corresponds to $b\sim 2$ if linearly evolved to $z=0$
\cite{fry1996}. The sample has an angular
abundance of ${\rm N}/\Omega \sim 17$ ${\rm halos}/{\rm deg}^2$ what
gives a shot-noise contribution, $\Omega/{\rm N}$, of order $1.8\times 10^{-5} {\rm
  sr}^2$. This corresponds to $\bar{n} \,b^2 \,C_\ell\sim 1$ at $\ell \sim 200$.

Top panel of Fig.~\ref{fig:shotnoise-1} shows the r.m.s. error measured in
the ensemble of $392$ mock redshift bins while lines are the
predictions from Eq.~(\ref{eq:CovW}) using the exact $C_{\ell}$ integration and including Poisson
shot-noise (solid blue) or neglecting it (dashed blue). As we see the
presence of shot-noise increases the error by $~20\%$ (in this case),
but this can be very well modeled by the addition of the simple
Poisson shot-noise $1/{\bar n}$ contribution to $C_{\ell}$. 
Bottom panel shows the singular values of the covariance matrix
SVD. Almost none of the singular values is well recovered when
shot-noise is neglected. 

\begin{figure}
\begin{center}
\includegraphics[width=0.4\textwidth]{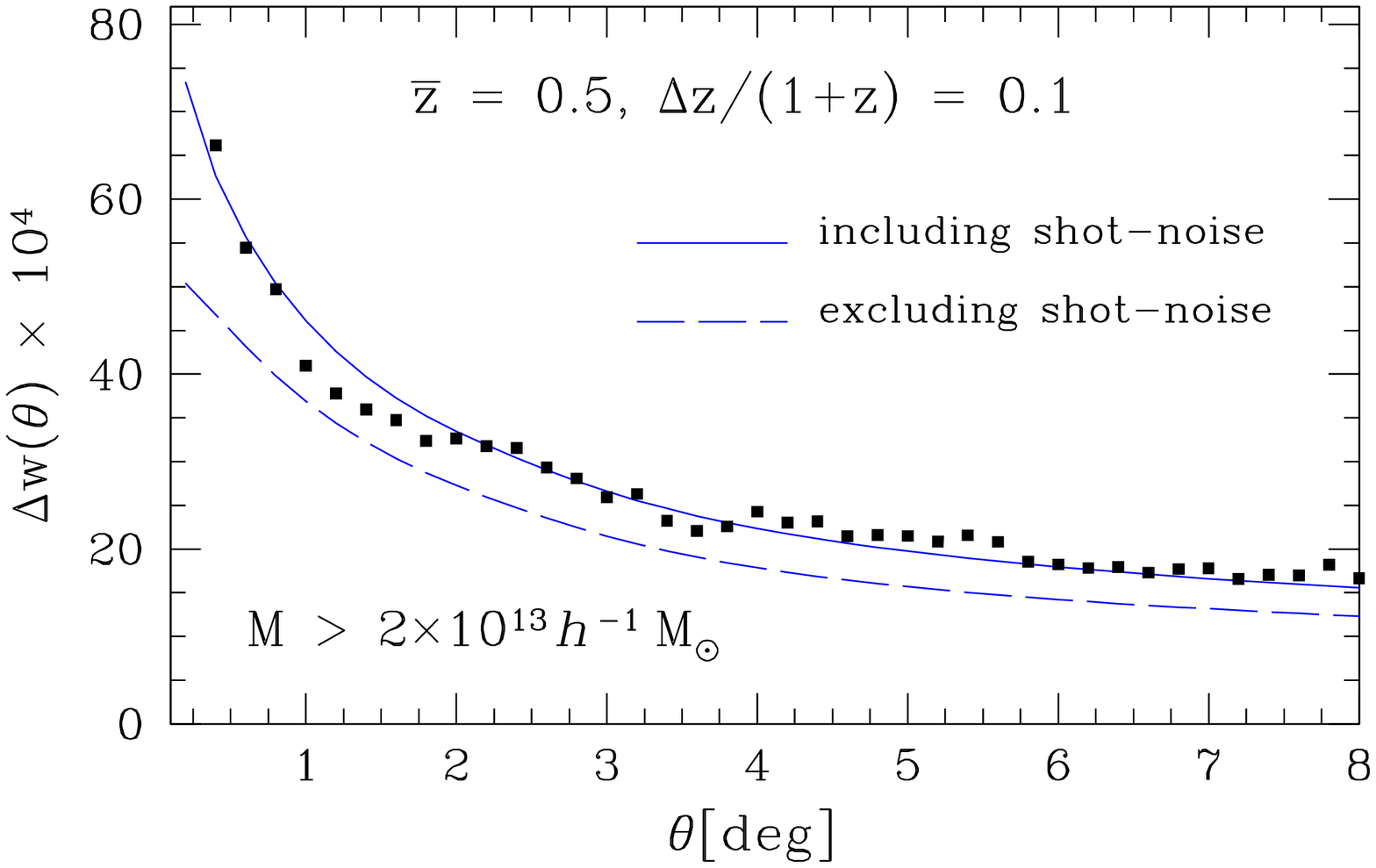} \\
\includegraphics[width=0.4\textwidth]{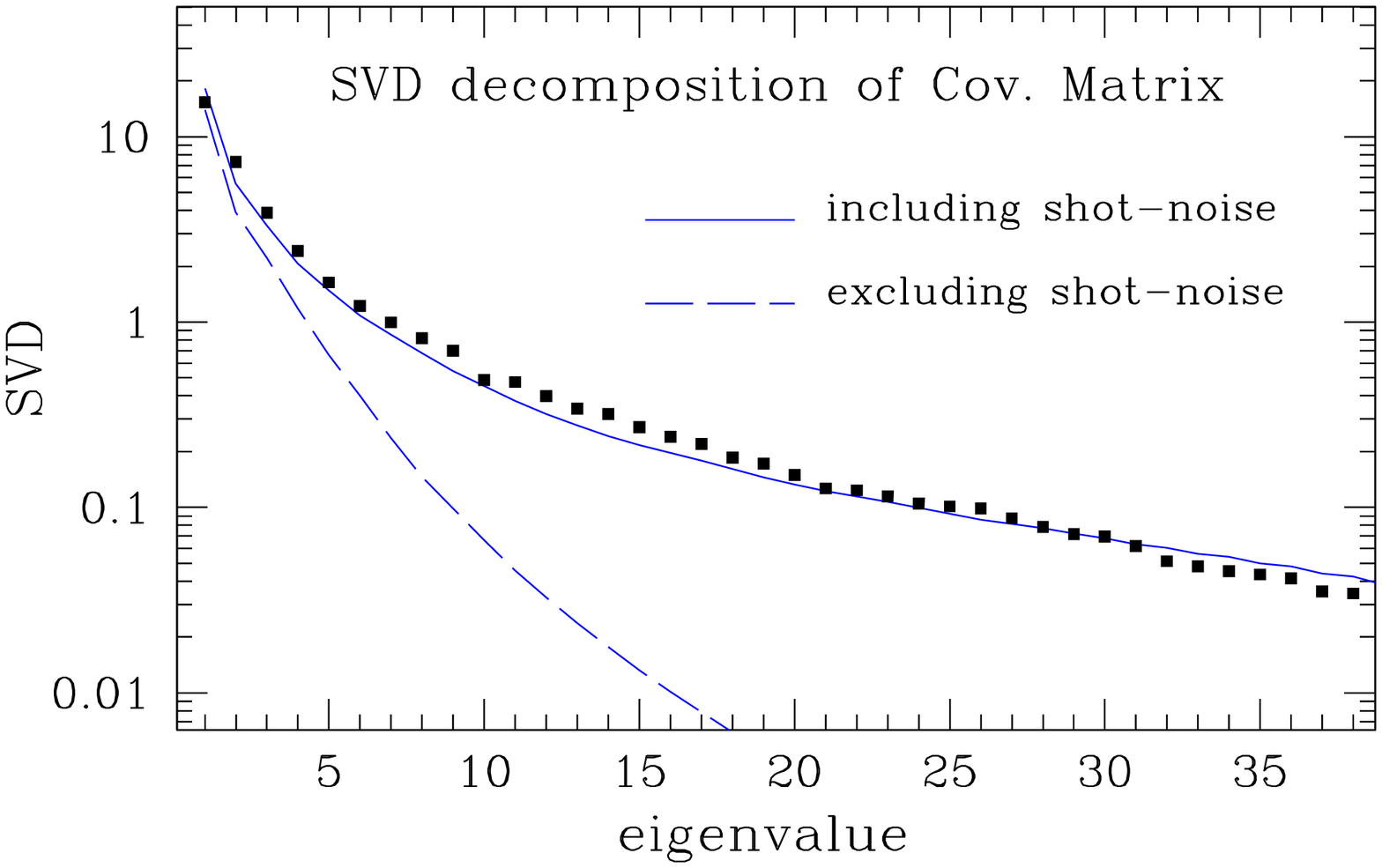}
\caption{{\it The effect of shot-noise at $z=0.5$.} Diagonal error (top panel)
  and Singular Value Decomposition of the Covariance Matrix (bottom
  panel) for the bin $z=0.5$ and $\Delta z/(1+z)=0.1$ for halos with
  $M>2\times 10^{13}\Msun$. Conclusions shared with Fig.~\ref{fig:shotnoise-1}. We used 175 mocks for this test.}
\label{fig:shotnoise-2}
\end{center}\end{figure}

In turn, Fig.~\ref{fig:shotnoise-2} corresponds to halos with $M \ge 2 \times
10^{13} \Msun$ at higher redshift, in the bin ${\bar z}=0.5-\Delta
z/(1+z)=0.1$ (the halo mass-cut was chosen slightly higher at higher redshift to
resemble a flux-limited survey). In this case the bias is
$b(z=0.5)=2.93$ ($b\sim 2.3$ at $z=0$) with an angular abundance of $\sim 11$ ${\rm
  halos}/{\rm deg}^2$ ($\bar{n} \,b^2 \,C_\ell\sim 1$ at $\ell \sim 200$).
The shot-noise increases the error by $\sim 30\%$ on top of the
sampling variance (solid vs. dashed lines in the top panel) and plays
a crucial role in the SVD eigenvalues (bottom panel). For this case
the SVD was done using $40$ bins of width $0.2^{\circ}$ in the range $[0.2^{\circ}-8^{\circ}]$.

Remarkably, in both cases (Figs.~\ref{fig:shotnoise-1} and \ref{fig:shotnoise-2}), the inclusion of the
simple Poisson white-noise on top of the sampling variance in
Eq.~(\ref{eq:CovW}) accounts very well for all singular values as well
as the diagonal error (solid lines).

Arguably, the halo samples used in this test are too much dominated by
shot-noise. We have also done the exercise of gradually increasing the
relative contribution of shot-noise in front of sampling variance by 
increasingly under-sampling the dark matter field. As expected, the
error is always recovered by the model just by increasing the Poisson term
in Eq.~(\ref{eq:CovW}) accordingly.

\subsection{Independence on bin-size}

When estimating a covariance matrix one important issue to be
considered is whether the binning used for the measurements have an impact on the recovered
correlations. Hence, in this section we aim at showing that our estimates for the
covariance matrix are not affected by simple rebinning of the measurements,
and so our model remains applicable in a general sense.

The selection of bin size, and binning strategy (e.g. linear
vs. logarithmic), is particularly important when analyzing real data
because one might be able to optimize the recovered model constrains by a
better split between true signal from noise. This is not our goal but
still is appropriate to discuss our binning choices.

In general, a too narrow bin could lead to noisy (due to shot-noise
in the bin) and very correlated errors.
Too wide would smooth and ``diagonalize'' the covariance but at the
expense of leaving only few data points where to compare with the model.
In turn, a logarithmic instead of linear binning is preferably when
dealing with a large dynamic range in scales.
For the problem considered in this paper, the angular correlation at
large scales, we are never beyond a factor of $10$ in dynamic range in
$\theta$, see for instance Fig.~\ref{fig:angular_correlation}. In turn we want to sample
this range uniformly, which inclined ourselves to linear binning.

In addition, since we have a feature in our signal that
we want to resolve  (the BAO bump) we need several
  bins across it which sets our maximum bin-size. We then chose to have
  $\sim 5 - 10$ bins across it which led to $\Delta \theta =
0.3^{\circ}$ at $z=0.3$, $\Delta \theta =
0.2^{\circ}$ at $z=0.5-0.73$ and $\Delta \theta =
0.1^{\circ}$ at $z=1.1$. Still, we could have selected thinner bins.

\begin{figure}
\begin{center}
\includegraphics[width=0.34\textwidth]{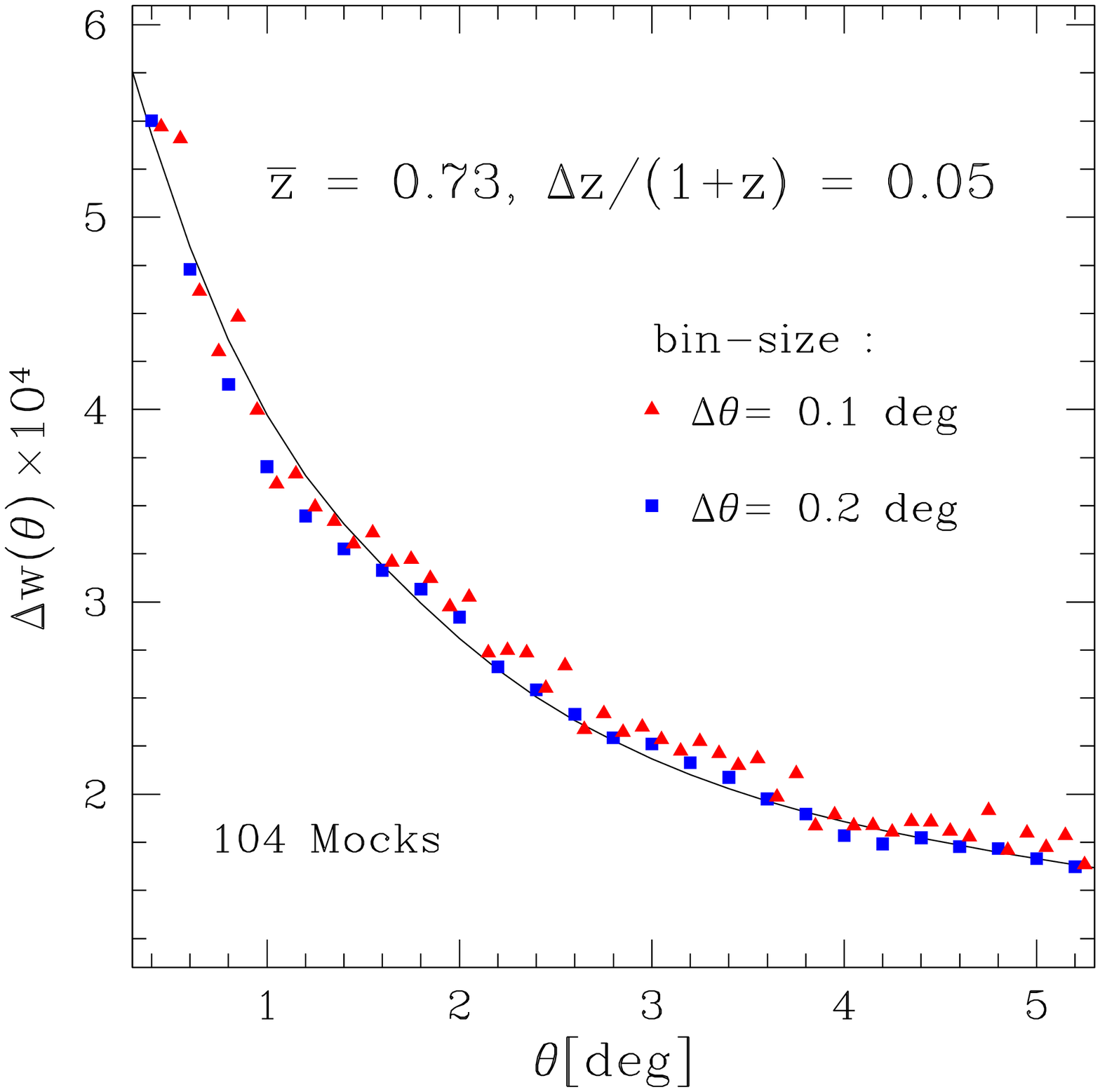} \\
\includegraphics[width=0.34\textwidth]{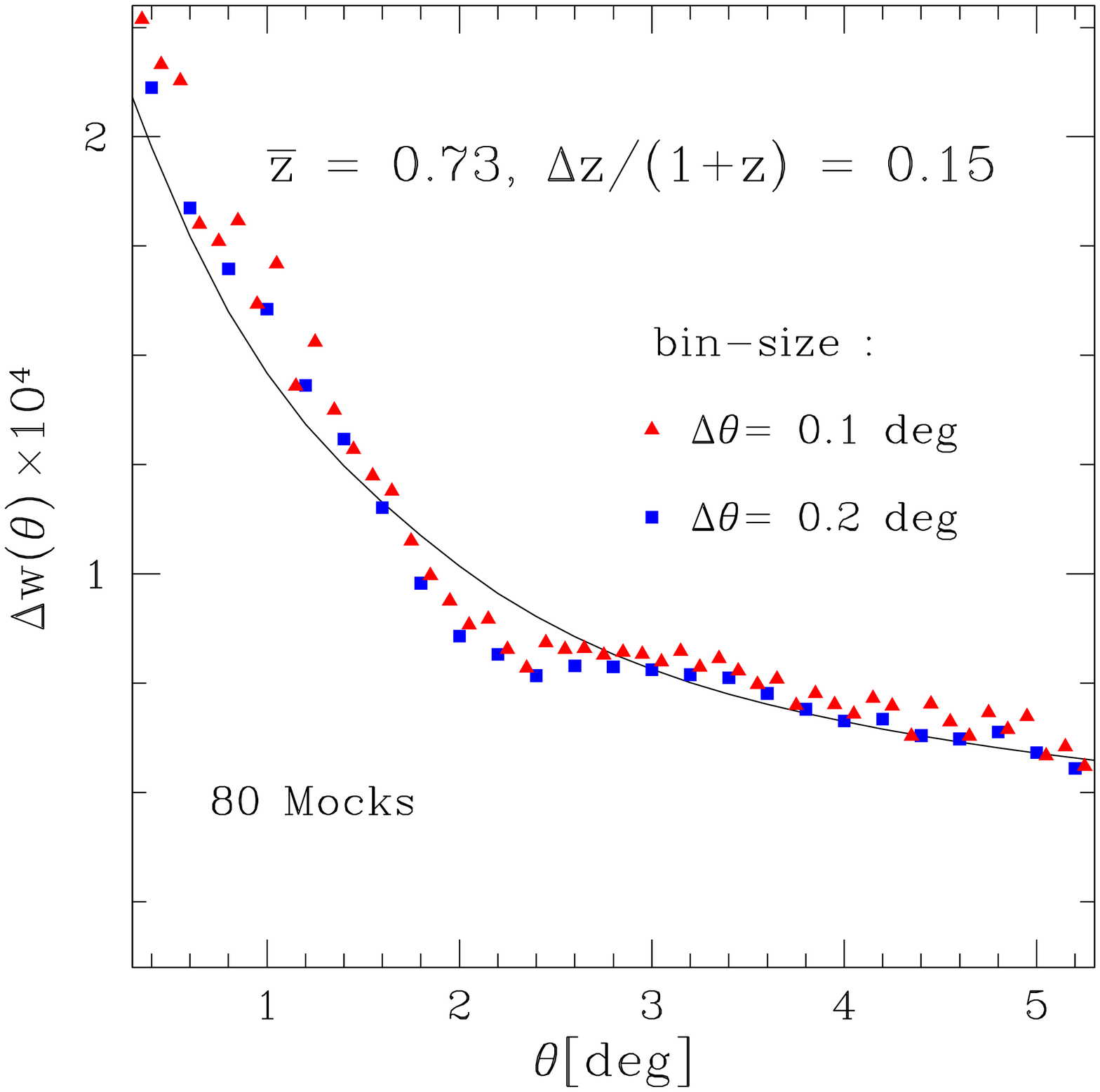} \\
\includegraphics[width=0.34\textwidth]{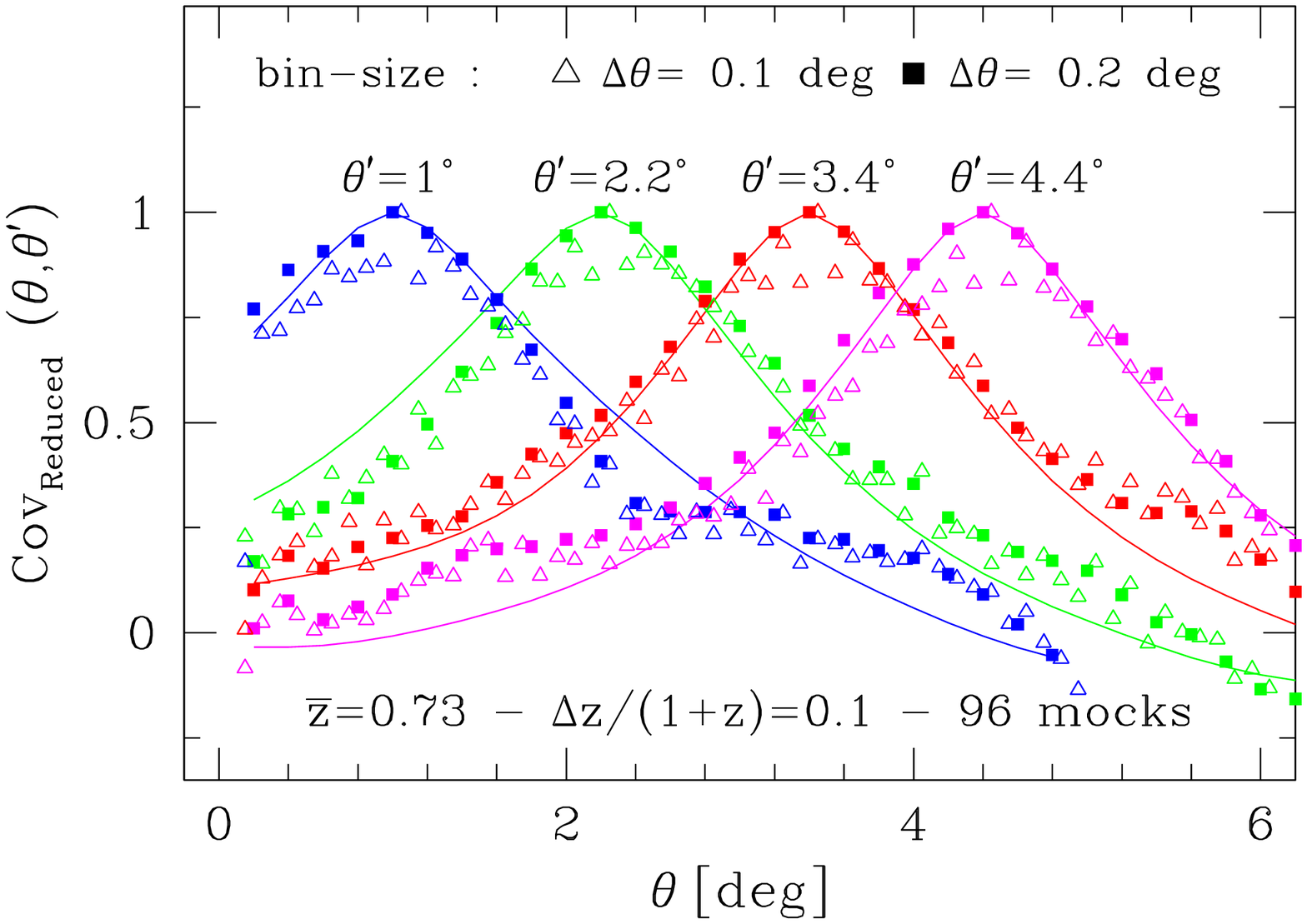} \\
\caption{{\it Independence on angular bin-size}. The figure shows how
  the diagonal error (top two panels) and reduced covariance (bottom
  panel) vary when one halves the angular bin-size. Recovered
  correlations becomes noiser but no appreciable systematic difference with the
  model is found.}
\label{fig:binsize}
\end{center}\end{figure}

Therefore, in Fig.~\ref{fig:binsize} we show the measured
diagonal error and the reduced covariance (i.e. same as Figs.
\ref{fig:ewtheta} and \ref{fig:cov_row}) for
the mock configuration centered at ${\bar z}=0.73$ (in real space) for a bin-size half the
one used throughout the paper. Top two panels correspond to the
diagonal error while bottom to rows of the reduced covariance
matrix. Both set of measurements agree with each other. The ones with
smaller $\Delta \theta$ seem slightly above in diagonal error and
below in reduced covariance. But, as expected, they also show more noisy.
In all, our model (shown in solid lines) describes both equally well.

\section{Conclusions}
\label{sec:conclusions}

The field of large scale cosmological structure will
undergo an unprecedented era in the immediate future with several large
observational  campaigns proposed or under implementation.
Many of these surveys, such as DES, PanStarrs and LSST, will use photometric techniques to estimate the
radial position of galaxies instead of measuring their full spectra, which is more time-demanding
. This gain allows to survey wider areas and fainter objects, but at
the exchange of increasing the uncertainties in the true redshifts and 
degrading the radial clustering amplitude. The proposal is then to
split the data into redshift bins and exploit the information available
from angular correlation functions, provided with an accurate
determination of the measurement errors. Yet, this goal can only be
accomplished if we develop accurate models for the signal and its
errors that take into account all relevant effects and are robustly tested in
realistic scenarios. 
In this paper we addresseded this issue in a comprehensive way.

We first developed an extensive set of mock catalogues (in the form of
redshift bins) reproducing the angular coverage and radial
distribution of a photometric survey like DES. We did it in increasing
steps of realism, first in real space, then including redshift
distortions or photometric errors and finally altogether.
For this we used a large
N-body simulation (of $\sim 450 \Gpccube$ simulated volume) provided
by the MICE collaboration ({\tt http://www.ice.cat/mice}).  
These mocks can be regarded as independent realizations as their volume
overlap  is minimal and therefore provide a unique statistical framework for model
testing (see Table~\ref{Table:mocks}). We claim they are also
equivalent to a light-cone analysis as we are doing narrow redshift
bins, which involve negligible evolution.

We next put forward a model for the angular correlation
function $w(\theta)$ accounting for all the relevant effects,
namely bin projection, nonlinear gravitational evolution, linear bias, redshift
space distortions and photo-z errors. 

An exhaustive comparison of our model for $w(\theta)$ against the
mock measurements showed a remarkably good agreement for a wide range
of $\theta$ and scenario (real space, photo-z, RSD and photo-z + RSD), validating
the treatment of the different effects and opening the door to the use
of this probe for real data analysis. Nonlinear gravitational
evolution produces minor distortions in the correlation pattern after
the bin projection.  In turn, analysis of
halo angular clustering showed a very good consistency with a linear
bias assumption. The interplay of photo-z and redshift distortions is the
most important consideration regarding the shape of $w(\theta)$.

Redshift space distortions introduces a large and scale dependent
enhancement of $w(\theta)$, that can reach a factor of a few at BAO
scales (see Fig.~\ref{fig:wtheta_photoz_zspace}).
For our widest bin (where the effect should be least important) it
still rises the amplitude of $w(\theta)$ by $\sim 50\%$ at
$\theta_{BAO}$ (with respect to real space). 
Conversely, photo-z effects lower the clustering amplitude by
extending the effective bin projection. 
 For example, for the widest
bin mentioned before (with $\Delta z \sim 4 \sigma_z$) we find that the two effects counter-act each
other at $\theta_{BAO}$, but leave a scale dependent signal towards smaller angles.
For narrower bins photo-z dominates, but redshift distortions is
certainly not negligible. These trends were concluded from both, our
photo-z + RSD mocks, and the analytical model.

In turn, we showed that the Limber approximation should not be used in
precision analysis of large scale clustering as it leads to the
incorrect shape of $w(\theta)$ in the full range of interesting
scales, and severely misestimates the amplitude of the $C_\ell$ spectra
for $\ell \simlt 40-50$. This is convincingly shown in
Figs.~\ref{fig:Cl},~\ref{fig:wtheta} and \ref{fig:w2Limber}. Another
interesting issue considered was the impact of uncertainties in the
true redshift distribution of objects. This showed an important aspect
since it can lead to several percent changes in both the shape of
$w(\theta)$ and the BAO peak position given the accuracy of present
photo-z estimators.

We would like to highlight that, in the process of describing $w(\theta)$,
we have also investigated a model for the 3-d matter correlation
function that is able to reproduce the clustering signal in a broad
range of scales and redshifts with only 2 parameters. This, discussed
in detail in Appendix \ref{Appendix:A}, can be of grand interest for
future spectroscopic surveys such as BOSS, Hetdex and WiggleZ.

We have made an equally exhaustive effort in modeling and testing the full error matrix
characterizing the measurements of $w(\theta)$. The covariance matrix
is often estimated from the data itself, using internal or re-sampling
methods such as Jack-knife or bootstraping. However, their limitation is still
a matter of some debate \cite{norberg09}. Having a full theoretical model
is thus very suitable for present and future analysis.

We took into account partial sky coverage by assuming that
${\rm Cov}(\theta,\theta^{\prime})$ scales as $f_{sky}^{-1}$. The full
sky situation is then easily treated by translating errors from harmonic
space, where the covariance matrix is diagonal and proportional to
$C_\ell$. Through the angular power spectra we included the same
effects considered for $w(\theta)$ into ${\rm
  Cov}(\theta,\theta^{\prime})$ (photo-z, redshift distortions,
bias) for a typical survey with $f_{sky}=1/8$.

Our modeling of errors recovers the correct variance in $w(\theta)$ as
measured in the mocks for a wide range of bin configurations
, from low to high redshift ($z\sim 0.3-1.1$) and
from thin to thick bins ($100\Mpc-550\Mpc$). And this conclusion extend to the more
realistic cases where we included photo-z effects and redshift space
distortions. In addition we used different halo samples to study cases where the shot-noise was comparable
or larger than the sample variance component of the error. This regime
was also nicely described analytically by adding a standard
Poisson shot-noise contribution to the variance of the $C_\ell$ spectra.

Moreover, and thanks to the large number of mocks constructed, we 
measured the full covariance matrix with high precision in different
configurations. We find that at least $150-200$ mocks are necessary for a well
defined reduced covariance, but this is discussed more properly
in \pcite{cabre10}. Remarkably in all cases tested, the modeling recovers very
accurately the true error matrix.

In a parallel line of research we have tested the recovery of
cosmological parameters using our model for $w(\theta)$ and the
theoretical expression of the covariance matrix. And compared this
with the same analysis but using the {\it true covariance} as measured in the mock
ensembles. Indeed, the best-fit values and errors contours coincide for
both approaches \cite{cabre10}
giving very encouraging prospects for the use of
our analytical expressions in real data analysis or in realistic
forecasts of upcoming photometric surveys. 


\section*{Acknowledgments}
We would like to thank the Large Scale Structure
Working group of the Dark Energy Survey for motivating the core of this
work. We are particularly thankful to Ashley Ross and Will Percival for stimulating
discussions regarding redshift distortions, Pablo Fosalba for
sharing  his experience in angular statistics and Jorge Carretero for
technical support with the simulation. Funding for this project was partially provided by the Spanish Ministerio de Ciencia e Innovacion (MICINN),
projects 200850I176, 
AYA2009- 13936-C06
Consolider-Ingenio CSD2007-00060, research project 2005SGR00728
from Generalitat de Catalunya and the Juan de la Cierva MEC
program. The mock catalogues essential to this paper were possible thanks to an
N-body simulation provided by the MICE collaboration.

The ensemble of mock redshift bins taking into account photo-z effects
and redshift distortions, as described in Table~\ref{Table:mocks}, will be
publicly available at {\tt http://www.ice.cat/mice}.

\appendix

\section{Spatial clustering}
\label{Appendix:A}

In Sec.~\ref{sec:nonlinear_matter_clustering} we presented a parametric model for the spatial correlation
function and argued that a single set of best-fit parameters could be
used to describe the clustering from low to high redshift.
In this appendix we show how this model performs against measurements of 3-d clustering in N-body
simulations, in particular {\it as a function of redshift}.

\subsubsection{Nonlinear Gravitational Clustering}

\begin{figure}
\begin{center}
\includegraphics[width=0.45\textwidth]{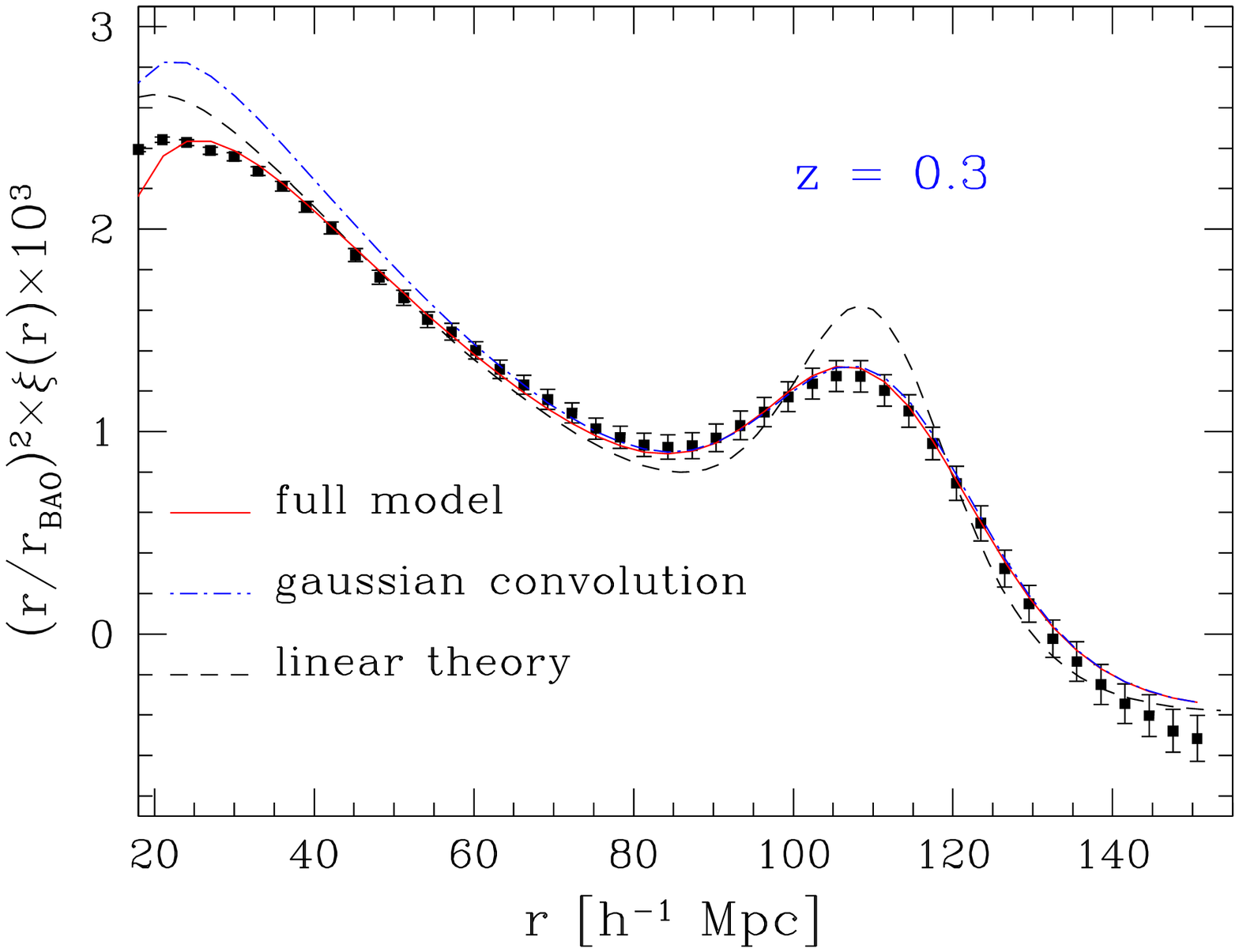} \\
\includegraphics[width=0.45\textwidth]{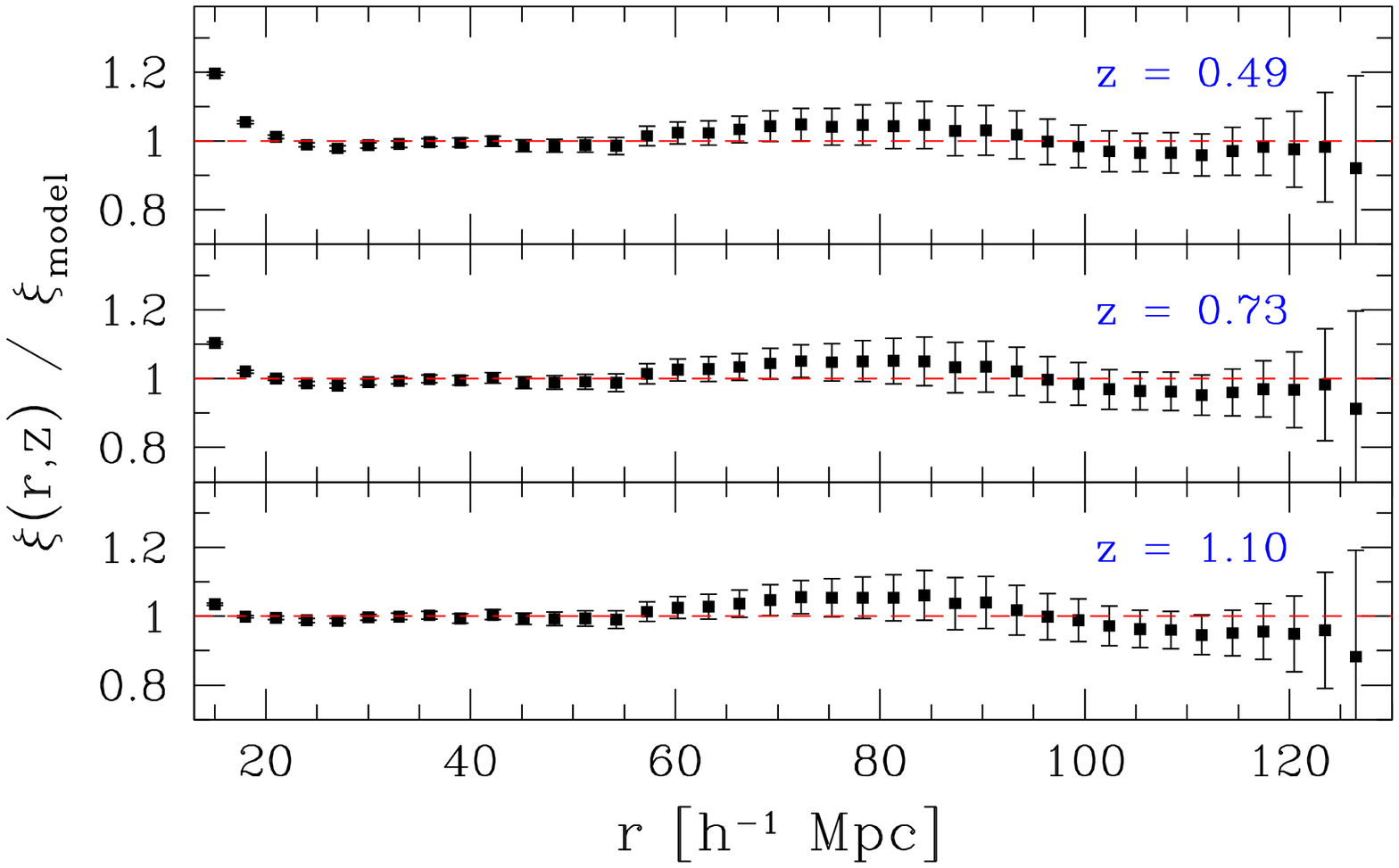}
\caption{{\it Spatial Correlation Function:} measured in the $z=0.3$
  comoving output of a large volume N-body run (MICE7680) compared with the parametric model given
in Eq.~(\ref{eq:parametric_model}) with best-fit parameters
$s_{bao}=5.54\Mpc$ and $A_{mc}=1.55$ (Top panel). The length scale used in
the y-axis is $r_{BAO}=110 \Mpc$. Bottom panels display the
ratio of the measured correlation function (at the given redshift)
to our model correlation, obtained after scaling
with redshift the $z=0.3$ best-fit values.} 
\label{fig:xi_z0p3}
\end{center}\end{figure}

The top panel of Fig.~\ref{fig:xi_z0p3} shows the spatial
correlation function measured in MICE7680 at $z=0.3$ compared with the parametric model given in Eq.~(\ref{eq:xiz}) for
best-fit parameters $s_{bao}=5.54 \Mpc$ and $A_{mc}=1.55$. To
simplify the measurement we only 
used a cubic sub-volume of the full comoving output, of length ${\rm L}_{\rm
box}=2560\Mpc$. Error bars were obtained from the scatter among $125$
Jack-knife volumes of the full box. We see that the model performs
very well down to scales $\sim 20\Mpc$ at this redshift. Notice how the convolution of
linear theory with a Gaussian smoothing (dot dashed line) leads to a slight increase in amplitude
above the measurements for $r \simlt 60\Mpc$. Measurements follow linear theory at these
scales. The effect of the mode-coupling term in Eqs.~(\ref{eq:parametric_model},\ref{eq:xiz}) is mainly to correct for
this mismatch.

From theoretical grounds we expect the smoothing length to be determined
by the amplitude of large-scale velocity flows and therefore given by \cite{bharadwaj96,crocce06,eisenstein07,matsubara08},
\beq
s_{bao} \sim [(1/3) \int d^3q P_{\rm Lin}(q)/q^2]^{1/2} = 5.42 \, {\rm
  Mpc}\,{\it h}^{-1} 
\label{eq:sbao}
\eeq
which is within $3\%$ agreement with the recovered best-fit value. In
turn, the lowest order estimation for $A_{mc}$ using perturbation
theory yields $34/21 \sim 1.62$ independently of cosmology and
redshift \cite{crocce08}, which is
also very close to our best-fit value $1.55$. 

Hence, it is possible to use these theoretical estimates as a
starting point in the modeling instead of using fitting parameters.
Notice however that, if one extrapolates this model to describe the
clustering of tracers, the actual values for $s_{bao}$ and $A_{mc}$ might
 depend on the particular tracer under study (e.g. on halo mass) \cite{sanchez08}. 
In addition, the rather large error bars obtained from present day
data do not put severe constraints on the values of $s_{bao}$ and
$A_{mc}$ \cite{sanchez09,percival10}.

As discussed in Sec.~\ref{sec:nonlinear_matter_clustering} one interesting aspect of a theoretical model when it comes
to analyze data split into continuous redshift bins is the possibility
to understand the redshift evolution of the nuisance parameters
involved, which in turn allows one to fit the least number of
parameters (i.e. just at one fiducial redshift). With this approach in
mind we now show that our model can indeed describe the 3-d clustering
from low to high redshift using a single set of best-fit
parameters. And this property will remain 
after the projection into angular correlations in Eq.~(\ref{eq:wtheta}).

The bottom panel of Fig. ~\ref{fig:xi_z0p3} shows our model in
Eq.~(\ref{eq:xiz}) against measurements of $\xi$ in the comoving
outputs of MICE7680 at $z=0.5,\,0.73$ and $1.1$. 
The values for $s_{bao}$ and $A_{mc}$ were taken from a best-fit
analysis to $\xi(r)$ at $z=0.3$. In all cases the agreement is very
good, similar to that at $z=0.3$.

Notice that conclusion in this appendix relate to 3-d clustering
in general, regardless of photo-z, hence are
also relevant for spectroscopic surveys such as BOSS \footnote{\tt http://www.sdss3.org/}, Hetdex
\footnote{\tt http://hetdex.org/} and WiggleZ \footnote{\tt http://wigglez.swin.edu.au/}.

\section{The Limber approximation}
\label{Appendix:B}

\subsection{$C_\ell$ power spectrum : exact evaluation and Limber formula}
\label{Appendix:B1}

In this appendix we discuss the way we implement the numerical
integrations that lead to the exact $C_{\ell}$ spectra in
Eqs.~(\ref{eq:cl},\ref{eq:psi}), and how 
this exact result compares with the widely used Limber approximation \cite{limber1953}.

Let us first recall the derivation of the exact expression of the angular power spectrum in
terms of the spatial one. This is done by
expanding the density field in Eq.~(\ref{eq:projected-density}) in Fourier Series and
subsequently expanding the plane wave into spherical harmonics
\footnote{$e^{i k r \hat{\bf k}\cdot \hat{\bf n}}=4\pi \sum_{\ell \ge
    0}\sum_{m=-\ell}^{\ell} i^{\ell} j_{\ell}(k r) Y_{\ell m}(\hat{\bf
    k})Y^{\star}_{\ell m}(\hat{\bf n})$}. After some straightforward
manipulation this leads to,
\beq
a_{\ell m}= 4 \pi i^{\ell} \int dz \, \phi(z) \int \frac{d^3k}{(2\pi)^3}
\delta(\vk,z) j_{\ell}(k\,r(z)) Y^{\star}_{\ell m}(\hat{\bf k}),
\label{eq:alm}
\eeq
where $j_{\ell}$ are the spherical Bessel functions of order
${\ell}$. Inserting Eq.~(\ref{eq:alm}) into Eq.~(\ref{eq:Clspectra})
and using the orthogonality relation of the spherical harmonics gives Eqs.~(\ref{eq:cl},\ref{eq:psi}),
\begin{eqnarray}
\indent \indent  C_{\ell, {\rm Exact}} &=& \frac{1}{2\pi^2} \int 4 \pi k^2 dk P(k)
\Psi^2_{\ell}(k) \nonumber \\
\Psi_{\ell}(k) &=& \int dz \phi(z) D(z) j_{\ell}(k r(z)) \nonumber
\end{eqnarray}

\begin{figure}
\begin{center}
\includegraphics[width=0.35\textwidth]{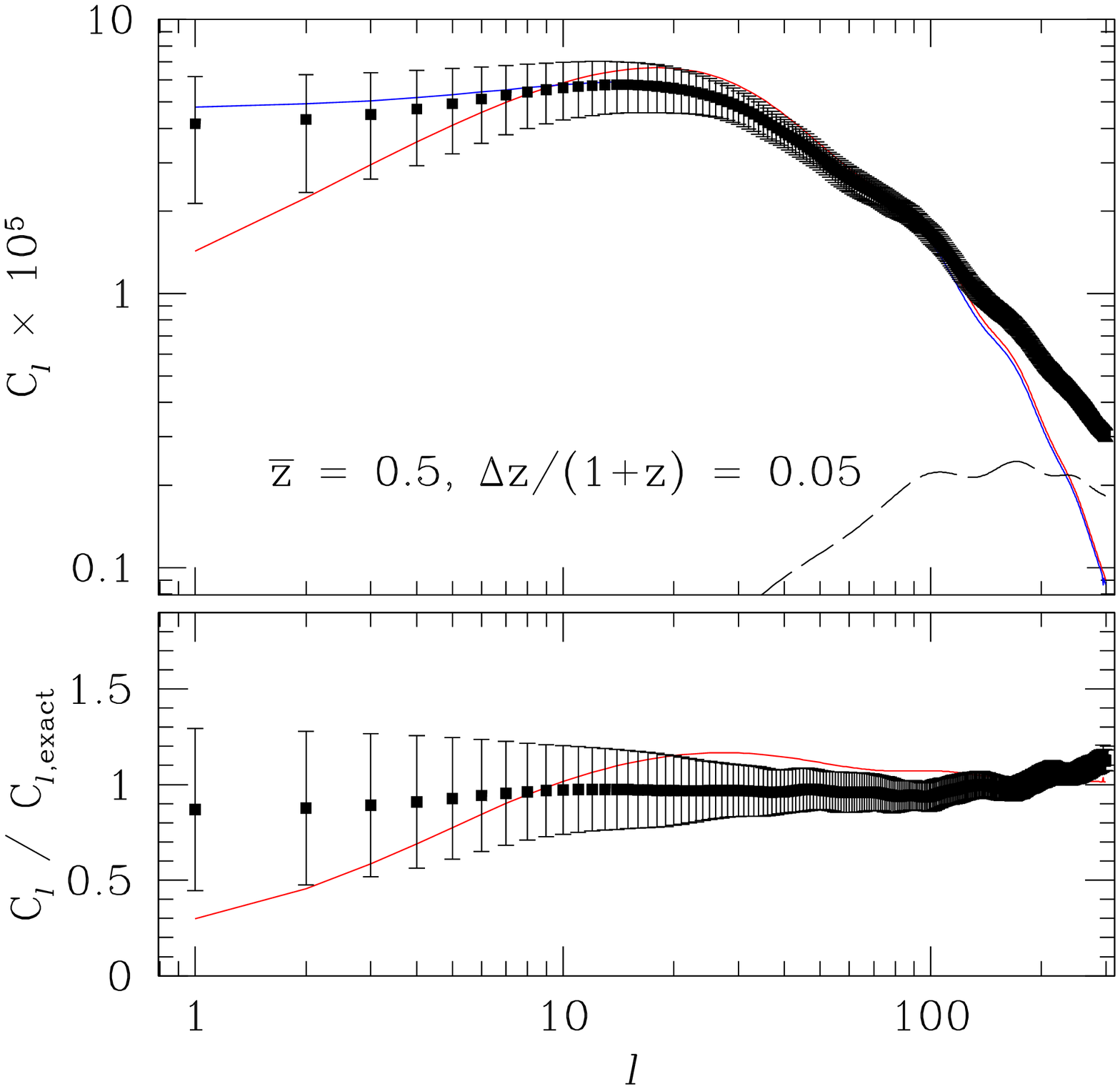}
\caption{Top Panel: Angular Power Spectrum measured in $324$ {\it Real
  Space} mock
  catalogues of mean redshift $\bar{z}=0.5$ and width $\Delta
  z/(1+z)=0.05$, compared with our theoretical
  estimate, which we brake into to additive contributions (Eq.~\ref{eq:clsplit}), low-$\ell$
  from the exact integration in Eq.~(\ref{eq:cl}) (solid blue line) and large-$\ell$ from the
  Limber approximation in Eq.~(\ref{eq:Cl-Limber}) (dashed black
  line). Solid red line shows the low-$\ell$ term computed with the Limber
  approximation, which leads to a severe
  under-estimation of the large angle power ($l \le 30$). This could
  critically impact the computation of $w(\theta)$ and its covariance 
  at large BAO scales. Bottom Panel: Ratio of the full
  $C_{\ell}$-Limber (red line) and the measurements to our estimation of the exact
(linear theory)  $C_{\ell}$ spectra. Displayed error bars correspond
to the ensemble r.m.s variance.}
\label{fig:Cl}
\end{center}\end{figure}

These integral expressions  are numerically
expensive to compute due to the oscillatory behavior of the spherical
Bessel functions $j_l(x)$ for $x \gg 1$. It is then desirable to seek
for ways to improve their convergence. The most popular short-cut to evaluate Eqs.~(\ref{eq:cl},\ref{eq:psi})
is the so-called Limber approximation \cite{limber1953,kaiser1992,kaiser1998}
that follows from the orthogonality relation of the spherical Bessel
functions (e.g. see \pcite{loverde08}),
\begin{eqnarray}
&& \int dk k^2 j_l(k r_1) j_l(k r_2) P(k) \\ && \approx \frac{\pi}{2} \frac{\delta^{\rm D}(r_1-r_2)}{r_1^2} P(k = \frac{l+1/2}{r}), \nonumber
\end{eqnarray}
that leads to the well-known expression,
\beq
C_{\ell,{\rm Limber}} = \int dz \phi^2(z) D^2(z) P(({\ell}+1/2)/r(z)) \frac{AM(z)}{r(z)^2}. \\
\label{eq:Cl-Limber}
\eeq
Strictly speaking this is valid for $\ell \gg 1$ (that is, small
angles), but in practice it coincides with the exact expression
already by $\ell \sim 50-100$ as discussed in Fig.~\ref{fig:Cl} (see
also \pcite{blake07,loverde08}). Therefore we proceed
as follows. First, let us concentrate on linear theory fluctuations.
Then, inspired by Eq.~(\ref{eq:parametric_model})
(and \pcite{crocce06,eisenstein07,matsubara08}) we decompose the {\it linear} power spectrum as
\beq
P(k) = P_{\rm Lin}(k) G(k) +  P_{\rm Lin}(k) (1-G(k)),
\label{eq:powersplit}
\eeq
with $G=\exp(-k^2 s_{bao}^2)$ and $s_{bao}$ taken from Eq.~(\ref{eq:sbao}). The first term dominates at large scales and accounts for the
degradation of BAO bump, leading also to the correct shape of the 3-d correlation
function in a broad range of scales (Fig~\ref{fig:xi_z0p3}). 
Since this is our regime of interest we
evaluate this piece using the exact integration, that can now be integrated
up to an upper bound such that $k_{\rm max} \, s_{bao} \sim 4-5$ where
$P_{\rm Lin}\exp(-k^2 s_{bao}^2)$ is already suppressed by several orders of magnitude.
The second term in Eq.~(\ref{eq:powersplit}) is non-negligible only at 
high-${\ell}$ therefore it can be computed using the Limber formula,
Eq.~(\ref{eq:Cl-Limber}). 
In all, we evaluate the exact $C_{\ell}$ as \footnote{The difference
  between our approach and a straightforward use of the exact
  integration for all the
  ${\ell}$ range is under $1\%$.}
\beq
C_{\ell} = C_{\ell,{\rm Exact}}(P_{\rm Lin} \times G) + C_{\ell, {\rm
    Limber}} (P_{\rm Lin}\times(1-G))
\label{eq:clsplit}
\eeq

\begin{figure}
\begin{center}
\includegraphics[width=0.35\textwidth]{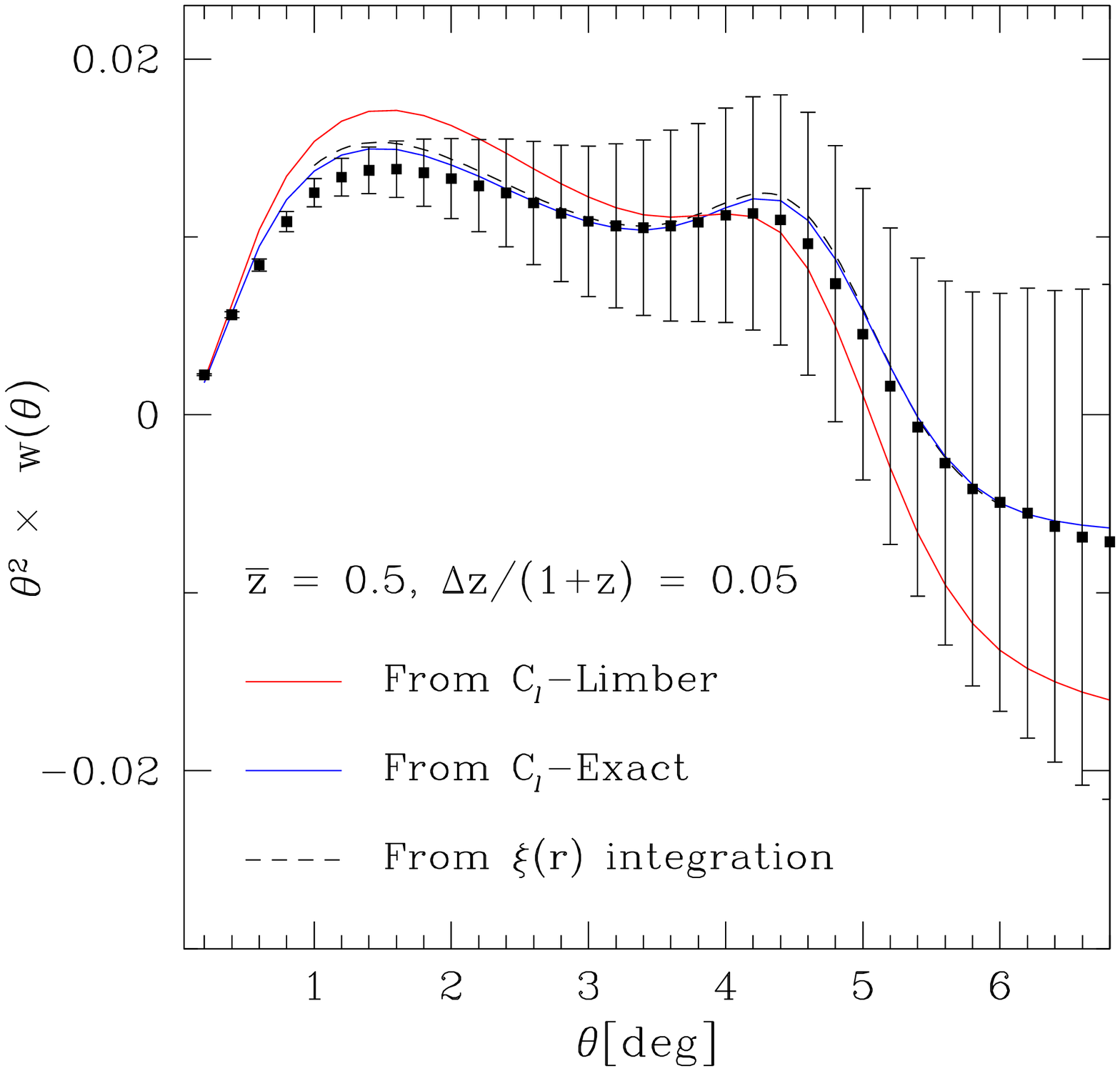}
\caption{{\it Angular Correlation Function} measured in the $324$ mock bins compared with the theoretical predictions  obtained from Eq.~(\ref{eq:wtheta}) (dashed line), or from $C_{\ell}$-space through the Legendre polynomials and Eq.~(\ref{eq:LegendreT}). Using the Limber approximation to compute the $C_{\ell}$ leads an incorrect shape for $w(\theta)$ (solid red line). In turn, the exact $C_{\ell}$ integration agrees as expected with Eq.~(\ref{eq:wtheta}) (solid blue line). Displayed error bars correspond
to the ensemble r.m.s variance.}
\label{fig:wtheta}
\end{center}\end{figure}

In the top panel of Fig.~\ref{fig:Cl} we show
the average $C_{\ell}$
power spectrum measured in
$324$ {\it real space} mocks centered at $\bar{z}=0.5$ with $\Delta
z/(1+z)=0.05$ (using SpICE, see \pcite{szapudi01}), and compare it with the exact integration (solid blue line) and
the Limber evaluation (solid red line) of the first term of Eq.~(\ref{eq:powersplit}).
Clearly, the Limber approximation fails by a large factor
at low-$l$, which is in broad agreement 
with the literature \cite{blake07,loverde08}. However, by $\ell \sim
60$ it fully coincides with the exact computation.
 The contribution of the second term in
Eq.~(\ref{eq:powersplit}) is only relevant for $\ell > 100$ (dashed
black line) and therefore it is safety computed using the Limber
approximation. We have checked that these conclusions remain true for
different $\bar{z}$ and $\Delta z$. Bottom panel of Fig. ~\ref{fig:Cl}
shows the full $C_{\ell}$-Limber (red plus dashed lines in top panel) and the measured spectra in ratio with our model for the
exact $C_{\ell}$ spectra, given by Eq.~(\ref{eq:clsplit}) (i.e. blue
plus dashed in top panel).

Notice that we have only attempted to evaluate the ${\it linear}$
spectrum. Nonlinear effects can be taken into account by replacing
$P_{\rm Lin}$ in Eq.~(\ref{eq:powersplit}) by fits to nonlinear power spectrum, such as {\tt
  halofit}, or exact analytical expressions
\cite{RPT,matarrese08,matsubara08,taruya09}. 
However we have found that for our purposes this gave no
additional contributions to the error estimations. In fact, the second term in Eq.~(\ref{eq:clsplit}) had a negligible impact in
the estimate of $\Delta w(\theta)$  for all cases we had
considered. It only gave a contribution when evaluating the covariance
matrix in cases where the shot-noise of the sample under consideration
is fully negligible (see Sec.~\ref{sec:impact_shot_noise}). We finish
by noting that linear redshift space distortions fully cancel for
large $\ell$ \cite{padmanabhan07},
therefore they only need to be accounted for in the first term of
Eq.~(\ref{eq:clsplit}) (following Eq.~\ref{eq:psir}). 

\subsection{Correlation Function : results in Real and Redshift Space}
\label{Appendix:B2}

The inaccuracy of the Limber approximation  also affects
the calculations in configuration space.
For instance, the way in which the disagreement in Fig~\ref{fig:Cl}
translates into Real Configuration Space is depicted in
Fig.~\ref{fig:wtheta}. Solid red and solid blue lines were obtained
using Eq.~(\ref{eq:LegendreT}) and $C_{\ell}$ from the Limber or the
Exact evaluation respectively (those shown in top panel of
Fig.~\ref{fig:Cl}). 
Dashed line follows from projecting the corresponding 3-d correlation function, as
in Eq.~(\ref{eq:wtheta}). Here, the failure of the Limber approximation to describe the BAO bump is more evident. The exact $C_{\ell}$ integration yields the same result as the one from Eq.~(\ref{eq:wtheta}), as expected.

These limitations are even more severe when redshift
distortions are taken into account.
Without lost of generality we will use a top-hat selection in the
following equations to illustrate the problem:
\beq
w(\theta) =  {1\over{\Delta_\chi^2}} \int_{\chi_{min}}^{\chi_{max}} d\chi_1
\int_{\chi_{min}}^{\chi_{max}} d\chi_2 ~\xi(r_1,r_2) 
\label{eq:ch1ch2}
\eeq
where $\Delta_\chi \equiv \chi_{max}-\chi_{min}$.
The correlation $\xi(r_1,r_2)$ is only a function of the relative
separation between $r_1$ and $r_2$.  Because of redshift distortions, 
this is in fact a function of $\pi$ and $\sigma$, the
 light-of-sight and transverse separation.  For small angles and
distance observer, i.e. the Limber approximation, we can take 
$\pi =\chi_2-\chi_1$ and 
$\sigma^2 = \chi_1\chi_2 \theta^2$ 
\footnote{For a non-flat cosmology
we need to replace $\chi$ by the comoving
angular diameter $D(\chi)$ distance here: $\sigma^2 = D(\chi_1) D(\chi_2) \theta^2$ 
}. We can then change variables in the above integrals
 from $\chi_1$ and $\chi_2$ to $\pi$ and $\sigma$
\beq
w(\theta) =  {2\over{\theta \Delta_\chi^2}} 
\int_{\sigma=\chi_{min}\theta}^{\sigma=\chi_{max}\theta} d\sigma
\int_{\pi=0}^{\pi=\Delta_\chi} d\pi ~\xips 
\label{eq:wt}
\eeq

This result is valid both in real and redshift space.
In real space, $\xips$ is replaced by the isotropic 
correlation $\xi(r)$   with $r^2=\pi^2+\sigma^2$.
In redshift space  we use the linear theory prediction 
(\pcite{kaiser1987,hamilton1992}):

\beq
\xi(\sigma,\pi)=\xi_0(s)P_0(\mu) + \xi_2(s)P_2(\mu)+\xi_4(s) P_4(\mu)
\label{eq:xisigpi}
\eeq
where $\pi$ and $\sigma$ represent the separation along and transverse to the
line-of-sight (l.o.s) and $\xi_\ell$ are the multi-poles of the
correlation function in terms  of Legendre polynomials $P_\ell$,
\beq
\xi_\ell(s) = \frac{2\ell+1}{2}\int_{-1}^{+1}\xi(\pi,\sigma)P_{\ell}(\mu)d\mu, 
\label{eq:xisigmapi}
\eeq
with $s=\sqrt{\sigma^2+\pi^2}$ and $\mu$ is the cosine angle with
the l.o.s. For the Kaiser model one has \cite{hamilton1992},
\begin{eqnarray}
\indent
\xi_0(s)&=&b^2\left(1+\frac{2\beta}{3}+\frac{\beta^2}{5}\right)
\xi(s) \label{eq:xi0} \\
\xi_2(s)&=&b^2\left(\frac{4\beta}{3}+\frac{4\beta^2}{7}\right)
\left[\xi(s)-\bar{\xi(s)}\right] \label{eq:xi2} \\
\xi_4(s)&=&b^2\frac{8\beta^2}{35}\left[\xi(s)+\frac{5}{2}\bar{\xi}(s)-\frac{7}{2}\bar{\bar{\xi}}(s)\right]
\label{eq:xi4}
\end{eqnarray}
where $b$ is the bias of the sample (assumed linear and local), $\beta
= f/b$, $f=\partial \ln D/\partial \ln a$ is the growth rate factor and
\begin{eqnarray}
\indent \indent \bar{\xi(r)}&=&\frac{3}{r^3}\int_0^r\xi(r^\prime){r^\prime}^2 dr^\prime , \\
\bar{\bar{\xi(r)}}&=&\frac{5}{r^5}\int_0^r\xi(r^\prime){r^\prime}^4 dr^\prime.
\end{eqnarray}

We will show next that Eq.~(\ref{eq:wt}) turns out to be 
a bad approximation to model BAO scales. But it provides
a good way to illustrate why redshift space distortions
are so important for the BAO detection for small (to moderate)
photo-z bin widths $\Delta_\chi$. For large values of $\Delta_\chi$ the 
integral reproduces the real space correlation (because the total
number of pairs are preserved by redshift space distortions), while
for $\Delta_\chi<500 \Mpc$ the $\pi$ integral is truncated
by the radial boundary of the top-hat window. Line-of-sight
pairs separated by $\pi>\Delta_\chi$ do not enter in the redshift
bin and are therefore not integrated. Smaller pairs are also affected
because  many of them are missing at the boundaries.
This missing pairs produce 
a distortion in the measured $w(\theta)$ as compared to real 
space correlation (see \pcite{fisher1994}, \pcite{padmanabhan07} and \pcite{nock10}).

\begin{figure}
\begin{center}
\includegraphics[width=0.35\textwidth]{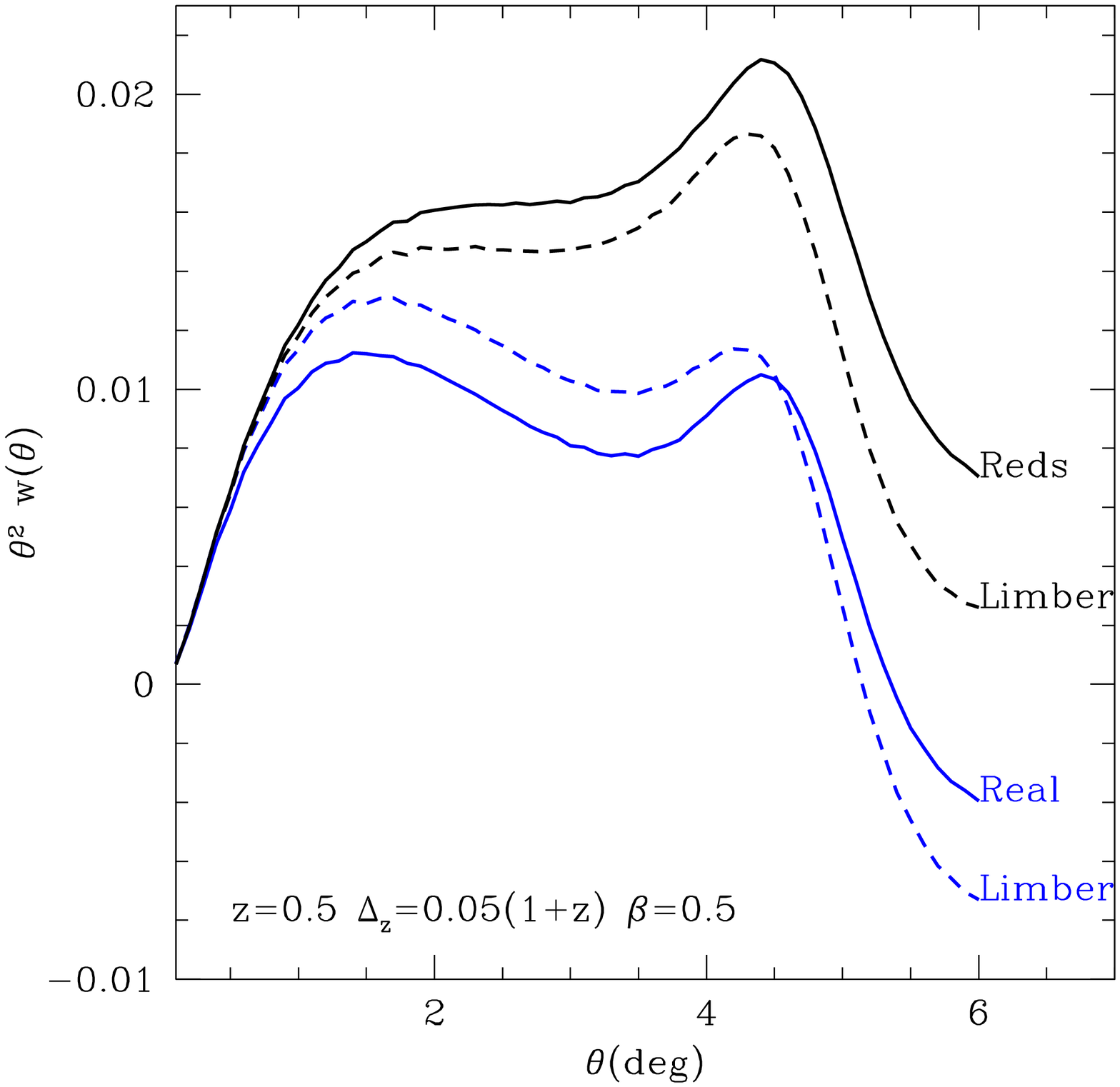}
\caption{
Angular correlation in the Limber approximation (dashed line)
in Eq.~(\ref{eq:wt}) in both real (bottom dashed line) and redshift space
(top dashed line). Continuous lines show the corresponding exact
calculations using Eq.~(\ref{eq:ch1ch2}).
}
\label{fig:w2Limber}
\end{center}\end{figure}

Both real and redshift results are quite inaccurate under the
Limber approximations. This is illustrated in 
Fig.~\ref{fig:w2Limber}, which 
compares the calculation of
the above integral in real and redshift space in the Limber 
approximation (dashed lines) with the corresponding exact results (continuous line)
by integrating Eq.~(\ref{eq:ch1ch2}). Note how the 
results in real space are very similar to the corresponding calculation in Fig.~\ref{fig:wtheta},
 based on power spectrum calculation. The results in redshift space
for the Limber approximations show even larger deviations than in real
space. The BAO peak, which shows around $\theta \simeq 4^{\circ}$ 
is clearly distorted by the Limber approximation. We conclude that the
Limber approximation is not good enough for precision BAO modeling
and we use the exact integration throughout this paper.

\section{Errors in Harmonic Space}
\label{Appendix:C}

In this appendix we comment briefly on the correlation between different
modes in Harmonic Space induced by a partial sky coverage. 

Equation (\ref{eq:CovW}) can be naively interpreted as if, in the
presence of a partial sky coverage, the errors in $C_{\ell}$ increase
by a factor $1/\sqrt{f_{sky}}$, that is
\beq
(\Delta C_\ell)^2\approx \frac{1}{f_{sky}}\frac{2}{2\ell+1}C^2_\ell,
\label{eq:naive}
\eeq
with the co-variance matrix remaining diagonal.
In reality, the presence of boundaries
{\it decreases} the diagonal error but it also introduces co-variance between
different ${\ell}$ modes (e.g. \pcite{cabre07} and references therein).

\begin{figure}
\begin{center}
\includegraphics[width=0.4\textwidth]{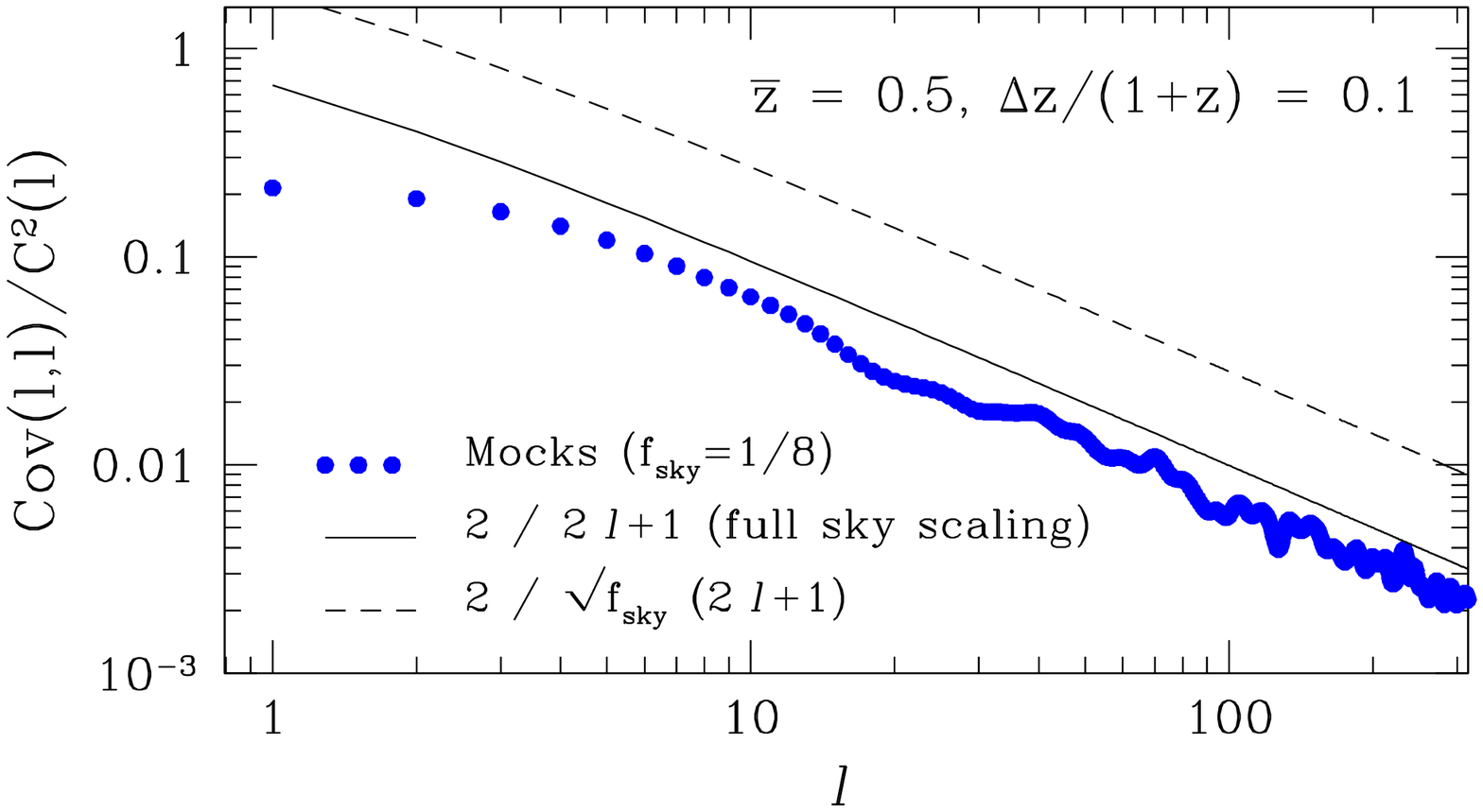}
\includegraphics[width=0.4\textwidth]{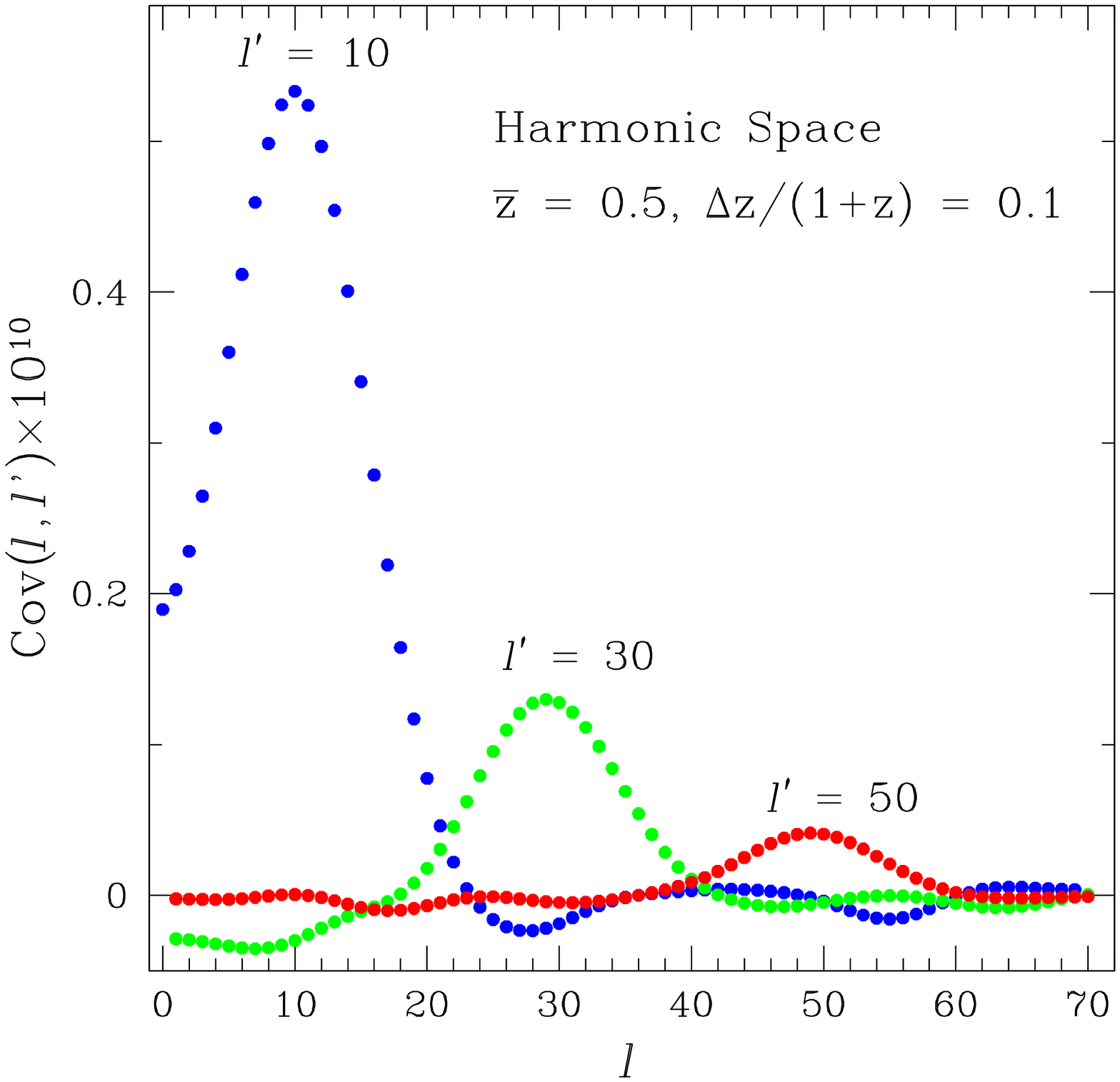}
\caption{{\it Correlations of Cl spectra induced by partial sky
    coverage}. The {\it total} error budget in a partial sky survey
  increases (roughly by a factor
  $1/f_{sky}$) compared to the full sky case. The covariance matrix
  is no longer diagonal (bottom panel) with the variance error smaller
  than its full sky value, by $30\%$ in this particular case, as shown
  in the top panel (see text
for details).}
\label{fig:Clcovariance}
\end{center}\end{figure}

This is clearly depicted in Figure \ref{fig:Clcovariance} where we show the error in
Harmonic space from $C_{\ell}$ measurements using $392$ mocks with
${\bar z}=0.5$ and $\Delta z/(1+z)=0.1$, and compared them with the expression above.
Top panel corresponds to the measured diagonal error, which indeed is much
smaller than that from the naive interpretation in
Eq.~(\ref{eq:naive}) given by the dashed line (even smaller than the full-sky limit shown in
solid line). Bottom panel, corresponding to ${\rm
  Cov}(\ell,\ell^{\prime})$ for fixed values of $\ell^{\prime}$, shows the emergence
of co-variance between different modes distributed in a range of
${\ell}$ values approximately given by $\pm 1/f_{sky}$.
In all, the ``total error'' is larger than in full-sky. 

This conclusion can be naively founded from the fact that
for a given value of ${\ell^{\prime}}$ the integral,
\beq
\int d\ell \, {\rm Cov}(\ell^{\prime},\ell) \approx
\frac{2}{f_{sky}(2\ell^{\prime}+1)} C^2_{\ell^{\prime}}
\eeq
what leads to the simple interpretation that, when the survey is
reduced from a full-sky limit by a fraction $f_{sky}$, the diagonal
covariance rises from its value $2/(2l+1)$ by a factor $1/f_{sky}$ but then
``leaks'' towards other ${\ell}$ modes resulting in a non-diagonal
error matrix. The final diagonal error is smaller than its ``full-sky''
value 
but is the off-diagonal elements of ${\rm Cov}(\ell,\ell^{\prime})$
what determine ${\rm Cov}(\theta,\theta^{\prime})$. Yet, we circumvent
the problem of computing non-diagonal components of ${\rm
  Cov}(\ell,\ell^{\prime})$ by assuming the scaling of the covariance
in Configuration space as $1/f_{sky}$.

A standard way to overcome the complex covariance in Harmonic Space is to bin
the measured $C_{\ell}$ spectra in such a way to make the covariance
matrix block-diagonal (e.g. \pcite{cabre07}). A simple rule of thumb discussed
in \pcite{cabre07} is to choose $\Delta \ell \, f_{sky}\sim 2$, which is
in very nice agreement with the width of the ${\rm Cov}(\ell,\ell^\prime)$ distribution in Fig.~\ref{fig:Clcovariance}.
 It remains to be studied whether
this have an impact in methods like BAO where one is after short-wavelength features on top of the
broad band $C_\ell$ shape.

\end{document}